\documentclass[twocolumn, tighten]{aastex7}

\newcommand\tgta{J0305m3150.C05}
\newcommand\tgtb{J1526m2050.C02}

\usepackage{times}

\usepackage{savesym}
\savesymbol{tablenum}
\usepackage{siunitx}
\restoresymbol{SIX}{tablenum}
\usepackage{enumitem}

\newcommand{\oiii}{\mbox{[\ion{O}{3}]}}

\newcommand{\ci}{\mbox{[\ion{C}{1}]}}
\newcommand{\cii}{\mbox{[\ion{C}{2}]}}

\newcommand\msun{\mbox{\si{M_\odot}}}
\newcommand\lsun{\mbox{\si{L_\odot}}}
\newcommand\smpy{\mbox{\si{M_\odot.yr^{-1}}}}

\shorttitle{NIRCam-Dark Starburst Galaxies at $z=6.6$}

\received{\today}
\submitjournal{ApJ}

\begin{document}

\title{
The Identification of Two JWST/NIRCam-Dark Starburst Galaxies at $z=6.6$ with ALMA
}

%

\correspondingauthor{Fengwu Sun}
\email{fengwu.sun@cfa.harvard.edu}


\author[0000-0002-4622-6617]{Fengwu Sun}
\affiliation{Center for Astrophysics $|$ Harvard \& Smithsonian, 60 Garden St., Cambridge, MA 02138, USA}
\email{fengwu.sun@cfa.harvard.edu}  

\author[0000-0001-5287-4242]{Jinyi Yang}
\affiliation{Department of Astronomy, University of Michigan, 1085 S. University Ave., Ann Arbor, MI 48109, USA}
\email{yjykaren@gmail.com}  

\author[0000-0002-7633-431X]{Feige Wang}
\affiliation{Department of Astronomy, University of Michigan, 1085 S. University Ave., Ann Arbor, MI 48109, USA}
\email{fgwang.astro@gmail.com}

\author[0000-0002-2929-3121]{Daniel J.\ Eisenstein}
\affiliation{Center for Astrophysics $|$ Harvard \& Smithsonian, 60 Garden St., Cambridge, MA 02138, USA}
\email{deisenstein@cfa.harvard.edu}

\author[0000-0002-2662-8803]{Roberto  Decarli}
\affiliation{INAF–Osservatorio di Astrofisica e Scienza dello Spazio, via Gobetti 93/3, I-40129, Bologna, Italy}
\email{roberto.decarli@inaf.it}

\author[0000-0003-3310-0131]{Xiaohui Fan}
\affiliation{Steward Observatory, University of Arizona, 933 N Cherry Avenue, Tucson, AZ 85721, USA}
\email{xiaohuidominicfan@gmail.com}

\author[0000-0003-2303-6519]{George H. Rieke}
\affiliation{Steward Observatory, University of Arizona, 933 N Cherry Avenue, Tucson, AZ 85721, USA}
\email{grieke@arizona.edu}

\author[0000-0002-2931-7824]{Eduardo Ba{\~n}ados}
\affiliation{Max-Planck-Institut f\"{u}r Astronomie, K\"{o}nigstuhl 17, 69117 Heidelberg, Germany}
\email{banados@mpia.de}

\author[0000-0001-8582-7012]{Sarah E.~I.~Bosman}
\affiliation{Institute for Theoretical Physics, Heidelberg University, Philosophenweg 12, D–69120, Heidelberg, Germany}
\affiliation{Max-Planck-Institut f\"{u}r Astronomie, K\"{o}nigstuhl 17, 69117 Heidelberg, Germany}
\email{bosman@thphys.uni-heidelberg.de}

\author[0000-0001-8467-6478]{Zheng Cai}
\affiliation{Department of Astronomy, Tsinghua University, Beijing 100084, China}
\email{zcai@mail.tsinghua.edu.cn}

\author[0000-0002-6184-9097]{Jaclyn B. Champagne}
\affiliation{Steward Observatory, University of Arizona, 933 N Cherry Avenue, Tucson, AZ 85721, USA}
\email{jbchampagne@arizona.edu}

\author[0000-0002-9090-4227]{Luis Colina}
\affiliation{Centro de Astrobiolog\'{\i}a (CAB), CSIC-INTA, Ctra. de Ajalvir km 4, Torrej\'on de Ardoz, E-28850, Madrid, Spain }
\email{colina@cab.inta-csic.es}

\author[0000-0003-2388-8172]{Francesco D'Eugenio}
\affiliation{Kavli Institute for Cosmology, University of Cambridge, Madingley Road, Cambridge, CB3 0HA, United Kingdom}
\affiliation{Cavendish Laboratory - Astrophysics Group, University of Cambridge, 19 JJ Thomson Avenue, Cambridge, CB3 0HE, United Kingdom}
\email{francesco.deugenio@gmail.com}

\author[0000-0001-7440-8832]{Yoshinobu Fudamoto}
\affiliation{Center for Frontier Science, Chiba University, 1-33 Yayoi-cho, Inage-ku, Chiba 263-8522, Japan}
\affiliation{Steward Observatory, University of Arizona, 933 N Cherry Avenue, Tucson, AZ 85721, USA}
\email{yoshinobu.fudamoto@gmail.com}

\author[0000-0001-6251-649X]{Mingyu Li} 
\affiliation{Department of Astronomy, Tsinghua University, Beijing 100084, China} 
\email{lmy22@mails.tsinghua.edu.cn}

\author[0000-0001-6052-4234]{Xiaojing Lin}
\affiliation{Department of Astronomy, Tsinghua University, Beijing 100084, China}
\affiliation{Steward Observatory, University of Arizona, 933 N Cherry Avenue, Tucson, AZ 85721, USA}
\email{xiaojinglin@arizona.edu}  

\author[0000-0003-3762-7344]{Weizhe Liu} 
\affiliation{Steward Observatory, University of Arizona, 933 N Cherry Avenue, Tucson, AZ 85721, USA}
\email{oscarlwz@gmail.com}  

\author[0000-0002-6221-1829]{Jianwei Lyu} 
\affiliation{Steward Observatory, University of Arizona, 933 N Cherry Avenue, Tucson, AZ 85721, USA}
\email{jianwei@arizona.edu}

\author[0000-0002-5941-5214]{Chiara Mazzucchelli}
\affiliation{Instituto de Estudios Astrof\'{\i}sicos, Facultad de Ingenier\'{\i}a y Ciencias, Universidad Diego Portales, Avenida Ejercito Libertador 441, Santiago, Chile}
\email{chiara.mazzucchelli@mail.udp.cl}

\author[0000-0002-5768-738X]{Xiangyu Jin}
\affiliation{Steward Observatory, University of Arizona, 933 N Cherry Avenue, Tucson, AZ 85721, USA}
\email{xiangyujin@arizona.edu}

\author[0000-0003-1470-5901]{Hyunsung D. Jun}
\affiliation{Department of Physics, Northwestern College, 101 7th St SW, Orange City, IA 51041, USA}
\email{hyunsung.jun@gmail.com}

\author[0000-0003-0111-8249]{Yunjing Wu}
\affiliation{Department of Astronomy, Tsinghua University, Beijing 100084, China}
\email{yunjingwu@arizona.edu}

\author[0000-0002-0123-9246]{Huanian Zhang} 
\affiliation{Department of Astronomy, Huazhong University of Science and Technology, Wuhan, Hubei 430074, China}
\email{huanian@hust.edu.cn}

\begin{abstract}

We analyze two dusty star-forming galaxies at $z=6.6$.
These galaxies are selected from the ASPIRE survey, a JWST Cycle-1 medium and ALMA Cycle-9 large program targeting 25 quasars and their environments at $z\simeq6.5 - 6.8$.
These galaxies are identified as companions to UV-luminous quasars and robustly detected in ALMA continuum and \cii\ emission, yet they are extraordinarily faint at the NIRCam wavelengths (down to $>28.0$\,AB mag in the F356W band).
They are more obscured than galaxies like Arp220, and thus we refer to them as ``NIRCam-dark'' starburst galaxies (star formation rate $\simeq 80 - 250$\,\smpy).
Such galaxies are typically missed by (sub)-millimeter blank-field surveys.
From the star-formation history (SFH), we show that the NIRCam-dark galaxies are viable progenitors of massive quiescent galaxies at $z\gtrsim4$ and descendants of UV-luminous galaxies at $z>10$. 
Although it is hard to constrain their number density from a quasar survey, we conclude that NIRCam-dark galaxies can be as abundant as $n\sim10^{-5.5}$\,Mpc$^{-3}$ assuming a light halo occupation model.
If true, this would equal to $\sim$30\% of the number densities of both the quiescent galaxies at $z\gtrsim4$ and UV-luminous galaxies at $z>10$.
We further predict that analogs at $z\sim8$ should exist according to the SFH of early massive quiescent galaxies. 
However, they may fall below the current detection limits of wide JWST and ALMA surveys, thus remaining ``JWST-dark''.
To fully trace the evolution of massive galaxies and dust-obscured cosmic star formation at $z\gtrsim8$, wide-field JWST/NIRCam imaging and slitless spectroscopic surveys of early protoclusters are essential.

\end{abstract}


\section{Introduction}
\label{sec:01_intro}

Dust plays a pivot role in obscuring the majority of the cosmic star formation at redshift $z\simeq1-4$ (see a review by \citealt{md14}).
Since the end of the last century, dusty star-forming galaxies (DSFGs) with IR luminosities resembling those of local ultra-luminous infrared galaxies (ULIRGs; $L_\mathrm{IR}\simeq 10^{12} - 10^{13}$\,\lsun) have been found in abundance across the bulk of the cosmic history (e.g., \citealt{smail97, hughes98, vieira10, oliver12} and reviews by, e.g., \citealt{hodge20}). 
These galaxies, also  known as submillimeter galaxies (SMGs), undergo vigorous starbursts with star formation rates (SFR) greater than 100\,\smpy, while only emitting a very small fraction of their light in the rest-frame UV/optical because of the strong dust obscuration.
When they exhaust their cold molecular gas reservoirs, DSFGs are believed to evolve into massive quiescent galaxies at lower redshifts \citep[e.g.,][]{toft14}.

Among all DSFGs discovered at (sub)-millimeter wavelengths, the optically faint DSFG population has been of particular interest.
Simply placing the same DSFG from $z=0.5$ to higher redshifts, the brightness of the galaxy will drop quickly in the optical--IR, but remain almost unchanged at millimeter wavelengths because of the strong negative K correction.
Therefore, a low flux density ratio between the rest-frame optical and far-IR is indicative of a high photometric redshift.
Indeed, one of the first DSFGs discovered at 850\,\micron, HDF850.1 \citep{hughes98}, remains undetected with HST from the optical to near-IR.
The redshift of HDF850.1 is confirmed at $z=5.18$ through CO and \cii\ spectroscopy with millimeter interferometry \citep{walter12}.
After more than two decades since its discovery, the stellar component of HDF850.1 was finally detected at 1--5\,\micron\ with JWST/NIRCam \citep{sunf24, hd25}.
Similar galaxies have been referred to as ``HST-dark'', ``near-IR-dark'', ``H-dropout'', or ``H-faint'' galaxies in a plethora of literature.
These galaxies are found to have a wide range of redshift ($z \simeq 2-6$), stellar age and dust attenuation distribution, and they consist of 15-20\% of DSFGs selected through ALMA millimeter observations \citep[e.g.,][]{chenc15, simpson15, franco18, wangt19, yamaguchi19, gomez22a, fujimoto24b}.


So far, the vast majority of DSFGs selected through (sub-)millimeter surveys in blank fields are at $z\lesssim6$ \citep[e.g.,][]{simpson19, aravena20, dudzevic20, gomez22a, bingl23, fujimoto24b, long24b}.
At $z \gtrsim 6$, the Epoch of Reionization (EoR), the detections of 
DSFGs frequently rely on gravitational lensing \citep[e.g.,][]{riechers13, watson15, strandet17, zavala18, laporte21} and pointed ALMA observations of known luminous galaxies and quasars \citep[e.g.,][]{decarli17, mazzucchelli19, venemans19, hashimoto19, tamura19, harikane20, inami22, hygate23}, especially through the Cycle-7 ALMA large program REBELS \citep{bouwens22}.
Many of these pointed observations succeeded in detecting the dust-obscured star formation not only in the targeted galaxies, but also their companions that could be totally obscured in the HST bands, and sometimes even Spitzer/IRAC 3.6/4.5\,\micron\ \citep[e.g.,][]{fudamoto21, fujimoto24b}.

These studies highlight the importance of dust-obscured star formation even in the EoR, although the measurements of obscured SFR densities based on these pointed (and thus biased) observations remain highly challenging \citep[e.g., see][]{algera23, vanleeuwen24}.
If heavily obscured galaxies frequently exist at these redshifts but below the detection limit of wide-field imaging surveys, our understanding of massive galaxy assembly would be highly incomplete. 
In fact, JWST imaging and spectroscopic observations of early massive quiescent galaxies frequently imply the existence of starburst progenitors at $z \simeq 6 - 10$ with SFR above 100\,\smpy\ \citep[e.g.,][]{carnall23a, glazebrook24, degraaff25a, baker25a}.
Some of these progenitors may have undergone highly efficient starbursts, converting the majority of accreted baryons to stars \citep[e.g.,][]{carnall24, turner25a}.
Associated with the starburst, these galaxies may be heavily dust-obscured and thus remain as an undetected population with JWST wide-field imaging surveys \citep[e.g.,][]{williams24, glazebrook24}.

In this work, we present the identification and analyses of two NIRCam-dark galaxies at $z=6.6$ with ALMA.
These two dusty starburst galaxies are selected through the ASPIRE JWST and ALMA survey \citep{wangf23, wangfprep}.
In Section~\ref{sec:02_obs}, we describe the observations and data processing techniques. 
We present the source selection, photometry and SED modeling in Section~\ref{sec:03_res}.
We discuss the implications of these galaxies to the overall picture of massive galaxy evolution in Section~\ref{sec:04_disc}.
Our conclusions are summarized in Section~\ref{sec:05_sum}.
Throughout this work, we assume a flat $\Lambda$CDM cosmology with $H_0= 70$\,\si{km.s^{-1}.Mpc^{-1}} and $\Omega_\mathrm{M} = 0.3$. 
AB magnitude system \citep{oke83} is adopted to describe source brightness in the optical and near-IR. 
We also assume a \citet{chabrier03} initial mass function. 
We define the IR luminosity ($L_\mathrm{IR}$) as the integrated luminosity over a rest-frame wavelength range from 8 to 1000\,\micron.

\section{Observation and Data Processing}
\label{sec:02_obs}

\subsection{JWST/NIRCam}
\label{ss:02a_jwst}

JWST/NIRCam \citep{rieke23} three-band imaging data of 25 luminous quasars at $z=6.5-6.8$ were obtained through the Cycle-1 ASPIRE program (GO-2078; PI: F.\ Wang; \citealt{wangf23}).
The exposure time with each filter is 1417\,s (F115W), 2834\,s (F200W) and 1417\,s (F356W), respectively.
The NIRCam data processing has been described in detail with previous papers from the ASPIRE collaboration \citep{wangf23,yangj23}.
Briefly, the data were processed through a modified JWST calibration pipeline \citep{jwst} with the reference file \verb|jwst_1080.pmap| (including JWST Cycle-1 NIRCam flux calibrations).
This includes a few commonly adopted customized treatments, for example, $1/f$ noise subtraction, iterative sky background removal and WCS alignment to Gaia DR3 \citep{gaiadr3} or the DESI Legacy Imaging Survey \citep{dey19} if not enough Gaia stars are found within the Field of View (FoV).

We also note that ASPIRE also obtained NIRCam wide-field slitless spectroscopy (WFSS) of the targeted quasar fields in the F356W band.
These data were not directly used in our analyses (but see Section~\ref{ss:04b_number}).

\subsection{ALMA}
\label{ss:02b_alma}

ALMA 1.2\,mm mosaics of all 25 ASPIRE quasar fields were obtained through a 100-hr ALMA Cycle-9 large program (PI: F. Wang; Program ID: 2022.1.01077.L).
In each quasar field, we map the 1\farcm2$\times$1\farcm1 region centered on the quasar at the \cii\,158\,\micron\ wavelength, and therefore search for companion galaxies at the quasar redshifts (within a velocity offset of $\Delta v \sim \pm2200$\,\si{km.s^{-1}}).

The ALMA data reduction has been described in detail by \citet{sunf25a} through a standard \textsc{casa} \verb|v6.4.1.12| pipeline \citep{casa22}.
To avoid the artificial boost of continuum flux densities from \cii\ emitters at quasar redshifts, we first flagged the spectral channels that have a velocity offset smaller than 500\,\si{km\,s^{-1}} from the quasar \cii\ line center.
To optimize the detection of sources with both compact and extended dust continuum or \cii\ emissions, we produced ALMA mosaics at both native resolution (Briggs weighting \texttt{robust=0.5}, no \textit{uv} tapering) and tapered resolution (\texttt{robust=2.0}, uv tapered with a Gaussian kernel of FWHM=1\arcsec0).
The synthesized beam FWHM is 0\farcs65$\pm$0\farcs05 and 1\farcs34$\pm$0\farcs05 for native and tapered continuum image mosaics, respectively.
The typical continuum rms noises (before primary beam response correction) of our ALMA mosaics are 0.031$\pm$0.004 mJy\,beam$^{-1}$ and 0.034$\pm$0.004 mJy\,beam$^{-1}$ at native and tapered resolution, respectively.

We also include archival ALMA measurements of our targets at Band 3, 4, 5 and 7.
Band-3, 4, 5 observations were obtained through program 2017.1.00139.S, 2017.1.01532.S and 2019.1.00147.S \citep{venemans19,pensabene21,lij22}.
We directly use these published photometric measurements.
Band 7 observations were obtained through 2021.1.00443.S (PI: J.\ Spilker), and the data were processed through a similar routine as that of the ASPIRE-ALMA data.

\begin{figure*}[!t]
\centering
\includegraphics[width=\linewidth]{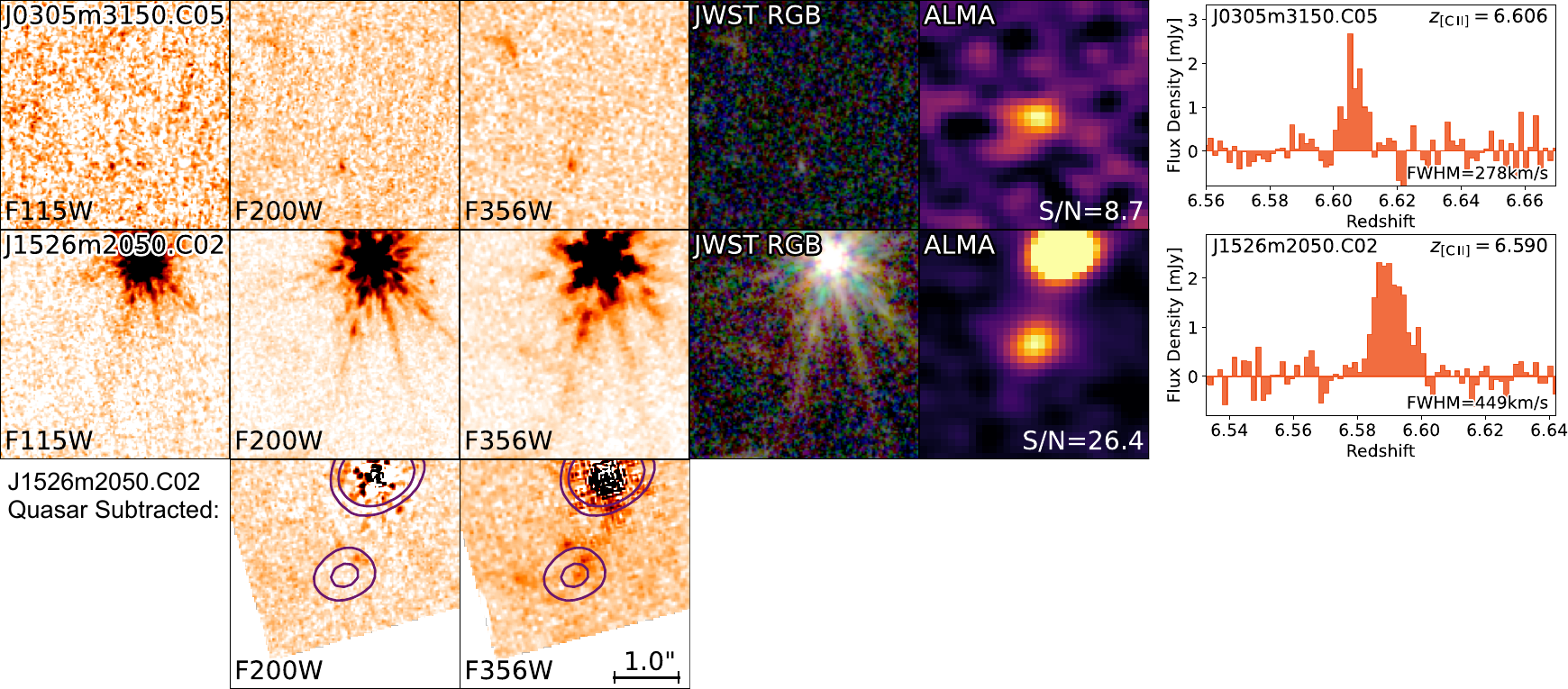}
\caption{JWST NIRCam images (F115W, F200W, F356W), ALMA 1.2\,mm continuum images and \cii\,158\,\micron\ spectra of two NIRCam-dark galaxies at $z_\mathrm{spec}\sim6.6$.
Because the luminous quasar J1526--2050 is visible to the northwest of J1526m2050.C02 (1\farcs6 separation), we also show the quasar-subtracted F200W and F356W image of J1526m2020.C02 in the third row, with ALMA continuum contours (purple) at 10 and 20$\sigma$ overlaid.
Image size is 4\arcsec$\times$4\arcsec.
These galaxies are selected as companions of quasars with ALMA, but remain undetected or faint at JWST/NIRCam wavelengths.
}
\label{fig:img}
\end{figure*}

\section{Results}
\label{sec:03_res}

\subsection{The identification of NIRCam-dark galaxies}
\label{ss:03a_id}

\citet{sunf25a} reported the detection of 117 continuum sources at primary beam response $\geq0.25$ (over 35\,arcmin$^2$) through the ASPIRE-ALMA survey.
Among these 1.2-mm continuum sources, 23 sources are quasars and the remaining sources are classified as DSFGs at cosmological distances.
Six of these DSFGs are at $z>6$, all confirmed as quasar companions through ALMA \cii\ detection or JWST/NIRCam grism spectroscopy of the \oiii\ doublets.

Among all ASPIRE-ALMA continuum sources, we identify two galaxies detected at 1.2\,mm at high significance ($8.7$ and $26.4\sigma$) but lacking obvious NIRCam counterparts.
The NIRCam and ALMA images of these two galaxies are shown in Figure~\ref{fig:img}.
Both galaxies are spectroscopically confirmed as companions to quasars at $z\sim6.6$ through ALMA \cii\ spectroscopy.
The first source, \tgta\ ($z=6.606$), was known as a companion galaxy to quasar J0305--3150 at $z=6.614$ through previous ALMA studies \citep[dubbed as C3; e.g.,][]{venemans19,venemans20,lij22,wangf23} with a velocity offset of $\Delta v = -290\pm30$\,\si{km.s^{-1}} and angular offset of 6\farcs9.
The second source, \tgtb\ ($z=6.590$), was also known as a companion galaxy to quasar J1526--2050 (or PSO\,J231--20, $z=6.587$) through previous ALMA studies \citep[][]{decarli17,mazzucchelli19,pensabene21} with a velocity offset of $\Delta v = +137\pm30$\,\si{km.s^{-1}} and angular offset of 1\farcs6.
The detections with multiple ALMA datasets at various frequencies validate the fidelity of these sources. 

We conduct photometry of the two sources in available JWST/NIRCam and ALMA data.
Because \tgtb\ is close to the luminous quasar, we subtract a  point spread function (PSF) model 
from the quasar image. The PSF model is constructed using a PSF star library in the ASPIRE fields. 
The detailed method and validation of the PSF model and subtraction will be presented in a forthcoming paper from the ASPIRE collaboration \citep{yangjprep}. 
The quasar-subtracted NIRCam images in the F200W and F356W bands are shown in the third row of Figure~\ref{fig:img}.

Compared with the ALMA continuum contours, we find faint diffuse emission (diameter $\sim1\arcsec$) at the location of \tgtb\ in the F356W band.
This is somewhat expected, because the surface brightness dimming is as a function of $(1+z)^{-4}$, and DSFGs at lower redshifts are known to host more extended stellar continuum (attenuated) than the dust continuum \citep[e.g.,][]{chenc15,hodge16}.
Therefore, to avoid substantial flux loss of diffuse emission, we obtain photometry using a circular aperture of radius $r=0\farcs5$.
The photometric uncertainties are estimated using random apertures in source-free regions within 10\arcsec\ from the targets.
We find that \tgta\ remains undetected in all NIRCam images ($>$28.0\,AB mag in F356W, $3\sigma$ limit), and the residuals of \tgtb\ are only detected in the F356W band (26.5$\pm$0.1\,AB mag).

These two DSFGs are extraordinarily faint at the NIRCam wavelengths.
For comparison, all 289 DSFGs in \citet{mcKinney25a} that fall in COSMOS-Web footprint \citep{casey23} are brighter than 26\,AB mag in the F444W band if measured with the same $r=0\farcs5$ aperture size.
To our knowledge, DSFGs similar to \tgta\ or \tgtb\ do not exist in ALMA blank-field surveys in the Hubble Ultra-Deep Field (HUDF) / GOODS-S region either \citep[e.g.,][]{dunlop17, hatsukade18, franco18, gl20, gomez22a}, as all secure ($>6\sigma$) ALMA continuum sources have been detected by NIRCam \citep[e.g.,][]{boogaard24,hill24} as long as they fall in the JADES footprints \citep{eisenstein23a}.
Therefore, we refer to these galaxies as ``NIRCam-dark'' ($S_\mathrm{3.6\,\mu m}/S_\mathrm{1.2\,mm} < 10^{-4}$) starburst galaxies in this paper.

ALMA Band-6/7 flux densities are measured from the peak of the \textit{uv}-tapered images as described by \citet{sunf25a}.
ALMA flux densities in other bands are taken from literature \citep{pensabene21,lij22}.
We also measure the ALMA 1.17\,mm continuum sizes using \textsc{casa} \textsc{imfit} function in the image plane with synthesized beam deconvolved. 
Both sources have compact dust continuum emission with circularized effective radius $R_\mathrm{e,circ} \sim 0\farcs19$ ($\sim1$\,kpc at $z=6.6$).
Table~\ref{tab:01_phot} summarizes the JWST and ALMA photometry of the two NIRCam-dark DSFGs.

\begin{table}[!t]
\footnotesize
\caption{The properties of the two NIRCam-dark galaxies. \label{tab:01_phot}}
\centering
\begin{tabular}{lrr}
\hline\hline
& \tgta & \tgtb \\\hline
R.A [deg] & 46.31826 & 231.65781 \\
Decl. [deg] & --31.84859 & --20.83397 \\
$z_\mathrm{spec}$ & 6.606 & 6.590 \\
Offset from quasar [\arcsec] & 6.9 & 1.6 \\
$R_\mathrm{e,circ}$ [\arcsec] & 0.19$\pm$0.08 & 0.19$\pm$0.05  \\\hline
NIRCam F115W [nJy] & $<$56.4 & $<$81.6 \\
NIRCam F200W [nJy] & $<$24.6 & $<$40.8 \\
NIRCam F356W [nJy] & 11.6$\pm$7.7 & 87.8$\pm$8.7 \\
ALMA 0.893\,mm [mJy] & 0.64$\pm$0.04 & 1.71$\pm$0.04 \\
ALMA 1.17\,mm [mJy] & 0.32$\pm$0.04 & 1.04$\pm$0.04 \\\hline
$\log[M_\mathrm{star} / \si{M_\odot}]$ & 10.0$\pm$0.5 & 10.5$\pm$0.5 \\
$\log[\mathrm{SFR} / \si{M_\odot.yr^{-1}}]$ & 1.9$\pm$0.1 & 2.4$\pm$0.1 \\
$A_V$ [mag] & 7.5$\pm$3.7 & 3.5$\pm$0.4 \\
$\log[L_\mathrm{IR} / \si{L_\odot}]$ & 11.9$\pm$0.1 & 12.4$\pm$0.1 \\
$\log[M_\mathrm{gas} / \si{M_\odot}]$ & 9.5$\pm$0.2 & 10.4$\pm$0.2 \\
$\log[M_\mathrm{dust} / \si{M_\odot}]$ & 7.6$\pm$0.3 & 8.3$\pm$0.1 \\
$T_\mathrm{dust}$ [K] & 44.7$\pm$10.9 & 36.5$\pm$2.4 \\\hline
\end{tabular}
\tablecomments{The circularized effective radii ($R_\mathrm{e,circ}$) are measured at ALMA 1.17\,mm (dust continuum). The molecular gas masses ($M_\mathrm{gas}$) are from \citet[\tgta]{lij22} and \citet[\tgtb]{pensabene21}. Other ALMA photometry is also available therein. Upper limits are at $3\sigma$.
}
\end{table}

\begin{figure*}[!ht]
\centering
\includegraphics[width=0.49\linewidth]{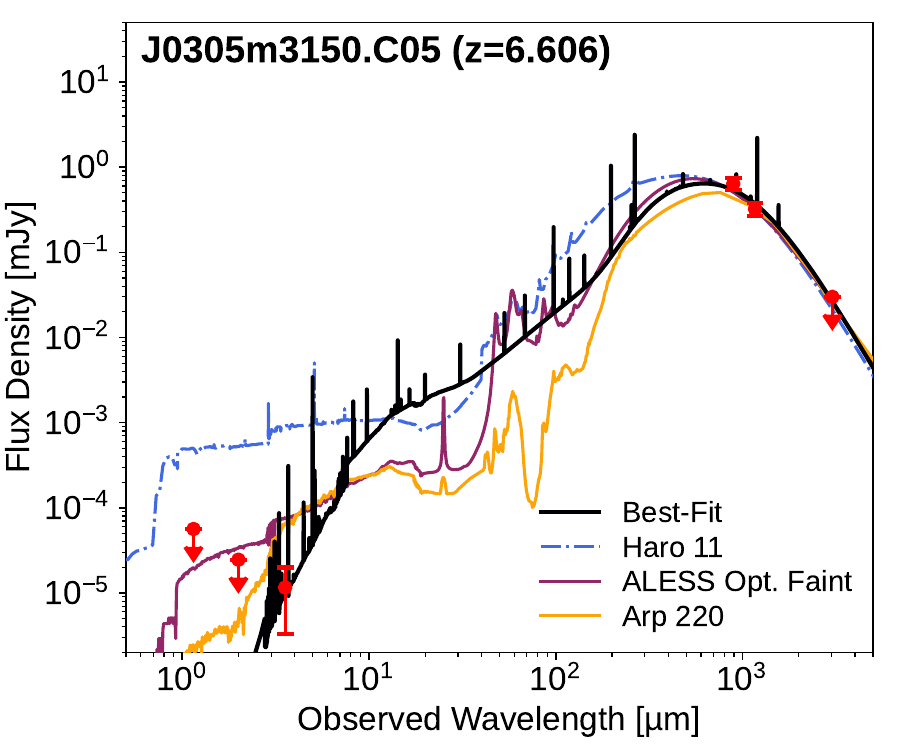}
\includegraphics[width=0.49\linewidth]{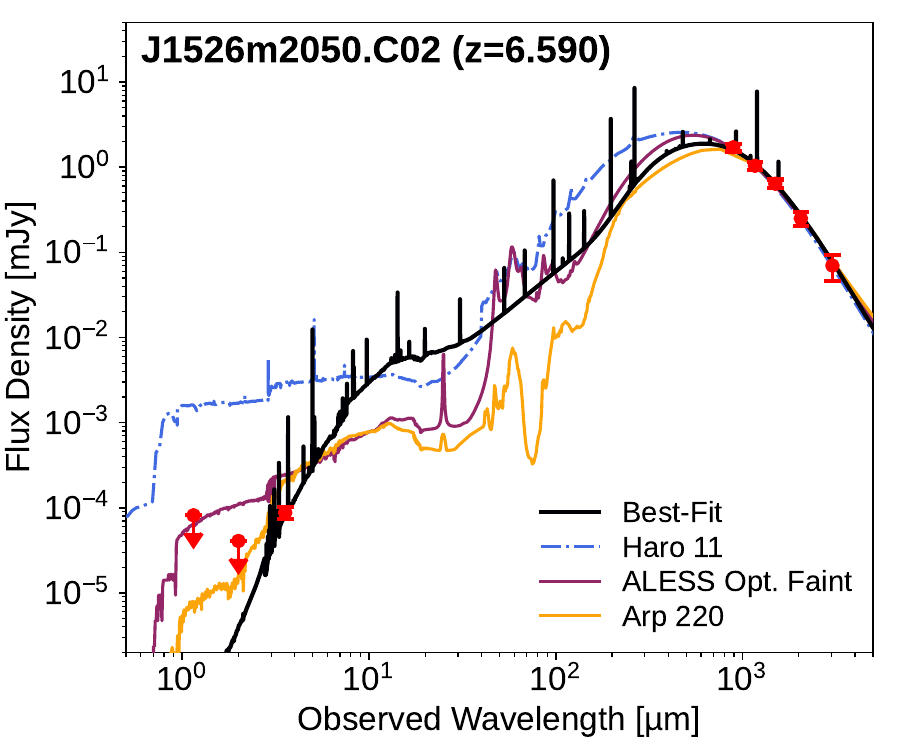}
\caption{SEDs of the two NIRCam-dark galaxies (red circles). 
The best-fit SED models obtained by \textsc{cigale} are shown in the black lines. 
The SED templates of Haro\,11 (nearby starburst galaxy, blue; \citealt{lyuj16}), Arp\,220 (nearby ULIRG, orange; \citealt{Silva98}) and ALESS optically faint DSFGs at $z\sim3$ (purple; \citealt{dacunha15}), all redshifted and scaled to the ALMA 1.2-mm\ flux densities, are shown for comparison.
}
\label{fig:sed}
\end{figure*}

\subsection{SED modeling and physical properties}
\label{ss:03b_sed}

We model the spectral energy distributions (SED) of the two NIRCam-dark galaxies with \textsc{cigale}, an energy-balance parametric SED-fitting software \citep{cigale09,cigale19}.
We assume a commonly used delayed-$\tau$ star-formation history (SFH), in which $\mathrm{SFR}(t) \propto t \exp(-t / \tau)$ and $\tau$ is the peak time of star formation.
We allow the age of main stellar population to be 20--500\,Myr and $\tau$ to be 20--2000\,Myr. 
An optional late starburst is allowed in the last 20\,Myr of SFH, which could produce up to 10\% of the total stellar mass.
We use \citet{bc03} stellar population synthesis models, and assume a solar metallicity ($Z_{\odot}$) given the strong presence of dust.
We adopt a modified \citet{calzetti00} attenuation curve, and allow the variation of the power-law slope by $\pm$0.2 and $A_V$ effectively at 0--15.
Nebular emission is included assuming an electron density of $n_e = 100$\,\si{cm^{-3}} and ionization parameter $\log U=-4 \sim -2$.
For simplicity, we assume a mid-to-far-IR dust continuum model from \citet{casey12}, parameterized by the dust temperature ($T_\mathrm{dust}$; 30--60\,K),
dust emissivity ($\beta_\mathrm{em}$; 1.6--2.0) and mid-IR power-law slope ($\alpha_\mathrm{MIR}$, fixed at 2.0).
Although the \citet{casey12} model does not include mid-IR polycyclic aromatic hydrocarbon (PAH) features, the inclusion of a mid-IR power-law slope can well reproduce the luminosity excess from PAHs from a comparison of widely used empirical SED templates \citep[e.g.,][]{chary01, rieke09} as tested by \citet{sunf25a}.

We note that an active galactic nucleus (AGN) component is not included in our SED modeling. 
Both sources remain undetected in the X-ray with Chandra \citep[][]{connor20, wangf21}, disfavoring the presence of unobscured AGNs.
The AGN contributions to the ALMA flux densities are also likely insignificant (see the discussion in \citealt{sunf25a}).

Figure~\ref{fig:sed} shows the best-fit SED models of the two NIRCam-dark galaxies.
For comparison, we also plot the SED templates of (\romannumeral1) the nearby starburst galaxy Haro\,11 (\citealt{lyuj16}, whose far-IR SED may resemble those of luminous DSFGs at $z\gtrsim6$ as argued by \citealt{derossi18}), (\romannumeral2) the nearby ULIRG Arp\,220 \citep[][including both the obscured nucleus and unobscured stellar component]{Silva98} and (\romannumeral3) the optically faint DSFGs at $z\sim3$ selected through the ALESS survey \citep{dacunha15}, all normalized to the ALMA 1.2-mm flux densities.
We find that the two NIRCam-dark galaxies appear to be more obscured in rest-frame optical even compared with Arp\,220 or $z\sim3$ optically faint DSFGs.
In particular, $z\sim3$ optically faint DSFGs in \citet{dacunha15} will remain above our F200W and F356W detection limit if they are redshifted to $z=6.6$.
The SED fitting implies a substantial $V$-band dust attenuation ($A_V$ up to 7.5$\pm$3.7\,mag for \tgta) for these NIRCam-dark DSFGs.

The multiple-wavelength ALMA observations offer useful constraints to the far-IR SEDs.
With IR luminosities $L_\mathrm{IR}\simeq10^{11.9} - 10^{12.4}$\,\lsun, we infer a dust-obscured SFR of 80--250\,\smpy\ for the two sources \citep{ke12}.
These are consistent with the SFR inferred from the \cii\ luminosities \citep[][]{pensabene21,lij22}.
In contrast, the constraints on their stellar mass are rather poor. 
We infer stellar mass $\log(M_\mathrm{star}/M_\mathrm{\odot})$ of 10.0$\pm$0.5 and 10.5$\pm$0.5 for both targets respectively, and the large errors are natural consequence of poor constraints in the rest-frame UV/optical through Bayesian inference.
The inferred physical properties of the two galaxies are also presented in Table~\ref{tab:01_phot}.

The molecular gas masses ($M_\mathrm{gas}$) of the two galaxies have been studied through \ci\ and CO observations by \citet{pensabene21} and \citet{lij22}.
We find comparable $M_\mathrm{gas}$ and $M_\mathrm{star}$ for the two galaxies, indicating high gas fractions associated with the starbursts.
We further model the far-IR SEDs of the two galaxies with a modified blackbody model assuming a dust absorption coefficient as $\kappa = 0.40 \times (\nu / 250)^{\beta_\mathrm{em}}$ in unit of \si{cm^{2}.g^{-1}}, where $\nu$ is the rest-frame frequency in GHz and $\beta_\mathrm{em}$ is fixed at 1.8 following previous works \citep[e.g.,][]{ds17, sunf22a}.
We also consider the CMB heating effect following \citet{dacunha13}, which is found to be small at the derived $T_\mathrm{dust}$.
The derived dust masses and temperatures are also presented in Table~\ref{tab:01_phot}.
Although the constraints are poor for \tgta\ because of limited ALMA coverage, both sources exhibit normal dust temperature and gas-to-dust ratio $\delta_\mathrm{GDR}$ around 100.

\section{Discussion}
\label{sec:04_disc}

\begin{figure*}
\centering
\includegraphics[width=\linewidth]{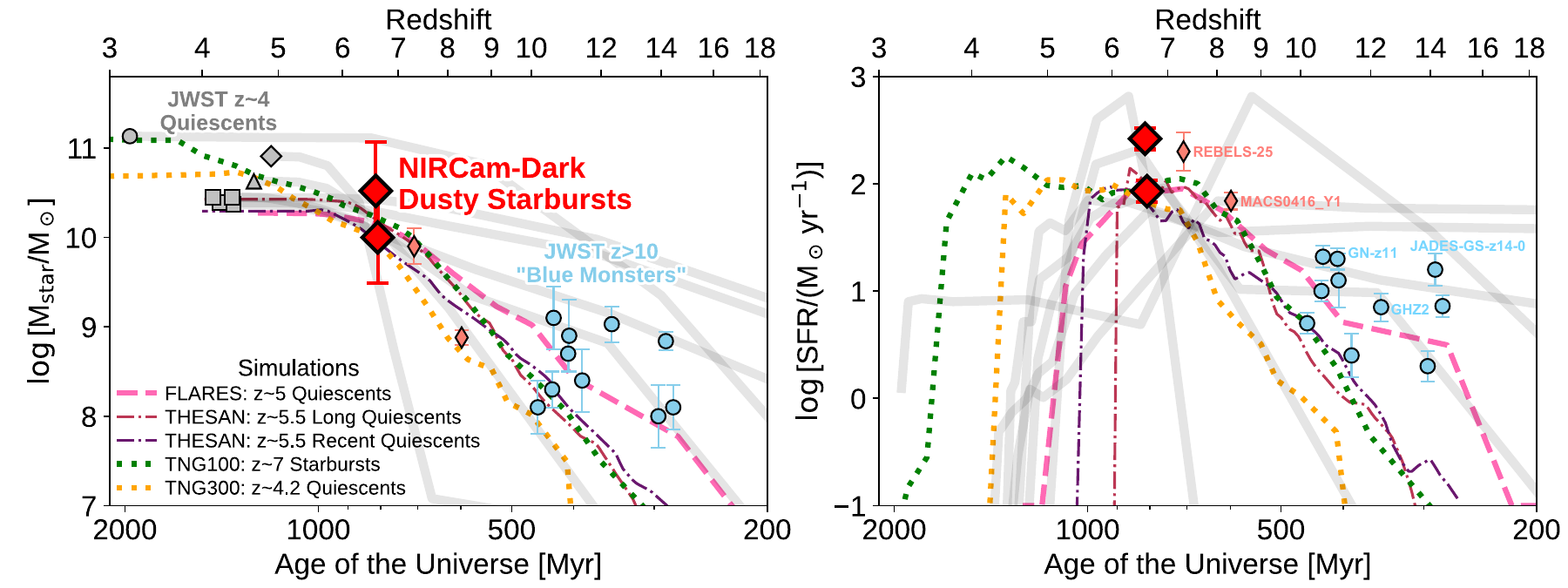}
\caption{Stellar mass (left) and SFR (right) versus redshift for NIRCam-dark dusty starburst galaxies (red diamonds) in the context of massive galaxy evolution.
For comparison we show the star-formation histories of massive quiescent galaxies at $z\sim4$ confirmed with JWST spectroscopy (gray circle: \citealt{glazebrook24} but using the SFH from \citealt{turner25a}; diamond: \citealt{degraaff25a} with solar-metallicity SFH model; triangle: \citealt{carnall23a} but using the SFH from \citealt{jiz24a}; squares: \citealt{baker25a}).
We also highlighted two ALMA-confirmed dusty starbursts known at $z>7$ prior to JWST in coral diamonds, including REBELS-25 \citep[$z=7.3$;][]{hygate23} and MACS0416\_Y1 \citep[$z=8.3$;][]{tamura19}.
JWST-confirmed luminous galaxies at $z>10$ with low dust content (also known as ``blue monsters''; e.g., \citealt{ziparo23}) are shown in blue circles \citep[including][]{arrabalharo23,bunker23,carniani24,castellano24,hsiao24a,kokorev25a,naidu25b,zavala25}.
The median evolution track of early massive galaxies in cosmological simulations are shown in colored lines (including FLARES, \citealt{lovell23}; IllustrisTNG, \citealt{pillepich18}, \citealt{nelson19} and \citealt{hartley23}; THESAN, \citealt{kannan22} and \citealt{chittenden25}).
NIRCam-dark dusty starburst galaxies presented by this study are candidates for the progenitors of massive quiescent galaxies at $z\sim4$ and descendants of ``blue monsters'' at $z>10$.
}
\label{fig:sfh}
\end{figure*}

\subsection{Placing NIRCam-dark starbursts in the context of massive galaxy evolution}
\label{ss:04a_evo}

These two NIRCam-dark galaxies are among the highest-redshift ALMA continuum sources selected by the ASPIRE-ALMA survey.
The redshifts of our targets are also higher than the vast majority of DSFGs selected from millimeter surveys in blank fields ($z\lesssim6$; see references in Section~\ref{sec:01_intro}).
Therefore, a low flux density ratio of $S_\mathrm{3.6\micron}/S_\mathrm{1.2\,mm}$ is somewhat expected at such high redshifts \citep[e.g.,][]{yamaguchi19}, leading to the faintness of these galaxies at NIRCam wavelengths.

However, from the viewpoint of massive galaxy evolution, the identification of NIRCam-dark galaxies in the EoR could be particularly important.
Figure~\ref{fig:sfh} shows the stellar mass and SFR of NIRCam-dark galaxies compared with (\romannumeral1) the SFH of massive quiescent galaxies at $z\gtrsim4$ that have been frequently confirmed with JWST spectroscopy \citep[e.g.,][]{carnall23a, glazebrook24, baker25a, degraaff25a}, and (\romannumeral2) $z>10$ luminous galaxies with low dust content confirmed by JWST (e.g., \citealt{arrabalharo23, bunker23, carniani24, castellano24, kokorev25a, naidu25b, zavala25}; also known as ``blue monsters'', e.g., \citealt{ziparo23}).
From the SFH modeled from JWST spectrophotometry, it is evident that many of the massive quiescent galaxies at $z\gtrsim4$ are expected to be massive ($M_\mathrm{star}\gtrsim10^{9}$\,\msun) star-forming (SFR\,$\gtrsim$\,10\,\smpy) galaxies at $z\sim12$, matching the observables of the ``blue monsters'' at this epoch.
The inferred SFH of these early quiescent galaxies typically peak at $z\simeq 6 - 8$, with SFR at $\gtrsim100$\,\smpy.

We find that the stellar mass and SFR of NIRCam-dark DSFGs at $z\sim7$ exactly match the starburst progenitors of massive quiescent galaxies at $z\sim4$ (Figure~\ref{fig:sfh}).
The compact sizes of their dust continuum emission (and therefore the molecular gas reservoir; $R_\mathrm{e,circ} \sim $1\,kpc) also resemble the compact stellar sizes of massive quiescent galaxies at $z\sim4$ \citep[$R_\mathrm{e,circ} \lesssim $1\,kpc; e.g.,][]{jiz24b}.
We therefore argue that NIRCam-dark DSFGs are likely the progenitors of $z\sim4$ massive quiescent galaxies and descendants of $z>10$ ``blue monsters'', or at least representing one viable intermediate galaxy population that can bridge the two massive galaxy populations before and after the EoR.

From cosmological simulations, it is also clear that the progenitors of $z\gtrsim4$ massive quiescent galaxies should be massive gas-rich starburst galaxies at $z\sim7$, matching the $M_\mathrm{star}$ and SFR of NIRCam-dark galaxies in our sample (e.g., see the SFHs in Figure~\ref{fig:sfh} simulated by FLARES, \citealt{lovell23}; IllustrisTNG, \citealt{pillepich18,nelson19}, see also \citealt{hartley23}; THESAN, \citealt{kannan22,chittenden25}).

However, we also note that many starburst galaxies with SFR\,$\gtrsim100$\,\smpy\ at $z\sim7$ may quench at later epochs (e.g., $z\sim3$ from TNG100), possibly depending on the quenching physical mechanisms. 
The obscuration of starbursts at $z\sim7$ also depends on complicated physics of dust production \citep[e.g.,][]{lesniewska19,choban25}, geometry and sometimes viewing angles \citep{cochrane24}, many of which are not clear from either observational or theoretical perspectives.
Specifically, the observed high dust-to-stellar mass ratio $\log(M_\mathrm{dust} / M_\mathrm{star}) \sim -2.3 \pm 0.5$ as observed for the two galaxies are perhaps particularly interesting.
The observed $M_\mathrm{dust} / M_\mathrm{star}$ approaches the expectation from the ``maximal dust model'' at $z\sim7$ through core-collapse supernovae dust production by \citet{dayal22}, and match the \citet{popping17b} model prediction at solar metallicity.
Nevertheless, future deeper JWST imaging observations, in particular with multi-band MIRI imaging, are needed to provide tighter constraints on the $M_\mathrm{star}$ of the two objects and thus models of dust production (and destruction) for massive starburst galaxies in the EoR.



\begin{figure*}[!t]
\centering
\includegraphics[width=0.49\linewidth]{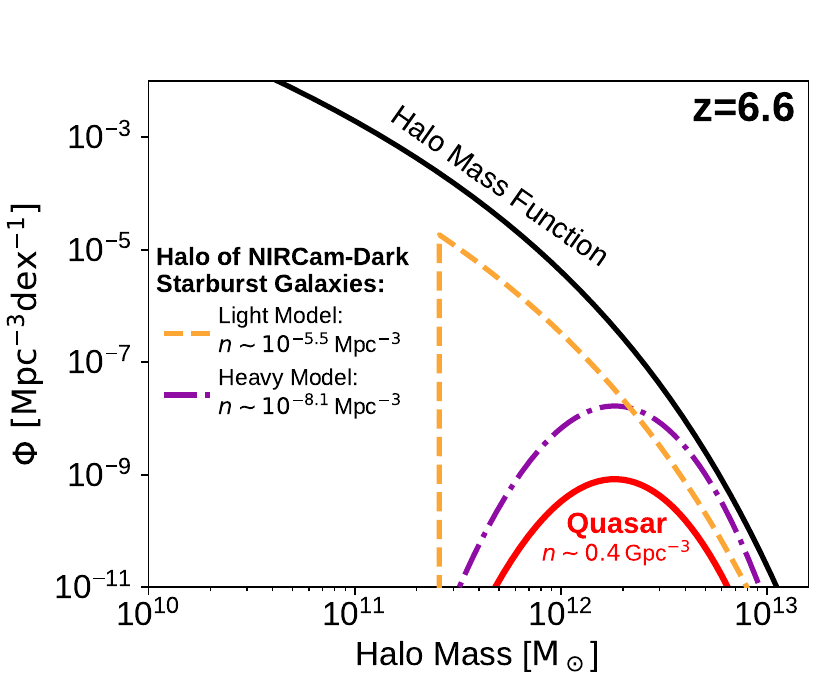}
\includegraphics[width=0.49\linewidth]{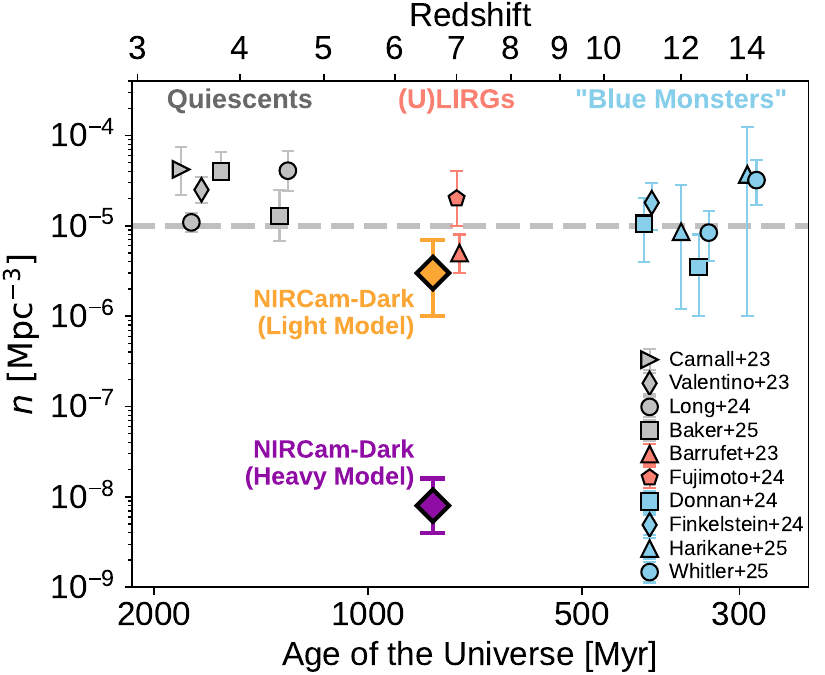}
\caption{
The number density constraint of NIRCam-dark galaxies. \textbf{Left}: The host halo mass function models of NIRCam-dark starburst galaxies at $z=6.6$.
We consider two models for their host mass distribution, including a light model (dashed orange line) and a heavy model (dashed-dotted purple line; Section~\ref{ss:04b_number}).
These two models predict completely different number densities of NIRCam-dark galaxies, while both of them match the existing observations.
For comparison we show the quasar host HMF (red) and overall HMF (black) at the same redshift.
\textbf{Right}: Comparison of the number densities with massive quiescent galaxies at $z\sim4$ \citep{carnall23b, valentino23, long24a, baker25a}, (U)LIRGs at $z\sim7$ \citep[][see Section~\ref{ss:04b_number} for details]{barrufet23a, fujimoto24b} and ``blue monsters'' at $z>10$ \citep[][computed from their measured UVLFs in the brightest bins, typically at $M_\mathrm{UV} \sim -21$\,mag]{donnan24, finkelstein24, harikane25, whitler25}.
The derived number densities of NIRCam-dark galaxies based on the light and heavy halo occupation model are shown as orange and purple diamonds, respectively.
}
\label{fig:n}
\end{figure*}

\subsection{The number density of NIRCam-dark starbursts}
\label{ss:04b_number}

To assess whether NIRCam-dark starbursts are possible progenitors of $z\gtrsim4$ massive quiescent galaxies and descendants of $z>10$ blue monsters, we analyze the number densities of these galaxy populations.
The number density of massive quiescent galaxies at $z\gtrsim4$ has been observed at $n \simeq 10^{-5} - 3\times10^{-5}$\,\si{Mpc^{-3}} \citep[e.g.,][]{carnall23b, valentino23, long24a, baker25a}.
On the other hand, JWST also measures the number density of luminous Lyman-break galaxies (absolute UV magnitude $M_\mathrm{UV} \sim -21$, corresponding to a SFR\,$\sim$\,10\,\smpy) at $z\sim12$ as $n \sim 10^{-5}$\,\si{Mpc^{-3}} \citep[e.g.,][]{donnan24, finkelstein24, robertson24, harikane25, naidu25b, whitler25}.
Therefore, if the abundance of NIRCam-dark DSFGs is also found to be $\sim10^{-5}$\,\si{Mpc^{-3}}, it would provide strong evidence to the evolutionary connection.

However, it is not easy to constrain the number density of NIRCam-dark galaxies because they are selected as companions of luminous quasars.
Luminous quasars at $z\gtrsim6$ are now confirmed to generally reside in massive dark matter halos thanks to JWST/NIRCam observations, in particularly with WFSS \citep[e.g.,][]{kashino23, wangf23, eilers24, champagne25, pudoka25}.
Through the galaxy auto-correlation function, quasar-galaxy cross-correlation function and comparison with dark-matter-only simulations, the host halo masses of quasars are measured as $\log(M_\mathrm{halo})\gtrsim12.30$ from the EIGER survey \citep[six quasars at $z\sim6$;][]{eilers24} and similarly with the ASPIRE survey \citep[25 quasars at $z\simeq6.5-6.8$;][]{wangfprep}.
Such massive halos are rare at $z\sim6.6$ ($n<10^{-7}$\,\si{Mpc^{-3}} based on the halo mass function, HMF model of \citealt{murray13}), although the number density is still much higher than that of quasars at this epoch ($n=0.39\pm0.11$\,\si{Gpc^{-3}}; \citealt{wangf19a}).

\citet{pizzati24a} developed an empirical quasar population model that successfully reproduces the clustering properties and luminosity functions (LF) of quasars based on N-body simulations.
Following this model, the luminosity of a quasar (or galaxy) is dependent on the halo mass, and the probability distribution function that a halo could host a detectable quasar (or galaxy), $P(M_\mathrm{halo})$, can be approximated as a Gaussian function:
\begin{equation}
    P(M_\mathrm{halo}) = P_\mathrm{max} \exp\{-\frac{[\log (M_\mathrm{halo}/ M_c)]^2}{2 \sigma_m^2}\}
    \label{eq:p_qso}
\end{equation}
where $P_\mathrm{max}$ is the maximum likelihood of halo occupation, $M_c$ is the characteristic halo mass, and $\sigma_m$ is the mass dispersion.
Taking the HMF at $z=6.6$ \citep[]{murray13} and the shape of quasar-host HMF (RMS of $M_\mathrm{halo}$) inferred by \citet{pizzati24a} for EIGER quasars \citep{eilers24}, we reproduce the observed quasar number density at $z\sim6.6$ with the parameter set $P_\mathrm{max} = 0.02$, $\log (M_c / M_\odot) = 12.73$, $\sigma_\mathrm{m} = 0.22$, and the resultant quasar host HMF is shown as the solid red line in Figure~\ref{fig:n} (left panel).
At the median $\log(M_\mathrm{halo}/M_\odot)\sim12.3$ of quasars, the fraction of halos that host a detectable quasar is very low ($\sim3\times10^{-3}$).

Quasars are already rare objects residing in these massive halos, which likely indicates a low duty cycle in the UV-bright phase \citep[$f_\mathrm{duty} \lesssim 1\%$; e.g.,][]{eilers24, pizzati24a, wangfprep}.
With ASPIRE, we identify two NIRCam-dark galaxies in 25 massive halo environments at $z\simeq6.5 - 6.8$.
Therefore, we conclude that the occurrence rate of NIRCam-dark galaxies ($f = 2/25=0.08_{-0.05}^{+0.10}$; error from Poisson statistics) must be much higher than that of the quasars, otherwise we would not be able to identify any NIRCam-dark galaxies with ASPIRE.

We explore the possible halo occupation models of NIRCam-dark galaxies in Figure~\ref{fig:n} (left panel). 
The first model, dubbed ``heavy model'', assumes that NIRCam-dark galaxies share the same halo characteristics ($M_c$ and $\sigma_m$, Equation~\ref{eq:p_qso}) but with much larger $P_\mathrm{max}$ and thus higher occurrence rate.
We note that the large halo mass of a quasar also implies a virial radius $R_{200} \sim 50$ proper kpc, large enough to encompass the observed NIRCam-dark companions.
We simulate the $M_\mathrm{halo}$ mass distribution of quasars from the aforementioned quasar host HMF, apply $P(M_\mathrm{halo})$ and match the observed NIRCam-dark occurrence rate as observed by ASPIRE.
We obtain $P_\mathrm{max} = 0.37_{-0.23}^{+0.46}$ from this heavy model, suggesting a high occupation fraction of NIRCam-dark galaxies in the most massive halos at this epoch.
According to this model, the predicted number density of NIRCam-dark galaxies is $n = 8_{-5}^{+10} \times 10^{-9}$\,\si{Mpc^{-3}}, about 20$\times$ of that of UV-bright quasars but still much lower than that of $z\gtrsim4$ massive quiescent galaxies or $z>10$ blue monsters (Figure~\ref{fig:n}, right panel).

We also test an alternative model, dubbed ``light model'', in which the occupation fraction of NIRCam-dark galaxies is constant above certain $M_\mathrm{halo}$ threshold.
Based on TNG100, we find that most of the $z\sim6.5$ halos that host galaxies at SFR\,$>50$\,\smpy\ have a ``cutoff'' halo masses at $\log(M_\mathrm{halo}/M_\odot) \geq {11.4}$, and thus we adopt the threshold.
The halo occupation fraction is therefore the observed NIRCam-dark occurrence rate ($0.08_{-0.05}^{+0.10}$).
At above the halo mass threshold, the expected number density of NIRCam-dark galaxy is $n = 3_{-2}^{+4} \times 10^{-6}$\,\si{Mpc^{-3}}.
Under such a light model, the number density of NIRCam-dark galaxies will be about $30_{-20}^{+40}\%$ of those of $z\gtrsim4$ massive quiescent galaxies and $z>10$ blue monsters (Figure~\ref{fig:n}, right panel).

We conclude that the number density of NIRCam-dark galaxies at $z \sim 7$ remains highly unconstrained: it may be as low as $n \sim 10^{-8.1}$\,\si{Mpc^{-3}}, but could also be as high as $\sim 10^{-5.5}$\,\si{Mpc^{-3}}.
From an abundance-matching perspective, this implies that a substantial fraction of the progenitors of massive quiescent galaxies at $z \gtrsim 4$, as well as the descendants of $z > 10$ blue monsters, could be NIRCam-dark at $z \sim 7$.
This hypothesis will need to be tested through both observations and simulations, including studies of dust production and obscuration mechanisms.

We also emphasize that our models assume independent halo occupations for quasars and NIRCam-dark galaxies. 
If the two populations are correlated or anti-correlated, the number density estimates could be even more uncertain.
We argue that the large $M_\mathrm{dust}$, high $A_v$ and high $M_\mathrm{dust}/M_\mathrm{star}$ of NIRCam-dark galaxies imply efficient dust production over extended period (e.g., $\sim300$\,Myr; \citealt{tamura19})--significantly longer than both the Salpeter timescale of quasar and the major merger timescale predicted by cosmological simulations, which are $\sim$\,50\,Myr at this epoch \citep[e.g.,][]{snyder17}.
Therefore, while interactions between quasar host galaxies and NIRCam-dark galaxies may occur in the present or upcoming epoch, their halo occupations can still be treated as independent.

We also note that, according to the IRLF at $z \sim 7$ from \citet{fujimoto24b}, the integrated number density of $z \sim 7$ DSFGs with SFR\,$\simeq 50$--$500$\,\smpy\ is $n \sim 2 \times 10^{-5}$\,\si{Mpc^{-3}}.
This exceeds the number density of $n \sim 5 \times 10^{-6}$\,\si{Mpc^{-3}} derived from the $z \sim 7$ IRLF of \citet{barrufet23a} over the same SFR range, which is based on REBELS-ALMA observations of the $L_\mathrm{IR}$ of UV-luminous galaxies and the UVLF as a proxy.
Although caution should be taken given the substantial uncertainties in the IRLFs at $z > 7$, if NIRCam-dark galaxies follow a ``light model'' of halo occupation, they may represent a substantial fraction (likely $\gtrsim 30\%$) of the (U)LIRG population and the obscured cosmic star formation rate density at $z \sim 7$.
These galaxies are likely to be missed in most blank-field surveys due to their rarity and faintness in the near-IR.
Indeed, the NIRCam-dark fraction of quasar companion DSFGs detected by ASPIRE-ALMA is 2/6\,=\,33\%, consistent with the fraction inferred above.

Finally, we note that NIRCam-dark galaxies at $z\simeq6 - 7$ may reside above the detection limit of wide ALMA surveys like GOODS-ALMA 2.0 \citep[][$\sim$72\,arcmin$^2$ down to $S_\mathrm{1.1mm} \sim 0.34$\,mJy]{gomez22a}, but the expected number of detections is only $N\sim0.5$ (and $N\sim1.5$ for general DSFGs with SFR\,$\simeq50-500$\,\smpy). Therefore, a non-detection so far is fully consistent with the cosmic shot noise.

\begin{figure*}[!th]
\centering
\includegraphics[width=\linewidth]{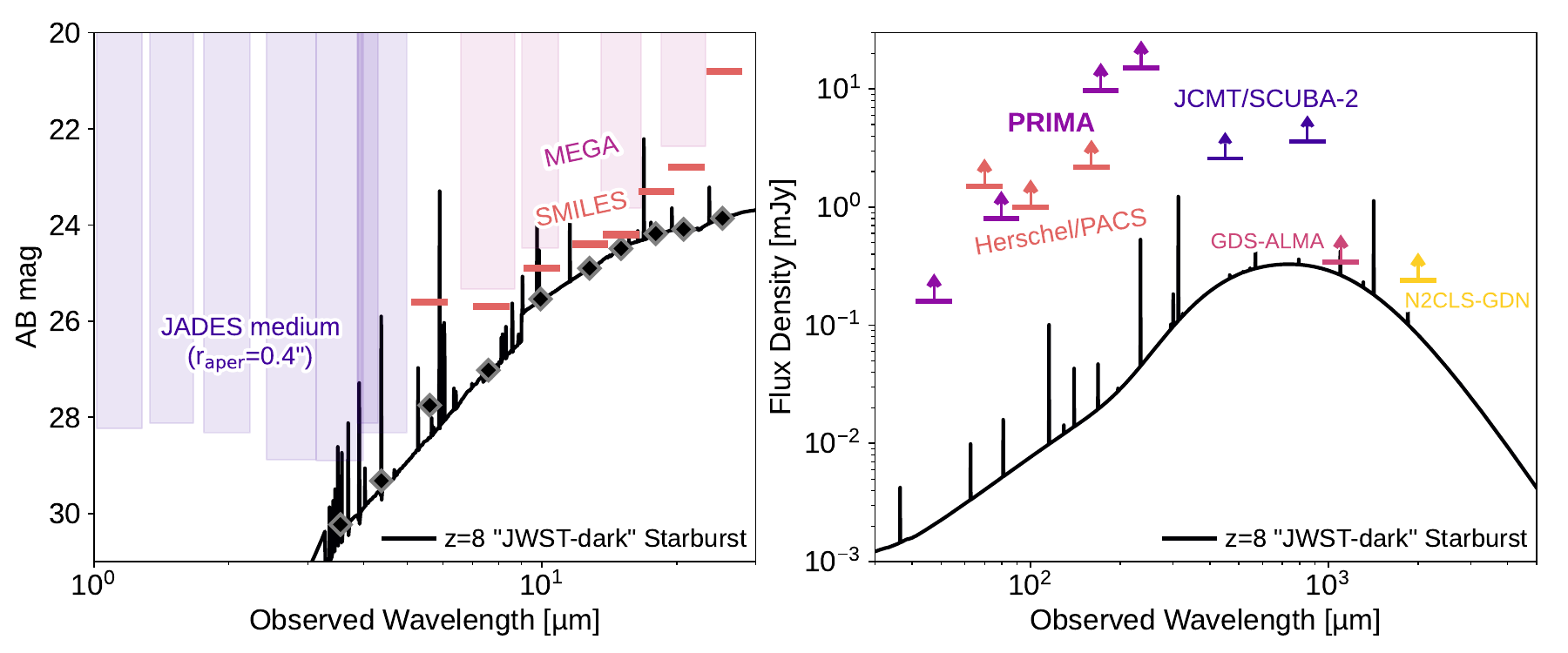}
\caption{The best-fit SED model of J0305m3150.C05 but redshifted to $z=8$ and scaled to a SFR of 60\,\smpy\ (solid black lines).
\textbf{Left}: the near-to-mid-infrared SED compared with the $5\sigma$ detection limits of representative JWST NIRCam and MIRI imaging surveys, including JADES-medium at 1--5\,\micron\ \citep[indigo;][]{eisenstein23a,deugenio25a}, SMILES at 5--25\,\micron\ \citep[coral;][]{riekeg24,alberts24a} and MEGA at 7--21\,\micron\ \citep[purple;][]{backhaus25}.
The gray-edged diamonds denote the brightnesses of the SED template in the corresponding JWST bands.
\textbf{Right}: the far-infrared SED compared with the $5\sigma$ detection limits of Herschel/PACS at 70--160\,\micron\ (from the PEP survey; \citealt{lutz11}), the planned PRIMA/PRIMAger at 47--235\,\micron\ (including realistic confusion noise; \citealt{donnellan24}), JCMT/SCUBA-2 at 450 and 850\,\micron\ \citep[e.g.,][]{limc20a}, the GOODS-ALMA v2.0 survey at 1.1\,mm \citep[][]{gomez22a}, the N2CLS survey in the GOODS-N field at 2\,\micron\ \citep{bingl23}.
Such a predicted ``JWST-dark'' galaxy at $z=8$, if it exists, will remain undetected with all aforementioned surveys or planned facilities.
}
\label{fig:z8}
\end{figure*}

\subsection{Multi-wavelength detectability}
\label{ss:04c_det}

We further investigate the multi-wavelength detectability of NIRCam-dark galaxies in wide-area infrared and millimeter surveys, to assess whether NIRCam-dark galaxies are indeed a population missed by most of the high-redshift galaxy surveys obtained so far.
The NIRCam imaging depth of the ASPIRE survey is slightly shallower than the average depth of CEERS \citep{finkelstein25}, but is deeper than wider surveys like COSMOS-Web \citep{casey23} and COSMOS-3D (JWST-GO-5893, PI: Kakiichi, K.; cf.\ \citealt{linx25b}).
At $z = 6.6$, the F356W brightness of \tgtb\ (and likely \tgta) can be detected by surveys like JADES-medium \citep[][$\gtrsim 125$\,arcmin$^2$]{eisenstein23a, deugenio25a} down to $\sim29$\,AB mag with aperture size of $r=0\farcs4$ ($5\sigma$).

Indeed, a NIRCam-dark MIRI-only source was reported by \citet{pg24a} in the HUDF, and two heavily reddened NIRCam-faint MIRI sources at photometric redshifts $z \sim 8$ have also been reported by \citet{akins23a} in the COSMOS field.
We note that these MIRI sources are rather compact, making them resembling reddened AGNs instead of DSFGs (similar to \textit{Virgil} at $z=6.6$, \citealt{rinaldi25a} and other HST-dark MIRI sources at $z>6$ in \citealt{williams24}).
NIRCam-dark ALMA-only sources have also been reported by \citet{fujimoto23a} through a deep 1.2-mm survey in the Abell\,2744 lensing cluster field and \citet{bakx24} as \cii\ emitters in quasar fields \citep{venemans20}.
However, we caution the relatively low S/N of these sources ($\lesssim5$ for continuum and $\lesssim6$ for emission-line scan within the ALMA spectral cubes).
Further deeper imaging and spectroscopic confirmation are necessary to confirm the fidelity, redshifts and nature of these sources.
We also remark that SPT0311-58, a extraordinarily luminous lensed DSFG at $z=6.90$ \citep{strandet17} can be classified as a NIRCam-dark galaxy according to the MIRI 10-\micron\ detection \citep{am23a} but non-detection at $\leq5$\micron\ with NIRSpec \citep{arribas24}, although the contamination from the foreground lens may complicate the interpretation.

At higher redshifts, the IR detectability of NIRCam-dark galaxies (and DSFGs in general) may rapidly decrease because of the surface brightness dimming and K correction.
Figure~\ref{fig:z8} shows the SED model of \tgta\ but redshifted to $z=8$, scaled to a SFR of 60\,\smpy.
The existence of galaxies with such SFR at $z\sim8$ is expected from the SFH of massive quiescent galaxies at $z\gtrsim4$, while most of the galaxies at $z\gtrsim8$ are observed at $M_\mathrm{UV} \lesssim -22$\,mag (one of the brightest is EGSz8p7 at z=8.68; \citealt{zitrin15, finkelstein24, harikane24a}), corresponding to unobscured SFR $\lesssim30$\,\smpy.
In other words, dust-obscured star-forming galaxies at $z>8$ are clearly missing from the current JWST observations (or at least spectroscopic confirmation); one example of such a galaxy is MACS0416\_Y1 at $z=8.31$ where the majority of the SFR is dust-obscured \citep{tamura19}. 
We also note that MACS0416\_Y1 likely resides in a complex galaxy overdensity (and thus massive halo) at $z\simeq8.3-8.5$ in and around the MACS0416 cluster field \citep{maz24, fudamoto25a}.

We compare with the detection limits of multiple surveys at the JWST wavelengths and also far-infrared to millimeter wavelengths.
At NIRCam wavelengths, the $z=8$ galaxy template will fall below the $5\sigma$ detection limit of wide surveys like JADES-medium (assuming $r=0\farcs4$ aperture), not to mention shallower surveys like CEERS \citep{finkelstein25}, PRIMER (JWST-GO-1837, PI: J.\ Dunlop) and COSMOS-web \citep{casey23}.
At MIRI wavelengths, the galaxy will fall below the $5\sigma$ detection limit of wide surveys like SMILES \citep{alberts24a, riekeg24} and MEGA \citep{backhaus25}, which is also the case for other wide surveys including PRIMER, COSMOS-Web and MEOW (JWST-GO-5407; PI: G.\ Leung).
In these existing wide-area surveys, these galaxies will even appear ``JWST-dark''.


In the far-IR, by considering realistic confusion noise limit, such a ``JWST-dark'' galaxy at $z=8$ will fall below the detection limit of Herschel/PACS and SPIRE at 70--500\,\micron\ (e.g., \citealt{nguyen10,lutz11}), JCMT/SCUBA-2 at 450 and 850\,\micron\ (e.g., \citealt{limc20a}) and even the planned PRIMA/PRIMAger at 47--235\,\micron\ \citep[][]{donnan24}.
It will also fall below the confusion limit of single-dish millimeter facilities like IRAM30m/NIKA2 at 1.2 and 2\,mm \citep[e.g.,][]{bingl23}, and also wide-field ALMA Band-6 continuum surveys like GOODS-ALMA v2.0 \citep[$\sim$72\,arcmin$^2$;][]{gomez22a}, not to mention shallower surveys like (Ex)-MORA \citep{casey21, long24b} and Cycle-10 large program CHAMPS (2023.1.00180.L; PI: A.\ Faisst).
Therefore, we conclude that current wide-area extragalactic surveys are generally insensitive to ``JWST-dark'' galaxies at $z\sim8$.

If ``JWST-dark'' galaxies do exist at such an early epoch, how could we detect and confirm these galaxies?
In fact, from Figure~\ref{fig:z8}, it is evident that the galaxy brightness is not far from the current detection limits in certain bands, specifically NIRCam F444W, MIRI F1000W$\sim$F1500W and ALMA Band-6/7.
With a few hours of JWST integration and probably less than an hour of ALMA integration (especially with the upcoming wideband sensitivity upgrade; \citealt{carpenter23}), these ``JWST-dark'' galaxies can be detected and studied with JWST and ALMA.
We argue that the key is to locate the targets prior to the MIRI and ALMA follow-up observations with relatively small FoVs.
This requires the selection of massive dark matter halos at $z\sim8$ in the first place (similar to the ASPIRE approach), possibly through wide-field NIRCam photometry and slitless spectroscopy of emission-line galaxies at $z\gtrsim8$ as protocluster tracers \citep[e.g.,][]{helton24b, maz24, fudamoto25a, liq25}.

Through the JWST Cycle-3 large NIRCam WFSS campaigns like SAPPHIRES \citep{sunf25b}, NEXUS \citep{sheny24}, COSMOS-3D, and POPPIES (JWST-GO-5398, PI: Kartaltepe, J. \& Rafelski, M.), more protocluster candidates will be selected at $z\gtrsim8$.
The pure-parallel NIRCam WFSS programs (e.g., SAPPHIRES and POPPIES) are particularly important because they could overcome the strong cosmic variance through multiple independent sightlines (see \citealt{sunf25b}).
Combined with pointed NIRCam WFSS observations of massive galaxies (e.g., JWST-GO-6480, PI: Schouws, S.), these surveys will further unveil the existence of NIRCam-dark or even JWST-dark galaxies, measuring their clustering properties and halo occupation fraction (e.g., differentiating the light versus heavy model in Section~\ref{ss:04b_number}), and thus providing key insights into the evolution history of massive galaxies and the dust-obscured cosmic star-formation history deeply into the EoR.

\section{Summary}
\label{sec:05_sum}

We present a study of two NIRCam-dark dusty star-forming galaxies at $z=6.6$.
These two galaxies are identified as companions to quasars through the ASPIRE JWST and ALMA survey, which targeted 25 UV-luminous quasars and their environments at $z \simeq 6.5 - 6.8$.
The main results are summarized as follows:

\begin{enumerate}
\item We securely detect ($>8\sigma$) the dust continuum emissions of the two galaxies through multiple ALMA bands, and the redshifts are confirmed through \cii\ line spectroscopy, placing them as quasar companions at $z=6.6$.
In contrast to the vast majority of DSFGs found through ALMA surveys in blank fields, these two sources are extraordinarily faint at the NIRCam wavelengths (F356W\,$>$\,28.0\,mag for \tgta\ and 26.5$\pm$0.1\,mag for \tgtb).
We refer to them as NIRCam-dark galaxies because of their near-IR faintness ($S_\mathrm{3.6\,\mu m} / S_\mathrm{1.2\,mm} < 10^{-4}$).

\item We obtain physical SED modeling of the two NIRCam-dark galaxies. 
These galaxies are undergoing starbursts (SFR\,$\simeq 80 - 250$\,\smpy), appear more obscured than Arp\,220 (whole galaxy) and optically faint DSFGs at $z\sim3$ selected through the ALESS survey \citep{dacunha15}.
The stellar masses are poorly constrained ($\log(M_\mathrm{star}/M_\odot)\simeq 10.0 - 10.5$ with an error of 0.5\,dex) because of the near-IR faintness.

\item Given the mass, SFR and redshifts, we show that the NIRCam-dark galaxies are viable progenitors of massive quiescent galaxies at $z\gtrsim4$ according to their star formation histories.
They could also be the descendants of certain UV-luminous galaxies at $z>10$ (``blue monsters''), bridging the two populations of massive galaxies before and after the epoch of reionization through an evolutionary perspective.

\item We show that the number density of NIRCam-dark galaxies is highly unconstrained from biased surveys like ASPIRE ($n\simeq10^{-8.1} \sim 10^{-5.5}$\,\si{Mpc^{-3}}). 
However, based on a light halo occupation model that could match the ASPIRE observations, the number density of NIRCam-dark galaxies at $z\sim6.6$ could reach $n = 3_{-2}^{+4} \times 10^{-6}$\,\si{Mpc^{-3}}.
If true, this will equal to $\sim30$\% of the number density of massive quiescent galaxies at $z\gtrsim4$ and blue monsters at $z>10$, suggesting a substantial contribution to the dust-obscured cosmic star-formation rate density at $z\sim7$ through the ``NIRCam-dark'' population.

\item From the SFH of massive quiescent galaxies at $z\gtrsim4$, galaxies with similar SFR should also exist at $z\sim8$.
If a substantial fraction of them share a similar SED to that of \tgta, these galaxies will reside below the detection limits of most wide-area JWST, ALMA and other single-dish (sub)-millimeter surveys obtained so far, and thus remaining ``JWST-dark''.
To study these ``JWST-dark'' galaxies, large-area JWST/NIRCam imaging and WFSS surveys of early galaxy protoclusters ($z\gtrsim8$) are essential.
Deep JWST and ALMA follow-up observations of these protoclusters will likely detect such galaxies, offering deep insight to the dust-obscured cosmic star formation in the Epoch of Reionization.

\end{enumerate}

\section*{Acknowledgment}

F.S. acknowledges funding from JWST/NIRCam contract to the University of Arizona, NAS5-02105 and support for JWST program \#2883, 4924, 5105, 6434 provided by NASA through grants from the Space Telescope Science Institute, which is operated by the Association of Universities for Research in Astronomy, Inc., under NASA contract NAS 5-03127. 
F.W. acknowledges support from NSF award AST-2513040.
D.J.E’s contributions were supported by JWST/NIRCam contract to the University of Arizona, NAS5-02015, and as a Simons Foundation Investigator.
S.E.I.B.\ is supported by the Deutsche Forschungsgemeinschaft (DFG) under Emmy Noether grant number BO 5771/1-1.
L.C.\ acknowledges support by grant PIB2021-127718NB-100 from the Spanish Ministry of Science and Innovation/State Agency of Research MCIN/AEI/10.13039/501100011033 and by “ERDF A way of making Europe”
C.M.\ acknowledges support from Fondecyt Iniciacion grant 11240336 and the ANID BASAL project FB210003. 
F.S.\ thanks Tiger Hsiao and James Trussler for helpful discussions.

This paper makes use of the following ALMA data: ADS/JAO.ALMA\#2017.1.00139.S, 2017.1.01532.S, 2019.1.00147.S, 2021.1.00443.S and 2022.1.01077.L. ALMA is a partnership of ESO (representing its member states), NSF (USA) and NINS (Japan), together with NRC (Canada), MOST and ASIAA (Taiwan), and KASI (Republic of Korea), in cooperation with the Republic of Chile. The Joint ALMA Observatory is operated by ESO, AUI/NRAO and NAOJ.
The National Radio Astronomy Observatory is a facility of the National Science Foundation operated under cooperative agreement by Associated Universities, Inc.

This work is based on observations made with the NASA/ESA/CSA James Webb Space Telescope. The data were obtained from the Mikulski Archive for Space Telescopes at the Space Telescope Science Institute, which is operated by the Association of Universities for Research in Astronomy, Inc., under NASA contract NAS 5-03127 for JWST. These observations are associated with program \#2078.
Support for program \#2078 was provided by NASA through a grant from the Space Telescope Science Institute, which is operated by the Association of Universities for Research in Astronomy, Inc., under NASA contract NAS 5-03127.

\begin{contribution}

FS led the ALMA data processing, analyses of the targets and the paper writing.
JY led the NIRCam imaging data processing.
FW led the JWST and ALMA observing programs.
All co-authors contributed to the scientific interpretation of the results and helped to write the manuscript.


\end{contribution}

%
\facilities{ALMA, JWST(NIRCam)}

\software{\textsc{astropy} \citep{2013A&A...558A..33A,2018AJ....156..123A}, 
\textsc{casa} \citep{casa22},
\textsc{cigale} \citep{cigale09,cigale19},
\textsc{jwst} \citep{jwst},
\textsc{photutils} \citep{photutils}
}

\setcounter{figure}{0}
\renewcommand{\thefigure}{\thesection\arabic{figure}}




\bibliography{00_main}{}

\begin{thebibliography}{}
\expandafter\ifx\csname natexlab\endcsname\relax\def\natexlab#1{#1}\fi
\providecommand{\url}[1]{\href{#1}{#1}}
\providecommand{\dodoi}[1]{doi:~\href{http://doi.org/#1}{\nolinkurl{#1}}}
\providecommand{\doeprint}[1]{\href{http://ascl.net/#1}{\nolinkurl{http://ascl.net/#1}}}
\providecommand{\doarXiv}[1]{\href{https://arxiv.org/abs/#1}{\nolinkurl{https://arxiv.org/abs/#1}}}

\bibitem[{H.~B. {Akins} {et~al.}(2023){Akins}, {Casey}, {Allen}, {Bagley},
  {Dickinson}, {Finkelstein}, {Franco}, {Harish}, {Arrabal Haro}, {Ilbert},
  {Kartaltepe}, {Koekemoer}, {Liu}, {Long}, {McCracken}, {Paquereau},
  {Papovich}, {Pirzkal}, {Rhodes}, {Robertson}, {Shuntov}, {Toft}, {Yang},
  {Barro}, {Bisigello}, {Buat}, {Champagne}, {Cooper}, {Costantin}, {de La
  Vega}, {Drakos}, {Faisst}, {Fontana}, {Fujimoto}, {Gillman},
  {G{\'o}mez-Guijarro}, {Gozaliasl}, {Hathi}, {Hayward}, {Hirschmann},
  {Holwerda}, {Jin}, {Kocevski}, {Kokorev}, {Lambrides}, {Lucas}, {Magdis},
  {Magnelli}, {McKinney}, {Mobasher}, {P{\'e}rez-Gonz{\'a}lez}, {Rich},
  {Seill{\'e}}, {Talia}, {Urry}, {Valentino}, {Whitaker}, {Yung}, {Zavala},
  {Cosmos-Web Team}, \& {Ceers Team}}]{akins23a}
{Akins}, H.~B., {Casey}, C.~M., {Allen}, N., {et~al.} 2023,
  \bibinfo{title}{{Two Massive, Compact, and Dust-obscured Candidate z ≃ 8
  Galaxies Discovered by JWST},} \apj, 956, 61,
  \dodoi{10.3847/1538-4357/acef21}

\bibitem[{S. {Alberts} {et~al.}(2024){Alberts}, {Lyu}, {Shivaei}, {Rieke},
  {P{\'e}rez-Gonz{\'a}lez}, {Bonaventura}, {Zhu}, {Helton}, {Ji}, {Morrison},
  {Robertson}, {Stone}, {Sun}, {Williams}, \& {Willmer}}]{alberts24a}
{Alberts}, S., {Lyu}, J., {Shivaei}, I., {et~al.} 2024, \bibinfo{title}{{SMILES
  Initial Data Release: Unveiling the Obscured Universe with MIRI Multiband
  Imaging},} \apj, 976, 224, \dodoi{10.3847/1538-4357/ad7396}

\bibitem[{H.~S.~B. {Algera} {et~al.}(2023){Algera}, {Inami}, {Oesch},
  {Sommovigo}, {Bouwens}, {Topping}, {Schouws}, {Stefanon}, {Stark}, {Aravena},
  {Barrufet}, {da Cunha}, {Dayal}, {Endsley}, {Ferrara}, {Fudamoto},
  {Gonzalez}, {Graziani}, {Hodge}, {Hygate}, {de Looze}, {Nanayakkara},
  {Schneider}, \& {van der Werf}}]{algera23}
{Algera}, H. S.~B., {Inami}, H., {Oesch}, P.~A., {et~al.} 2023,
  \bibinfo{title}{{The ALMA REBELS survey: the dust-obscured cosmic star
  formation rate density at redshift 7},} \mnras, 518, 6142,
  \dodoi{10.1093/mnras/stac3195}

\bibitem[{J. {{\'A}lvarez-M{\'a}rquez} {et~al.}(2023){{\'A}lvarez-M{\'a}rquez},
  {Crespo G{\'o}mez}, {Colina}, {Neeleman}, {Walter}, {Labiano},
  {P{\'e}rez-Gonz{\'a}lez}, {Bik}, {Noorgaard-Nielsen}, {Ostlin}, {Wright},
  {Alonso-Herrero}, {Azollini}, {Caputi}, {Eckart}, {Le F{\`e}vre},
  {Garc{\'\i}a-Mar{\'\i}n}, {Greve}, {Hjorth}, {Ilbert}, {Kendrew}, {Pye},
  {Tikkanen}, {Topinka}, {van der Werf}, {Ward}, {van Dishoeck}, {G{\"u}del},
  {Henning}, {Lagage}, {Ray}, \& {Waelkens}}]{am23a}
{{\'A}lvarez-M{\'a}rquez}, J., {Crespo G{\'o}mez}, A., {Colina}, L., {et~al.}
  2023, \bibinfo{title}{{MIRI/JWST observations reveal an extremely obscured
  starburst in the z = 6.9 system SPT0311-58},} \aap, 671, A105,
  \dodoi{10.1051/0004-6361/202245400}

\bibitem[{M. {Aravena} {et~al.}(2020){Aravena}, {Boogaard},
  {G{\'o}nzalez-L{\'o}pez}, {Decarli}, {Walter}, {Carilli}, {Smail}, {Weiss},
  {Assef}, {Bauer}, {Bouwens}, {Cortes}, {Cox}, {da Cunha}, {Daddi},
  {D{\'\i}az-Santos}, {Inami}, {Ivison}, {Novak}, {Popping}, {Riechers}, {van
  der Werf}, \& {Wagg}}]{aravena20}
{Aravena}, M., {Boogaard}, L., {G{\'o}nzalez-L{\'o}pez}, J., {et~al.} 2020,
  \bibinfo{title}{{The ALMA Spectroscopic Survey in the Hubble Ultra Deep
  Field: The nature of the faintest dusty star-forming galaxies},} arXiv
  e-prints, arXiv:2006.04284.
\newblock \doarXiv{2006.04284}

\bibitem[{P. {Arrabal Haro} {et~al.}(2023){Arrabal Haro}, {Dickinson},
  {Finkelstein}, {Kartaltepe}, {Donnan}, {Burgarella}, {Carnall}, {Cullen},
  {Dunlop}, {Fern{\'a}ndez}, {Fujimoto}, {Jung}, {Krips}, {Larson}, {Papovich},
  {P{\'e}rez-Gonz{\'a}lez}, {Amor{\'\i}n}, {Bagley}, {Buat}, {Casey},
  {Chworowsky}, {Cohen}, {Ferguson}, {Giavalisco}, {Huertas-Company},
  {Hutchison}, {Kocevski}, {Koekemoer}, {Lucas}, {McLeod}, {McLure}, {Pirzkal},
  {Seill{\'e}}, {Trump}, {Weiner}, {Wilkins}, \& {Zavala}}]{arrabalharo23}
{Arrabal Haro}, P., {Dickinson}, M., {Finkelstein}, S.~L., {et~al.} 2023,
  \bibinfo{title}{{Confirmation and refutation of very luminous galaxies in the
  early Universe},} \nat, 622, 707, \dodoi{10.1038/s41586-023-06521-7}

\bibitem[{S. {Arribas} {et~al.}(2024){Arribas}, {Perna}, {Rodr{\'\i}guez Del
  Pino}, {Lamperti}, {D'Eugenio}, {P{\'e}rez-Gonz{\'a}lez}, {Jones}, {Crespo
  G{\'o}mez}, {Curti}, {Lim}, {{\'A}lvarez-M{\'a}rquez}, {Bunker}, {Carniani},
  {Charlot}, {Jakobsen}, {Maiolino}, {{\"U}bler}, {Willott}, {B{\"o}ker},
  {Chevallard}, {Circosta}, {Cresci}, {Kumari}, {Parlanti}, {Scholtz},
  {Venturi}, \& {Witstok}}]{arribas24}
{Arribas}, S., {Perna}, M., {Rodr{\'\i}guez Del Pino}, B., {et~al.} 2024,
  \bibinfo{title}{{GA-NIFS: The core of an extremely massive protocluster at
  the epoch of reionisation probed with JWST/NIRSpec},} \aap, 688, A146,
  \dodoi{10.1051/0004-6361/202348824}

\bibitem[{ {Astropy Collaboration} {et~al.}(2013){Astropy Collaboration},
  {Robitaille}, {Tollerud}, {Greenfield}, {Droettboom}, {Bray}, {Aldcroft},
  {Davis}, {Ginsburg}, {Price-Whelan}, {Kerzendorf}, {Conley}, {Crighton},
  {Barbary}, {Muna}, {Ferguson}, {Grollier}, {Parikh}, {Nair}, {Unther},
  {Deil}, {Woillez}, {Conseil}, {Kramer}, {Turner}, {Singer}, {Fox}, {Weaver},
  {Zabalza}, {Edwards}, {Azalee Bostroem}, {Burke}, {Casey}, {Crawford},
  {Dencheva}, {Ely}, {Jenness}, {Labrie}, {Lim}, {Pierfederici}, {Pontzen},
  {Ptak}, {Refsdal}, {Servillat}, \& {Streicher}}]{2013A&A...558A..33A}
{Astropy Collaboration}, {Robitaille}, T.~P., {Tollerud}, E.~J., {et~al.} 2013,
  \bibinfo{title}{{Astropy: A community Python package for astronomy},} \aap,
  558, A33, \dodoi{10.1051/0004-6361/201322068}

\bibitem[{ {Astropy Collaboration} {et~al.}(2018){Astropy Collaboration},
  {Price-Whelan}, {Sip{\H{o}}cz}, {G{\"u}nther}, {Lim}, {Crawford}, {Conseil},
  {Shupe}, {Craig}, {Dencheva}, {Ginsburg}, {VanderPlas}, {Bradley},
  {P{\'e}rez-Su{\'a}rez}, {de Val-Borro}, {Aldcroft}, {Cruz}, {Robitaille},
  {Tollerud}, {Ardelean}, {Babej}, {Bach}, {Bachetti}, {Bakanov}, {Bamford},
  {Barentsen}, {Barmby}, {Baumbach}, {Berry}, {Biscani}, {Boquien}, {Bostroem},
  {Bouma}, {Brammer}, {Bray}, {Breytenbach}, {Buddelmeijer}, {Burke},
  {Calderone}, {Cano Rodr{\'\i}guez}, {Cara}, {Cardoso}, {Cheedella}, {Copin},
  {Corrales}, {Crichton}, {D'Avella}, {Deil}, {Depagne}, {Dietrich}, {Donath},
  {Droettboom}, {Earl}, {Erben}, {Fabbro}, {Ferreira}, {Finethy}, {Fox},
  {Garrison}, {Gibbons}, {Goldstein}, {Gommers}, {Greco}, {Greenfield},
  {Groener}, {Grollier}, {Hagen}, {Hirst}, {Homeier}, {Horton}, {Hosseinzadeh},
  {Hu}, {Hunkeler}, {Ivezi{\'c}}, {Jain}, {Jenness}, {Kanarek}, {Kendrew},
  {Kern}, {Kerzendorf}, {Khvalko}, {King}, {Kirkby}, {Kulkarni}, {Kumar},
  {Lee}, {Lenz}, {Littlefair}, {Ma}, {Macleod}, {Mastropietro}, {McCully},
  {Montagnac}, {Morris}, {Mueller}, {Mumford}, {Muna}, {Murphy}, {Nelson},
  {Nguyen}, {Ninan}, {N{\"o}the}, {Ogaz}, {Oh}, {Parejko}, {Parley}, {Pascual},
  {Patil}, {Patil}, {Plunkett}, {Prochaska}, {Rastogi}, {Reddy Janga},
  {Sabater}, {Sakurikar}, {Seifert}, {Sherbert}, {Sherwood-Taylor}, {Shih},
  {Sick}, {Silbiger}, {Singanamalla}, {Singer}, {Sladen}, {Sooley},
  {Sornarajah}, {Streicher}, {Teuben}, {Thomas}, {Tremblay}, {Turner},
  {Terr{\'o}n}, {van Kerkwijk}, {de la Vega}, {Watkins}, {Weaver}, {Whitmore},
  {Woillez}, {Zabalza}, \& {Astropy Contributors}}]{2018AJ....156..123A}
{Astropy Collaboration}, {Price-Whelan}, A.~M., {Sip{\H{o}}cz}, B.~M., {et~al.}
  2018, \bibinfo{title}{{The Astropy Project: Building an Open-science Project
  and Status of the v2.0 Core Package},} \aj, 156, 123,
  \dodoi{10.3847/1538-3881/aabc4f}

\bibitem[{B.~E. {Backhaus} {et~al.}(2025){Backhaus}, {Kirkpatrick}, {Yang},
  {Troiani}, {Hamblin}, {Kartaltepe}, {Kocevski}, {Koekemoer}, {Lambrides},
  {Papovich}, \& {Ronayne}}]{backhaus25}
{Backhaus}, B.~E., {Kirkpatrick}, A., {Yang}, G., {et~al.} 2025,
  \bibinfo{title}{{MEGA Mass Assembly with JWST: The MIRI EGS Galaxy and AGN
  Survey},} arXiv e-prints, arXiv:2503.19078, \dodoi{10.48550/arXiv.2503.19078}

\bibitem[{W.~M. {Baker} {et~al.}(2025){Baker}, {Lim}, {D'Eugenio}, {Maiolino},
  {Ji}, {Arribas}, {Bunker}, {Carniani}, {Charlot}, {de Graaff}, {Hainline},
  {Looser}, {Lyu}, {Rinaldi}, {Robertson}, {Schaller}, {Schaye}, {Scholtz},
  {{\"U}bler}, {Williams}, {Willmer}, {Willott}, \& {Zhu}}]{baker25a}
{Baker}, W.~M., {Lim}, S., {D'Eugenio}, F., {et~al.} 2025, \bibinfo{title}{{The
  abundance and nature of high-redshift quiescent galaxies from JADES
  spectroscopy and the FLAMINGO simulations},} \mnras, 539, 557,
  \dodoi{10.1093/mnras/staf475}

\bibitem[{T.~J.~L.~C. {Bakx} {et~al.}(2024){Bakx}, {Algera}, {Venemans},
  {Sommovigo}, {Fujimoto}, {Carniani}, {Hagimoto}, {Hashimoto}, {Inoue},
  {Salak}, {Serjeant}, {Vallini}, {Eales}, {Ferrara}, {Fudamoto}, {Imamura},
  {Inoue}, {Knudsen}, {Matsuo}, {Sugahara}, {Tamura}, {Taniguchi}, \&
  {Yamanaka}}]{bakx24}
{Bakx}, T. J.~L.~C., {Algera}, H. S.~B., {Venemans}, B., {et~al.} 2024,
  \bibinfo{title}{{Gas conditions of a star-formation selected sample in the
  first billion years},} \mnras, 532, 2270, \dodoi{10.1093/mnras/stae1613}

\bibitem[{L. {Barrufet} {et~al.}(2023){Barrufet}, {Oesch}, {Bouwens}, {Inami},
  {Sommovigo}, {Algera}, {da Cunha}, {Aravena}, {Dayal}, {Ferrara}, {Fudamoto},
  {Gonzalez}, {Graziani}, {Hygate}, {de Looze}, {Nanayakkara}, {Pallottini},
  {Schneider}, {Stefanon}, {Topping}, \& {van der Werf}}]{barrufet23a}
{Barrufet}, L., {Oesch}, P.~A., {Bouwens}, R., {et~al.} 2023,
  \bibinfo{title}{{The ALMA REBELS Survey: the first infrared luminosity
  function measurement at z {\ensuremath{\sim}} 7},} \mnras, 522, 3926,
  \dodoi{10.1093/mnras/stad1259}

\bibitem[{L. {Bing} {et~al.}(2023){Bing}, {B{\'e}thermin}, {Lagache}, {Adam},
  {Ade}, {Ajeddig}, {Andr{\'e}}, {Artis}, {Aussel}, {Beelen}, {Beno{\^\i}t},
  {Berta}, {Billot}, {Bourrion}, {Calvo}, {Catalano}, {De Petris},
  {D{\'e}sert}, {Doyle}, {Driessen}, {Elbaz}, {Gkogkou}, {Gomez}, {Goupy},
  {Hanser}, {K{\'e}ruzor{\'e}}, {Kramer}, {Ladjelate}, {Liu}, {Leclercq},
  {Lestrade}, {Lustig}, {Mac{\'\i}as-P{\'e}rez}, {Maury}, {Mauskopf}, {Mayet},
  {Monfardini}, {Mu{\~n}oz-Echeverr{\'\i}a}, {Perotto}, {Pisano}, {Ponthieu},
  {Rev{\'e}ret}, {Rigby}, {Ritacco}, {Romero}, {Roussel}, {Ruppin}, {Schuster},
  {Sievers}, {Tucker}, \& {Zylka}}]{bingl23}
{Bing}, L., {B{\'e}thermin}, M., {Lagache}, G., {et~al.} 2023,
  \bibinfo{title}{{NIKA2 Cosmological Legacy Survey. Survey description and
  galaxy number counts},} \aap, 677, A66, \dodoi{10.1051/0004-6361/202346579}

\bibitem[{L.~A. {Boogaard} {et~al.}(2024){Boogaard}, {Gillman}, {Melinder},
  {Walter}, {Colina}, {{\"O}stlin}, {Caputi}, {Iani}, {P{\'e}rez-Gonz{\'a}lez},
  {van der Werf}, {Greve}, {Wright}, {Alonso-Herrero},
  {{\'A}lvarez-M{\'a}rquez}, {Annunziatella}, {Bik}, {Bosman}, {Costantin},
  {Crespo G{\'o}mez}, {Dicken}, {Eckart}, {Hjorth}, {Jermann}, {Labiano},
  {Langeroodi}, {Meyer}, {Moutard}, {Pei{\ss}ker}, {Pye}, {Rinaldi},
  {Tikkanen}, {Topinka}, \& {Henning}}]{boogaard24}
{Boogaard}, L.~A., {Gillman}, S., {Melinder}, J., {et~al.} 2024,
  \bibinfo{title}{{MIDIS: JWST/MIRI Reveals the Stellar Structure of
  ALMA-selected Galaxies in the Hubble Ultra Deep Field at Cosmic Noon},} \apj,
  969, 27, \dodoi{10.3847/1538-4357/ad43e5}

\bibitem[{M. {Boquien} {et~al.}(2019){Boquien}, {Burgarella}, {Roehlly},
  {Buat}, {Ciesla}, {Corre}, {Inoue}, \& {Salas}}]{cigale19}
{Boquien}, M., {Burgarella}, D., {Roehlly}, Y., {et~al.} 2019,
  \bibinfo{title}{{CIGALE: a python Code Investigating GALaxy Emission},} \aap,
  622, A103, \dodoi{10.1051/0004-6361/201834156}

\bibitem[{R.~J. {Bouwens} {et~al.}(2022){Bouwens}, {Smit}, {Schouws},
  {Stefanon}, {Bowler}, {Endsley}, {Gonzalez}, {Inami}, {Stark}, {Oesch},
  {Hodge}, {Aravena}, {da Cunha}, {Dayal}, {de Looze}, {Ferrara}, {Fudamoto},
  {Graziani}, {Li}, {Nanayakkara}, {Pallottini}, {Schneider}, {Sommovigo},
  {Topping}, {van der Werf}, {Algera}, {Barrufet}, {Hygate}, {Labb{\'e}},
  {Riechers}, \& {Witstok}}]{bouwens22}
{Bouwens}, R.~J., {Smit}, R., {Schouws}, S., {et~al.} 2022,
  \bibinfo{title}{{Reionization Era Bright Emission Line Survey: Selection and
  Characterization of Luminous Interstellar Medium Reservoirs in the z > 6.5
  Universe},} \apj, 931, 160, \dodoi{10.3847/1538-4357/ac5a4a}

\bibitem[{L. Bradley {et~al.}(2024)Bradley, Sip{\H o}cz, Robitaille, Tollerud,
  Vin{\'{\i}}cius, Deil, Barbary, Wilson, Busko, Donath, G{\"u}nther, Cara,
  Lim, Me{\ss}linger, Burnett, Conseil, Droettboom, Bostroem, Bray, Bratholm,
  Jamieson, Ginsburg, Barentsen, Craig, Pascual, Rathi, Perrin, Morris, \&
  Perren}]{photutils}
Bradley, L., Sip{\H o}cz, B., Robitaille, T., {et~al.} 2024,
  \bibinfo{title}{astropy/photutils: 1.12.0,}, 1.12.0 Zenodo,
  \dodoi{10.5281/zenodo.10967176}

\bibitem[{G. {Bruzual} \& S. {Charlot}(2003){Bruzual} \& {Charlot}}]{bc03}
{Bruzual}, G., \& {Charlot}, S. 2003, \bibinfo{title}{{Stellar population
  synthesis at the resolution of 2003},} \mnras, 344, 1000,
  \dodoi{10.1046/j.1365-8711.2003.06897.x}

\bibitem[{A.~J. {Bunker} {et~al.}(2023){Bunker}, {Saxena}, {Cameron},
  {Willott}, {Curtis-Lake}, {Jakobsen}, {Carniani}, {Smit}, {Maiolino},
  {Witstok}, {Curti}, {D'Eugenio}, {Jones}, {Ferruit}, {Arribas}, {Charlot},
  {Chevallard}, {Giardino}, {de Graaff}, {Looser}, {L{\"u}tzgendorf}, {Maseda},
  {Rawle}, {Rix}, {Del Pino}, {Alberts}, {Egami}, {Eisenstein}, {Endsley},
  {Hainline}, {Hausen}, {Johnson}, {Rieke}, {Rieke}, {Robertson}, {Shivaei},
  {Stark}, {Sun}, {Tacchella}, {Tang}, {Williams}, {Willmer}, {Baker}, {Baum},
  {Bhatawdekar}, {Bowler}, {Boyett}, {Chen}, {Circosta}, {Helton}, {Ji},
  {Kumari}, {Lyu}, {Nelson}, {Parlanti}, {Perna}, {Sandles}, {Scholtz},
  {Suess}, {Topping}, {{\"U}bler}, {Wallace}, \& {Whitler}}]{bunker23}
{Bunker}, A.~J., {Saxena}, A., {Cameron}, A.~J., {et~al.} 2023,
  \bibinfo{title}{{JADES NIRSpec Spectroscopy of GN-z11:
  Lyman-{\ensuremath{\alpha}} emission and possible enhanced nitrogen abundance
  in a z = 10.60 luminous galaxy},} \aap, 677, A88,
  \dodoi{10.1051/0004-6361/202346159}

\bibitem[{H. Bushouse {et~al.}(2023)Bushouse, Eisenhamer, Dencheva, Davies,
  Greenfield, Morrison, Hodge, Simon, Grumm, Droettboom, Slavich, Sosey, Pauly,
  Miller, Jedrzejewski, Hack, Davis, Crawford, Law, Gordon, Regan, Cara,
  MacDonald, Bradley, Shanahan, Jamieson, Teodoro, \& Williams}]{jwst}
Bushouse, H., Eisenhamer, J., Dencheva, N., {et~al.} 2023, \bibinfo{title}{JWST
  Calibration Pipeline,}, 1.10.2 Zenodo, \dodoi{10.5281/zenodo.7829329}

\bibitem[{D. {Calzetti} {et~al.}(2000){Calzetti}, {Armus}, {Bohlin}, {Kinney},
  {Koornneef}, \& {Storchi-Bergmann}}]{calzetti00}
{Calzetti}, D., {Armus}, L., {Bohlin}, R.~C., {et~al.} 2000,
  \bibinfo{title}{{The Dust Content and Opacity of Actively Star-forming
  Galaxies},} \apj, 533, 682, \dodoi{10.1086/308692}

\bibitem[{A.~C. {Carnall} {et~al.}(2023{\natexlab{a}}){Carnall}, {McLure},
  {Dunlop}, {McLeod}, {Wild}, {Cullen}, {Magee}, {Begley}, {Cimatti}, {Donnan},
  {Hamadouche}, {Jewell}, \& {Walker}}]{carnall23a}
{Carnall}, A.~C., {McLure}, R.~J., {Dunlop}, J.~S., {et~al.}
  2023{\natexlab{a}}, \bibinfo{title}{{A massive quiescent galaxy at redshift
  4.658},} \nat, 619, 716, \dodoi{10.1038/s41586-023-06158-6}

\bibitem[{A.~C. {Carnall} {et~al.}(2023{\natexlab{b}}){Carnall}, {McLeod},
  {McLure}, {Dunlop}, {Begley}, {Cullen}, {Donnan}, {Hamadouche}, {Jewell},
  {Jones}, {Pollock}, \& {Wild}}]{carnall23b}
{Carnall}, A.~C., {McLeod}, D.~J., {McLure}, R.~J., {et~al.}
  2023{\natexlab{b}}, \bibinfo{title}{{A surprising abundance of massive
  quiescent galaxies at 3 < z < 5 in the first data from JWST CEERS},} \mnras,
  520, 3974, \dodoi{10.1093/mnras/stad369}

\bibitem[{A.~C. {Carnall} {et~al.}(2024){Carnall}, {Cullen}, {McLure},
  {McLeod}, {Begley}, {Donnan}, {Dunlop}, {Shapley}, {Rowlands}, {Almaini},
  {Arellano-C{\'o}rdova}, {Barrufet}, {Cimatti}, {Ellis}, {Grogin},
  {Hamadouche}, {Illingworth}, {Koekemoer}, {Leung}, {Lovell},
  {P{\'e}rez-Gonz{\'a}lez}, {Santini}, {Stanton}, \& {Wild}}]{carnall24}
{Carnall}, A.~C., {Cullen}, F., {McLure}, R.~J., {et~al.} 2024,
  \bibinfo{title}{{The JWST EXCELS survey: too much, too young, too fast?
  Ultra-massive quiescent galaxies at 3 < z < 5},} \mnras, 534, 325,
  \dodoi{10.1093/mnras/stae2092}

\bibitem[{S. {Carniani} {et~al.}(2024){Carniani}, {Hainline}, {D'Eugenio},
  {Eisenstein}, {Jakobsen}, {Witstok}, {Johnson}, {Chevallard}, {Maiolino},
  {Helton}, {Willott}, {Robertson}, {Alberts}, {Arribas}, {Baker},
  {Bhatawdekar}, {Boyett}, {Bunker}, {Cameron}, {Cargile}, {Charlot}, {Curti},
  {Curtis-Lake}, {Egami}, {Giardino}, {Isaak}, {Ji}, {Jones}, {Kumari},
  {Maseda}, {Parlanti}, {P{\'e}rez-Gonz{\'a}lez}, {Rawle}, {Rieke}, {Rieke},
  {Del Pino}, {Saxena}, {Scholtz}, {Smit}, {Sun}, {Tacchella}, {{\"U}bler},
  {Venturi}, {Williams}, \& {Willmer}}]{carniani24}
{Carniani}, S., {Hainline}, K., {D'Eugenio}, F., {et~al.} 2024,
  \bibinfo{title}{{Spectroscopic confirmation of two luminous galaxies at a
  redshift of 14},} \nat, 633, 318, \dodoi{10.1038/s41586-024-07860-9}

\bibitem[{J. {Carpenter} {et~al.}(2023){Carpenter}, {Brogan}, {Iono}, \&
  {Mroczkowski}}]{carpenter23}
{Carpenter}, J., {Brogan}, C., {Iono}, D., \& {Mroczkowski}, T. 2023, in
  Physics and Chemistry of Star Formation: The Dynamical ISM Across Time and
  Spatial Scales, ed. V.~{Ossenkopf-Okada}, R.~{Schaaf}, I.~{Breloy}, \&
  J.~{Stutzki}, 304, \dodoi{10.48550/arXiv.2211.00195}

\bibitem[{ {CASA Team} {et~al.}(2022){CASA Team}, {Bean}, {Bhatnagar},
  {Castro}, {Donovan Meyer}, {Emonts}, {Garcia}, {Garwood}, {Golap}, {Gonzalez
  Villalba}, {Harris}, {Hayashi}, {Hoskins}, {Hsieh}, {Jagannathan},
  {Kawasaki}, {Keimpema}, {Kettenis}, {Lopez}, {Marvil}, {Masters},
  {McNichols}, {Mehringer}, {Miel}, {Moellenbrock}, {Montesino}, {Nakazato},
  {Ott}, {Petry}, {Pokorny}, {Raba}, {Rau}, {Schiebel}, {Schweighart},
  {Sekhar}, {Shimada}, {Small}, {Steeb}, {Sugimoto}, {Suoranta}, {Tsutsumi},
  {van Bemmel}, {Verkouter}, {Wells}, {Xiong}, {Szomoru}, {Griffith},
  {Glendenning}, \& {Kern}}]{casa22}
{CASA Team}, {Bean}, B., {Bhatnagar}, S., {et~al.} 2022, \bibinfo{title}{{CASA,
  the Common Astronomy Software Applications for Radio Astronomy},} \pasp, 134,
  114501, \dodoi{10.1088/1538-3873/ac9642}

\bibitem[{C.~M. {Casey}(2012){Casey}}]{casey12}
{Casey}, C.~M. 2012, \bibinfo{title}{{Far-infrared spectral energy distribution
  fitting for galaxies near and far},} \mnras, 425, 3094,
  \dodoi{10.1111/j.1365-2966.2012.21455.x}

\bibitem[{C.~M. {Casey} {et~al.}(2021){Casey}, {Zavala}, {Manning}, {Aravena},
  {B{\'e}thermin}, {Caputi}, {Champagne}, {Clements}, {Drew}, {Finkelstein},
  {Fujimoto}, {Hayward}, {Dekel}, {Kokorev}, {Lagos}, {Long}, {Magdis}, {Man},
  {Mitsuhashi}, {Popping}, {Spilker}, {Staguhn}, {Talia}, {Toft}, {Treister},
  {Weaver}, \& {Yun}}]{casey21}
{Casey}, C.~M., {Zavala}, J.~A., {Manning}, S.~M., {et~al.} 2021,
  \bibinfo{title}{{Mapping Obscuration to Reionization with ALMA (MORA): 2 mm
  Efficiently Selects the Highest-redshift Obscured Galaxies},} \apj, 923, 215,
  \dodoi{10.3847/1538-4357/ac2eb4}

\bibitem[{C.~M. {Casey} {et~al.}(2023){Casey}, {Kartaltepe}, {Drakos},
  {Franco}, {Harish}, {Paquereau}, {Ilbert}, {Rose}, {Cox}, {Nightingale},
  {Robertson}, {Silverman}, {Koekemoer}, {Massey}, {McCracken}, {Rhodes},
  {Akins}, {Allen}, {Amvrosiadis}, {Arango-Toro}, {Bagley}, {Bongiorno},
  {Capak}, {Champagne}, {Chartab}, {Ch{\'a}vez Ortiz}, {Chworowsky}, {Cooke},
  {Cooper}, {Darvish}, {Ding}, {Faisst}, {Finkelstein}, {Fujimoto}, {Gentile},
  {Gillman}, {Gould}, {Gozaliasl}, {Hayward}, {He}, {Hemmati}, {Hirschmann},
  {Jahnke}, {Jin}, {Khostovan}, {Kokorev}, {Lambrides}, {Laigle}, {Larson},
  {Leung}, {Liu}, {Liaudat}, {Long}, {Magdis}, {Mahler}, {Mainieri}, {Manning},
  {Maraston}, {Martin}, {McCleary}, {McKinney}, {McPartland}, {Mobasher},
  {Pattnaik}, {Renzini}, {Rich}, {Sanders}, {Sattari}, {Scognamiglio},
  {Scoville}, {Sheth}, {Shuntov}, {Sparre}, {Suzuki}, {Talia}, {Toft},
  {Trakhtenbrot}, {Urry}, {Valentino}, {Vanderhoof}, {Vardoulaki}, {Weaver},
  {Whitaker}, {Wilkins}, {Yang}, \& {Zavala}}]{casey23}
{Casey}, C.~M., {Kartaltepe}, J.~S., {Drakos}, N.~E., {et~al.} 2023,
  \bibinfo{title}{{COSMOS-Web: An Overview of the JWST Cosmic Origins Survey},}
  \apj, 954, 31, \dodoi{10.3847/1538-4357/acc2bc}

\bibitem[{M. {Castellano} {et~al.}(2024){Castellano}, {Napolitano}, {Fontana},
  {Roberts-Borsani}, {Treu}, {Vanzella}, {Zavala}, {Arrabal Haro},
  {Calabr{\`o}}, {Llerena}, {Mascia}, {Merlin}, {Paris}, {Pentericci},
  {Santini}, {Bakx}, {Bergamini}, {Cupani}, {Dickinson}, {Filippenko},
  {Glazebrook}, {Grillo}, {Kelly}, {Malkan}, {Mason}, {Morishita},
  {Nanayakkara}, {Rosati}, {Sani}, {Wang}, \& {Yoon}}]{castellano24}
{Castellano}, M., {Napolitano}, L., {Fontana}, A., {et~al.} 2024,
  \bibinfo{title}{{JWST NIRSpec Spectroscopy of the Remarkable Bright Galaxy
  GHZ2/GLASS-z12 at Redshift 12.34},} \apj, 972, 143,
  \dodoi{10.3847/1538-4357/ad5f88}

\bibitem[{G. {Chabrier}(2003){Chabrier}}]{chabrier03}
{Chabrier}, G. 2003, \bibinfo{title}{{Galactic Stellar and Substellar Initial
  Mass Function},} \pasp, 115, 763, \dodoi{10.1086/376392}

\bibitem[{J.~B. {Champagne} {et~al.}(2025){Champagne}, {Wang}, {Zhang}, {Yang},
  {Fan}, {Hennawi}, {Sun}, {Ba{\~n}ados}, {Bosman}, {Costa}, {Eilers},
  {Endsley}, {Jin}, {Jun}, {Li}, {Lin}, {Liu}, {Loiacono}, {Lupi},
  {Mazzucchelli}, {Pudoka}, {Protu{\v{s}}ov{\`a}}, {Rojas-Ruiz}, {Tee},
  {Trebitsch}, {Venemans}, {Zhuang}, \& {Zou}}]{champagne25}
{Champagne}, J.~B., {Wang}, F., {Zhang}, H., {et~al.} 2025, \bibinfo{title}{{A
  Quasar-anchored Protocluster at z = 6.6 in the ASPIRE Survey. I. Properties
  of [O III] Emitters in a 10 Mpc Overdensity Structure},} \apj, 981, 113,
  \dodoi{10.3847/1538-4357/adb1bd}

\bibitem[{R. {Chary} \& D. {Elbaz}(2001){Chary} \& {Elbaz}}]{chary01}
{Chary}, R., \& {Elbaz}, D. 2001, \bibinfo{title}{{Interpreting the Cosmic
  Infrared Background: Constraints on the Evolution of the Dust-enshrouded Star
  Formation Rate},} \apj, 556, 562, \dodoi{10.1086/321609}

\bibitem[{C.-C. {Chen} {et~al.}(2015){Chen}, {Smail}, {Swinbank}, {Simpson},
  {Ma}, {Alexander}, {Biggs}, {Brandt}, {Chapman}, {Coppin}, {Danielson},
  {Dannerbauer}, {Edge}, {Greve}, {Ivison}, {Karim}, {Menten}, {Schinnerer},
  {Walter}, {Wardlow}, {Wei{\ss}}, \& {van der Werf}}]{chenc15}
{Chen}, C.-C., {Smail}, I., {Swinbank}, A.~M., {et~al.} 2015,
  \bibinfo{title}{{An ALMA Survey of Submillimeter Galaxies in the Extended
  Chandra Deep Field South: Near-infrared Morphologies and Stellar Sizes},}
  \apj, 799, 194, \dodoi{10.1088/0004-637X/799/2/194}

\bibitem[{H.~G. {Chittenden} {et~al.}(2025){Chittenden}, {Glazebrook},
  {Nanayakkara}, {Kawinwanichakij}, {Lagos}, {Kimmig}, \&
  {Remus}}]{chittenden25}
{Chittenden}, H.~G., {Glazebrook}, K., {Nanayakkara}, T., {et~al.} 2025,
  \bibinfo{title}{{On the unique evolutionary mechanisms of massive quiescent
  galaxies in the epoch of reionisation},} arXiv e-prints, arXiv:2504.19696,
  \dodoi{10.48550/arXiv.2504.19696}

\bibitem[{C.~R. {Choban} {et~al.}(2025){Choban}, {Salim}, {Kere{\v{s}}},
  {Hayward}, \& {Sandstrom}}]{choban25}
{Choban}, C.~R., {Salim}, S., {Kere{\v{s}}}, D., {Hayward}, C.~C., \&
  {Sandstrom}, K.~M. 2025, \bibinfo{title}{{A dusty dawn: galactic dust buildup
  at z {\ensuremath{\gtrsim}} 5},} \mnras, 537, 1518,
  \dodoi{10.1093/mnras/staf118}

\bibitem[{R.~K. {Cochrane} {et~al.}(2024){Cochrane}, {Angl{\'e}s-Alc{\'a}zar},
  {Cullen}, \& {Hayward}}]{cochrane24}
{Cochrane}, R.~K., {Angl{\'e}s-Alc{\'a}zar}, D., {Cullen}, F., \& {Hayward},
  C.~C. 2024, \bibinfo{title}{{Disappearing Galaxies: The Orientation
  Dependence of JWST-bright, HST-dark, Star-forming Galaxy Selection},} \apj,
  961, 37, \dodoi{10.3847/1538-4357/ad02f8}

\bibitem[{T. {Connor} {et~al.}(2020){Connor}, {Ba{\~n}ados}, {Mazzucchelli},
  {Stern}, {Decarli}, {Fan}, {Farina}, {Lusso}, {Neeleman}, \&
  {Walter}}]{connor20}
{Connor}, T., {Ba{\~n}ados}, E., {Mazzucchelli}, C., {et~al.} 2020,
  \bibinfo{title}{{X-Ray Observations of a [C II]-bright, z = 6.59
  Quasar/Companion System},} \apj, 900, 189, \dodoi{10.3847/1538-4357/abaab9}

\bibitem[{E. {da Cunha} {et~al.}(2013){da Cunha}, {Groves}, {Walter},
  {Decarli}, {Weiss}, {Bertoldi}, {Carilli}, {Daddi}, {Elbaz}, {Ivison},
  {Maiolino}, {Riechers}, {Rix}, {Sargent}, \& {Smail}}]{dacunha13}
{da Cunha}, E., {Groves}, B., {Walter}, F., {et~al.} 2013, \bibinfo{title}{{On
  the Effect of the Cosmic Microwave Background in High-redshift
  (Sub-)millimeter Observations},} \apj, 766, 13,
  \dodoi{10.1088/0004-637X/766/1/13}

\bibitem[{E. {da Cunha} {et~al.}(2015){da Cunha}, {Walter}, {Smail},
  {Swinbank}, {Simpson}, {Decarli}, {Hodge}, {Weiss}, {van der Werf},
  {Bertoldi}, {Chapman}, {Cox}, {Danielson}, {Dannerbauer}, {Greve}, {Ivison},
  {Karim}, \& {Thomson}}]{dacunha15}
{da Cunha}, E., {Walter}, F., {Smail}, I.~R., {et~al.} 2015,
  \bibinfo{title}{{An ALMA Survey of Sub-millimeter Galaxies in the Extended
  Chandra Deep Field South: Physical Properties Derived from
  Ultraviolet-to-radio Modeling},} \apj, 806, 110,
  \dodoi{10.1088/0004-637X/806/1/110}

\bibitem[{P. {Dayal} {et~al.}(2022){Dayal}, {Ferrara}, {Sommovigo}, {Bouwens},
  {Oesch}, {Smit}, {Gonzalez}, {Schouws}, {Stefanon}, {Kobayashi}, {Bremer},
  {Algera}, {Aravena}, {Bowler}, {da Cunha}, {Fudamoto}, {Graziani}, {Hodge},
  {Inami}, {De Looze}, {Pallottini}, {Riechers}, {Schneider}, {Stark}, \&
  {Endsley}}]{dayal22}
{Dayal}, P., {Ferrara}, A., {Sommovigo}, L., {et~al.} 2022,
  \bibinfo{title}{{The ALMA REBELS survey: the dust content of z 7 Lyman break
  galaxies},} \mnras, 512, 989, \dodoi{10.1093/mnras/stac537}

\bibitem[{A. {de Graaff} {et~al.}(2025){de Graaff}, {Setton}, {Brammer},
  {Cutler}, {Suess}, {Labb{\'e}}, {Leja}, {Weibel}, {Maseda}, {Whitaker},
  {Bezanson}, {Boogaard}, {Cleri}, {De Lucia}, {Franx}, {Greene}, {Hirschmann},
  {Matthee}, {McConachie}, {Naidu}, {Oesch}, {Price}, {Rix}, {Valentino},
  {Wang}, \& {Williams}}]{degraaff25a}
{de Graaff}, A., {Setton}, D.~J., {Brammer}, G., {et~al.} 2025,
  \bibinfo{title}{{Efficient formation of a massive quiescent galaxy at
  redshift 4.9},} Nature Astronomy, 9, 280, \dodoi{10.1038/s41550-024-02424-3}

\bibitem[{M.~E. {De Rossi} {et~al.}(2018){De Rossi}, {Rieke}, {Shivaei},
  {Bromm}, \& {Lyu}}]{derossi18}
{De Rossi}, M.~E., {Rieke}, G.~H., {Shivaei}, I., {Bromm}, V., \& {Lyu}, J.
  2018, \bibinfo{title}{{The Far-infrared Emission of the First Massive
  Galaxies},} \apj, 869, 4, \dodoi{10.3847/1538-4357/aaebf8}

\bibitem[{R. {Decarli} {et~al.}(2017){Decarli}, {Walter}, {Venemans},
  {Ba{\~n}ados}, {Bertoldi}, {Carilli}, {Fan}, {Farina}, {Mazzucchelli},
  {Riechers}, {Rix}, {Strauss}, {Wang}, \& {Yang}}]{decarli17}
{Decarli}, R., {Walter}, F., {Venemans}, B.~P., {et~al.} 2017,
  \bibinfo{title}{{Rapidly star-forming galaxies adjacent to quasars at
  redshifts exceeding 6},} \nat, 545, 457, \dodoi{10.1038/nature22358}

\bibitem[{F. {D'Eugenio} {et~al.}(2025){D'Eugenio}, {Cameron}, {Scholtz},
  {Carniani}, {Willott}, {Curtis-Lake}, {Bunker}, {Parlanti}, {Maiolino},
  {Willmer}, {Jakobsen}, {Robertson}, {Johnson}, {Tacchella}, {Cargile},
  {Rawle}, {Arribas}, {Chevallard}, {Curti}, {Egami}, {Eisenstein}, {Kumari},
  {Looser}, {Rieke}, {Rodr{\'\i}guez Del Pino}, {Saxena}, {{\"U}bler},
  {Venturi}, {Witstok}, {Baker}, {Bhatawdekar}, {Bonaventura}, {Boyett},
  {Charlot}, {Danhaive}, {Hainline}, {Hausen}, {Helton}, {Ji}, {Ji}, {Jones},
  {Juod{\v{z}}balis}, {Maseda}, {P{\'e}rez-Gonz{\'a}lez}, {Perna},
  {Pusk{\'a}s}, {Shivaei}, {Silcock}, {Simmonds}, {Smit}, {Sun}, {Villanueva},
  {Williams}, \& {Zhu}}]{deugenio25a}
{D'Eugenio}, F., {Cameron}, A.~J., {Scholtz}, J., {et~al.} 2025,
  \bibinfo{title}{{JADES Data Release 3: NIRSpec/Microshutter Assembly
  Spectroscopy for 4000 Galaxies in the GOODS Fields},} \apjs, 277, 4,
  \dodoi{10.3847/1538-4365/ada148}

\bibitem[{A. {Dey} {et~al.}(2019){Dey}, {Schlegel}, {Lang}, {Blum}, {Burleigh},
  {Fan}, {Findlay}, {Finkbeiner}, {Herrera}, {Juneau}, {Landriau}, {Levi},
  {McGreer}, {Meisner}, {Myers}, {Moustakas}, {Nugent}, {Patej}, {Schlafly},
  {Walker}, {Valdes}, {Weaver}, {Y{\`e}che}, {Zou}, {Zhou}, {Abareshi},
  {Abbott}, {Abolfathi}, {Aguilera}, {Alam}, {Allen}, {Alvarez}, {Annis},
  {Ansarinejad}, {Aubert}, {Beechert}, {Bell}, {BenZvi}, {Beutler}, {Bielby},
  {Bolton}, {Brice{\~n}o}, {Buckley-Geer}, {Butler}, {Calamida}, {Carlberg},
  {Carter}, {Casas}, {Castander}, {Choi}, {Comparat}, {Cukanovaite}, {Delubac},
  {DeVries}, {Dey}, {Dhungana}, {Dickinson}, {Ding}, {Donaldson}, {Duan},
  {Duckworth}, {Eftekharzadeh}, {Eisenstein}, {Etourneau}, {Fagrelius},
  {Farihi}, {Fitzpatrick}, {Font-Ribera}, {Fulmer}, {G{\"a}nsicke},
  {Gaztanaga}, {George}, {Gerdes}, {Gontcho}, {Gorgoni}, {Green}, {Guy},
  {Harmer}, {Hernandez}, {Honscheid}, {Huang}, {James}, {Jannuzi}, {Jiang},
  {Joyce}, {Karcher}, {Karkar}, {Kehoe}, {Kneib}, {Kueter-Young}, {Lan},
  {Lauer}, {Le Guillou}, {Le Van Suu}, {Lee}, {Lesser}, {Perreault Levasseur},
  {Li}, {Mann}, {Marshall}, {Mart{\'\i}nez-V{\'a}zquez}, {Martini}, {du Mas des
  Bourboux}, {McManus}, {Meier}, {M{\'e}nard}, {Metcalfe},
  {Mu{\~n}oz-Guti{\'e}rrez}, {Najita}, {Napier}, {Narayan}, {Newman}, {Nie},
  {Nord}, {Norman}, {Olsen}, {Paat}, {Palanque-Delabrouille}, {Peng},
  {Poppett}, {Poremba}, {Prakash}, {Rabinowitz}, {Raichoor}, {Rezaie},
  {Robertson}, {Roe}, {Ross}, {Ross}, {Rudnick}, {Safonova}, {Saha},
  {S{\'a}nchez}, {Savary}, {Schweiker}, {Scott}, {Seo}, {Shan}, {Silva},
  {Slepian}, {Soto}, {Sprayberry}, {Staten}, {Stillman}, {Stupak}, {Summers},
  {Sien Tie}, {Tirado}, {Vargas-Maga{\~n}a}, {Vivas}, {Wechsler}, {Williams},
  {Yang}, {Yang}, {Yapici}, {Zaritsky}, {Zenteno}, {Zhang}, {Zhang}, {Zhou}, \&
  {Zhou}}]{dey19}
{Dey}, A., {Schlegel}, D.~J., {Lang}, D., {et~al.} 2019,
  \bibinfo{title}{{Overview of the DESI Legacy Imaging Surveys},} \aj, 157,
  168, \dodoi{10.3847/1538-3881/ab089d}

\bibitem[{T. {D{\'\i}az-Santos} {et~al.}(2017){D{\'\i}az-Santos}, {Armus},
  {Charmandaris}, {Lu}, {Stierwalt}, {Stacey}, {Malhotra}, {van der Werf},
  {Howell}, {Privon}, {Mazzarella}, {Goldsmith}, {Murphy}, {Barcos-Mu{\~n}oz},
  {Linden}, {Inami}, {Larson}, {Evans}, {Appleton}, {Iwasawa}, {Lord},
  {Sanders}, \& {Surace}}]{ds17}
{D{\'\i}az-Santos}, T., {Armus}, L., {Charmandaris}, V., {et~al.} 2017,
  \bibinfo{title}{{A Herschel/PACS Far-infrared Line Emission Survey of Local
  Luminous Infrared Galaxies},} \apj, 846, 32, \dodoi{10.3847/1538-4357/aa81d7}

\bibitem[{C.~T. {Donnan} {et~al.}(2024){Donnan}, {McLure}, {Dunlop}, {McLeod},
  {Magee}, {Arellano-C{\'o}rdova}, {Barrufet}, {Begley}, {Bowler}, {Carnall},
  {Cullen}, {Ellis}, {Fontana}, {Illingworth}, {Grogin}, {Hamadouche},
  {Koekemoer}, {Liu}, {Mason}, {Santini}, \& {Stanton}}]{donnan24}
{Donnan}, C.~T., {McLure}, R.~J., {Dunlop}, J.~S., {et~al.} 2024,
  \bibinfo{title}{{JWST PRIMER: a new multifield determination of the evolving
  galaxy UV luminosity function at redshifts z ≃ 9 - 15},} \mnras, 533, 3222,
  \dodoi{10.1093/mnras/stae2037}

\bibitem[{J.~M.~S. {Donnellan} {et~al.}(2024){Donnellan}, {Oliver},
  {B{\'e}thermin}, {Bing}, {Bolatto}, {Bradford}, {Burgarella}, {Ciesla},
  {Glenn}, {Pope}, {Serjeant}, {Shirley}, {Smith}, \& {Sorrell}}]{donnellan24}
{Donnellan}, J.~M.~S., {Oliver}, S.~J., {B{\'e}thermin}, M., {et~al.} 2024,
  \bibinfo{title}{{Overcoming confusion noise with hyperspectral imaging from
  PRIMAger},} \mnras, 532, 1966, \dodoi{10.1093/mnras/stae1539}

\bibitem[{U. {Dudzevi{\v{c}}i{\={u}}t{\.{e}}}
  {et~al.}(2020){Dudzevi{\v{c}}i{\={u}}t{\.{e}}}, {Smail}, {Swinbank}, {Stach},
  {Almaini}, {da Cunha}, {An}, {Arumugam}, {Birkin}, {Blain}, {Chapman},
  {Chen}, {Conselice}, {Coppin}, {Dunlop}, {Farrah}, {Geach}, {Gullberg},
  {Hartley}, {Hodge}, {Ivison}, {Maltby}, {Scott}, {Simpson}, {Simpson},
  {Thomson}, {Walter}, {Wardlow}, {Weiss}, \& {van der Werf}}]{dudzevic20}
{Dudzevi{\v{c}}i{\={u}}t{\.{e}}}, U., {Smail}, I., {Swinbank}, A.~M., {et~al.}
  2020, \bibinfo{title}{{An ALMA survey of the SCUBA-2 CLS UDS field: physical
  properties of 707 sub-millimetre galaxies},} \mnras, 494, 3828,
  \dodoi{10.1093/mnras/staa769}

\bibitem[{J.~S. {Dunlop} {et~al.}(2017){Dunlop}, {McLure}, {Biggs}, {Geach},
  {Micha{\l}owski}, {Ivison}, {Rujopakarn}, {van Kampen}, {Kirkpatrick},
  {Pope}, {Scott}, {Swinbank}, {Targett}, {Aretxaga}, {Austermann}, {Best},
  {Bruce}, {Chapin}, {Charlot}, {Cirasuolo}, {Coppin}, {Ellis}, {Finkelstein},
  {Hayward}, {Hughes}, {Ibar}, {Jagannathan}, {Khochfar}, {Koprowski},
  {Narayanan}, {Nyland}, {Papovich}, {Peacock}, {Rieke}, {Robertson},
  {Vernstrom}, {Werf}, {Wilson}, \& {Yun}}]{dunlop17}
{Dunlop}, J.~S., {McLure}, R.~J., {Biggs}, A.~D., {et~al.} 2017,
  \bibinfo{title}{{A deep ALMA image of the Hubble Ultra Deep Field},} \mnras,
  466, 861, \dodoi{10.1093/mnras/stw3088}

\bibitem[{A.-C. {Eilers} {et~al.}(2024){Eilers}, {Mackenzie}, {Pizzati},
  {Matthee}, {Hennawi}, {Zhang}, {Bordoloi}, {Kashino}, {Lilly}, {Naidu},
  {Simcoe}, {Yue}, {Frenk}, {Helly}, {Schaller}, \& {Schaye}}]{eilers24}
{Eilers}, A.-C., {Mackenzie}, R., {Pizzati}, E., {et~al.} 2024,
  \bibinfo{title}{{EIGER. VI. The Correlation Function, Host Halo Mass, and
  Duty Cycle of Luminous Quasars at z {\ensuremath{\gtrsim}} 6},} \apj, 974,
  275, \dodoi{10.3847/1538-4357/ad778b}

\bibitem[{D.~J. {Eisenstein} {et~al.}(2023){Eisenstein}, {Willott}, {Alberts},
  {Arribas}, {Bonaventura}, {Bunker}, {Cameron}, {Carniani}, {Charlot},
  {Curtis-Lake}, {D'Eugenio}, {Endsley}, {Ferruit}, {Giardino}, {Hainline},
  {Hausen}, {Jakobsen}, {Johnson}, {Maiolino}, {Rieke}, {Rieke}, {Rix},
  {Robertson}, {Stark}, {Tacchella}, {Williams}, {Willmer}, {Baker}, {Baum},
  {Bhatawdekar}, {Boyett}, {Chen}, {Chevallard}, {Circosta}, {Curti},
  {Danhaive}, {DeCoursey}, {de Graaff}, {Dressler}, {Egami}, {Helton},
  {Hviding}, {Ji}, {Jones}, {Kumari}, {L{\"u}tzgendorf}, {Laseter}, {Looser},
  {Lyu}, {Maseda}, {Nelson}, {Parlanti}, {Perna}, {Pusk{\'a}s}, {Rawle},
  {Rodr{\'\i}guez Del Pino}, {Sandles}, {Saxena}, {Scholtz}, {Sharpe},
  {Shivaei}, {Silcock}, {Simmonds}, {Skarbinski}, {Smit}, {Stone}, {Suess},
  {Sun}, {Tang}, {Topping}, {{\"U}bler}, {Villanueva}, {Wallace}, {Whitler},
  {Witstok}, \& {Woodrum}}]{eisenstein23a}
{Eisenstein}, D.~J., {Willott}, C., {Alberts}, S., {et~al.} 2023,
  \bibinfo{title}{{Overview of the JWST Advanced Deep Extragalactic Survey
  (JADES)},} arXiv e-prints, arXiv:2306.02465,
  \dodoi{10.48550/arXiv.2306.02465}

\bibitem[{S.~L. {Finkelstein} {et~al.}(2024){Finkelstein}, {Leung}, {Bagley},
  {Dickinson}, {Ferguson}, {Papovich}, {Akins}, {Arrabal Haro}, {Dav{\'e}},
  {Dekel}, {Kartaltepe}, {Kocevski}, {Koekemoer}, {Pirzkal}, {Somerville},
  {Yung}, {Amor{\'\i}n}, {Backhaus}, {Behroozi}, {Bisigello}, {Bromm}, {Casey},
  {Ch{\'a}vez Ortiz}, {Cheng}, {Chworowsky}, {Cleri}, {Cooper}, {Davis}, {de la
  Vega}, {Elbaz}, {Franco}, {Fontana}, {Fujimoto}, {Giavalisco}, {Grogin},
  {Holwerda}, {Huertas-Company}, {Hirschmann}, {Iyer}, {Jogee}, {Jung},
  {Larson}, {Lucas}, {Mobasher}, {Morales}, {Morley}, {Mukherjee},
  {P{\'e}rez-Gonz{\'a}lez}, {Ravindranath}, {Rodighiero}, {Rowland},
  {Tacchella}, {Taylor}, {Trump}, \& {Wilkins}}]{finkelstein24}
{Finkelstein}, S.~L., {Leung}, G. C.~K., {Bagley}, M.~B., {et~al.} 2024,
  \bibinfo{title}{{The Complete CEERS Early Universe Galaxy Sample: A
  Surprisingly Slow Evolution of the Space Density of Bright Galaxies at z
  {\ensuremath{\sim}} 8.5{\textendash}14.5},} \apjl, 969, L2,
  \dodoi{10.3847/2041-8213/ad4495}

\bibitem[{S.~L. {Finkelstein} {et~al.}(2025){Finkelstein}, {Bagley}, {Arrabal
  Haro}, {Dickinson}, {Ferguson}, {Kartaltepe}, {Kocevski}, {Koekemoer},
  {Lotz}, {Papovich}, {P{\'e}rez-Gonz{\'a}lez}, {Pirzkal}, {Somerville},
  {Trump}, {Yang}, {Yung}, {Fontana}, {Grazian}, {Grogin}, {Kewley},
  {Kirkpatrick}, {Larson}, {Pentericci}, {Ravindranath}, {Wilkins}, {Almaini},
  {Amor{\'\i}n}, {Barro}, {Bhatawdekar}, {Bisigello}, {Brooks}, {Buat},
  {Buitrago}, {Burgarella}, {Calabr{\`o}}, {Castellano}, {Cheng}, {Cleri},
  {Cole}, {Cooper}, {Cooper}, {Costantin}, {Cox}, {Croton}, {Daddi}, {Davis},
  {Dekel}, {Elbaz}, {Fern{\'a}ndez}, {Fujimoto}, {Gandolfi}, {Gardner},
  {Gawiser}, {Giavalisco}, {G{\'o}mez-Guijarro}, {Guo}, {Gupta}, {Hathi},
  {Harish}, {Henry}, {Hirschmann}, {Hu}, {Hutchison}, {Iyer}, {Jaskot}, {Jha},
  {Jung}, {Kassin}, {Kokorev}, {Kurczynski}, {Leung}, {Llerena}, {Long},
  {Lucas}, {Lu}, {McGrath}, {McIntosh}, {Merlin}, {Mobasher}, {Morales},
  {Napolitano}, {Pacucci}, {Pandya}, {Rafelski}, {Rodighiero}, {Rose},
  {Santini}, {Seill{\'e}}, {Simons}, {Shen}, {Straughn}, {Tacchella}, {Taylor},
  {Vanderhoof}, {Vega-Ferrero}, {Weiner}, {Willmer}, {Zhu}, {Bell}, {Wuyts},
  {Holwerda}, {Wang}, {Wang}, {Zavala}, \& {CEERS
  Collaboration}}]{finkelstein25}
{Finkelstein}, S.~L., {Bagley}, M.~B., {Arrabal Haro}, P., {et~al.} 2025,
  \bibinfo{title}{{The Cosmic Evolution Early Release Science Survey (CEERS)},}
  \apjl, 983, L4, \dodoi{10.3847/2041-8213/adbbd3}

\bibitem[{M. {Franco} {et~al.}(2018){Franco}, {Elbaz}, {B{\'e}thermin},
  {Magnelli}, {Schreiber}, {Ciesla}, {Dickinson}, {Nagar}, {Silverman},
  {Daddi}, {Alexander}, {Wang}, {Pannella}, {Le Floc'h}, {Pope}, {Giavalisco},
  {Maury}, {Bournaud}, {Chary}, {Demarco}, {Ferguson}, {Finkelstein}, {Inami},
  {Iono}, {Juneau}, {Lagache}, {Leiton}, {Lin}, {Magdis}, {Messias},
  {Motohara}, {Mullaney}, {Okumura}, {Papovich}, {Pforr}, {Rujopakarn},
  {Sargent}, {Shu}, \& {Zhou}}]{franco18}
{Franco}, M., {Elbaz}, D., {B{\'e}thermin}, M., {et~al.} 2018,
  \bibinfo{title}{{GOODS-ALMA: 1.1 mm galaxy survey. I. Source catalog and
  optically dark galaxies},} \aap, 620, A152,
  \dodoi{10.1051/0004-6361/201832928}

\bibitem[{Y. {Fudamoto} {et~al.}(2021){Fudamoto}, {Oesch}, {Schouws},
  {Stefanon}, {Smit}, {Bouwens}, {Bowler}, {Endsley}, {Gonzalez}, {Inami},
  {Labbe}, {Stark}, {Aravena}, {Barrufet}, {da Cunha}, {Dayal}, {Ferrara},
  {Graziani}, {Hodge}, {Hutter}, {Li}, {De Looze}, {Nanayakkara}, {Pallottini},
  {Riechers}, {Schneider}, {Ucci}, {van der Werf}, \& {White}}]{fudamoto21}
{Fudamoto}, Y., {Oesch}, P.~A., {Schouws}, S., {et~al.} 2021,
  \bibinfo{title}{{Normal, dust-obscured galaxies in the epoch of
  reionization},} \nat, 597, 489, \dodoi{10.1038/s41586-021-03846-z}

\bibitem[{Y. {Fudamoto} {et~al.}(2025){Fudamoto}, {Helton}, {Lin}, {Sun},
  {Behroozi}, {Hsiao}, {Egami}, {Bunker}, {Harikane}, {Ouchi}, {Liu}, {Liu},
  {Maiolino}, {Ji}, {Jin}, {Tee}, {Wang}, {Willmer}, {Xu}, \&
  {Zhu}}]{fudamoto25a}
{Fudamoto}, Y., {Helton}, J.~M., {Lin}, X., {et~al.} 2025,
  \bibinfo{title}{{SAPPHIRES: A Galaxy Over-Density in the Heart of Cosmic
  Reionization at $z=8.47$},} arXiv e-prints, arXiv:2503.15597,
  \dodoi{10.48550/arXiv.2503.15597}

\bibitem[{S. {Fujimoto} {et~al.}(2023){Fujimoto}, {Bezanson}, {Labbe},
  {Brammer}, {Price}, {Wang}, {Weaver}, {Fudamoto}, {Oesch}, {Williams},
  {Dayal}, {Feldmann}, {Greene}, {Leja}, {Whitaker}, {Zitrin}, {Cutler},
  {Furtak}, {Pan}, {Chemerynska}, {Kokorev}, {Miller}, {Atek}, {van Dokkum},
  {Juneau}, {Kassin}, {Khullar}, {Marchesini}, {Maseda}, {Nelson}, {Setton}, \&
  {Smit}}]{fujimoto23a}
{Fujimoto}, S., {Bezanson}, R., {Labbe}, I., {et~al.} 2023,
  \bibinfo{title}{{DUALZ: Deep UNCOVER-ALMA Legacy High-Z Survey},} arXiv
  e-prints, arXiv:2309.07834, \dodoi{10.48550/arXiv.2309.07834}

\bibitem[{S. {Fujimoto} {et~al.}(2024){Fujimoto}, {Kohno}, {Ouchi}, {Oguri},
  {Kokorev}, {Brammer}, {Sun}, {Gonz{\'a}lez-L{\'o}pez}, {Bauer}, {Caminha},
  {Hatsukade}, {Richard}, {Smail}, {Tsujita}, {Ueda}, {Uematsu}, {Zitrin},
  {Coe}, {Kneib}, {Postman}, {Umetsu}, {Lagos}, {Popping}, {Ao}, {Bradley},
  {Caputi}, {Dessauges-Zavadsky}, {Egami}, {Espada}, {Ivison}, {Jauzac},
  {Knudsen}, {Koekemoer}, {Magdis}, {Mahler}, {Mu{\~n}oz Arancibia}, {Rawle},
  {Shimasaku}, {Toft}, {Umehata}, {Valentino}, {Wang}, \& {Wang}}]{fujimoto24b}
{Fujimoto}, S., {Kohno}, K., {Ouchi}, M., {et~al.} 2024, \bibinfo{title}{{ALMA
  Lensing Cluster Survey: Deep 1.2 mm Number Counts and Infrared Luminosity
  Functions at z ≃ 1{\textendash}8},} \apjs, 275, 36,
  \dodoi{10.3847/1538-4365/ad5ae2}

\bibitem[{ {Gaia Collaboration} {et~al.}(2023){Gaia Collaboration},
  {Vallenari}, {Brown}, {Prusti}, {de Bruijne}, {Arenou}, {Babusiaux},
  {Biermann}, {Creevey}, {Ducourant}, \& et~al.}]{gaiadr3}
{Gaia Collaboration}, {Vallenari}, A., {Brown}, A.~G.~A., {et~al.} 2023,
  \bibinfo{title}{{Gaia Data Release 3. Summary of the content and survey
  properties},} \aap, 674, A1, \dodoi{10.1051/0004-6361/202243940}

\bibitem[{K. {Glazebrook} {et~al.}(2024){Glazebrook}, {Nanayakkara},
  {Schreiber}, {Lagos}, {Kawinwanichakij}, {Jacobs}, {Chittenden}, {Brammer},
  {Kacprzak}, {Labbe}, {Marchesini}, {Marsan}, {Oesch}, {Papovich}, {Remus},
  {Tran}, {Esdaile}, \& {Chandro-Gomez}}]{glazebrook24}
{Glazebrook}, K., {Nanayakkara}, T., {Schreiber}, C., {et~al.} 2024,
  \bibinfo{title}{{A massive galaxy that formed its stars at z
  {\ensuremath{\approx}} 11},} \nat, 628, 277,
  \dodoi{10.1038/s41586-024-07191-9}

\bibitem[{C. {G{\'o}mez-Guijarro} {et~al.}(2022){G{\'o}mez-Guijarro}, {Elbaz},
  {Xiao}, {B{\'e}thermin}, {Franco}, {Magnelli}, {Daddi}, {Dickinson},
  {Demarco}, {Inami}, {Rujopakarn}, {Magdis}, {Shu}, {Chary}, {Zhou},
  {Alexander}, {Bournaud}, {Ciesla}, {Ferguson}, {Finkelstein}, {Giavalisco},
  {Iono}, {Juneau}, {Kartaltepe}, {Lagache}, {Le Floc'h}, {Leiton}, {Lin},
  {Motohara}, {Mullaney}, {Okumura}, {Pannella}, {Papovich}, {Pope}, {Sargent},
  {Silverman}, {Treister}, \& {Wang}}]{gomez22a}
{G{\'o}mez-Guijarro}, C., {Elbaz}, D., {Xiao}, M., {et~al.} 2022,
  \bibinfo{title}{{GOODS-ALMA 2.0: Source catalog, number counts, and
  prevailing compact sizes in 1.1 mm galaxies},} \aap, 658, A43,
  \dodoi{10.1051/0004-6361/202141615}

\bibitem[{J. {Gonz{\'a}lez-L{\'o}pez} {et~al.}(2020){Gonz{\'a}lez-L{\'o}pez},
  {Novak}, {Decarli}, {Walter}, {Aravena}, {Boogaard}, {Popping}, {Weiss},
  {Assef}, {Bauer}, {Bouwens}, {Cortes}, {Cox}, {Daddi}, {da Cunha},
  {D{\'\i}az-Santos}, {Ivison}, {Magnelli}, {Riechers}, {Smail}, {van der
  Werf}, \& {Wagg}}]{gl20}
{Gonz{\'a}lez-L{\'o}pez}, J., {Novak}, M., {Decarli}, R., {et~al.} 2020,
  \bibinfo{title}{{The ALMA Spectroscopic Survey in the HUDF: Deep 1.2 mm
  continuum number counts},} arXiv e-prints, arXiv:2002.07199.
\newblock \doarXiv{2002.07199}

\bibitem[{Y. {Harikane} {et~al.}(2024){Harikane}, {Nakajima}, {Ouchi}, {Umeda},
  {Isobe}, {Ono}, {Xu}, \& {Zhang}}]{harikane24a}
{Harikane}, Y., {Nakajima}, K., {Ouchi}, M., {et~al.} 2024,
  \bibinfo{title}{{Pure Spectroscopic Constraints on UV Luminosity Functions
  and Cosmic Star Formation History from 25 Galaxies at z $_{spec}$ =
  8.61-13.20 Confirmed with JWST/NIRSpec},} \apj, 960, 56,
  \dodoi{10.3847/1538-4357/ad0b7e}

\bibitem[{Y. {Harikane} {et~al.}(2020){Harikane}, {Ouchi}, {Inoue}, {Matsuoka},
  {Tamura}, {Bakx}, {Fujimoto}, {Moriwaki}, {Ono}, {Nagao}, {Tadaki}, {Kojima},
  {Shibuya}, {Egami}, {Ferrara}, {Gallerani}, {Hashimoto}, {Kohno}, {Matsuda},
  {Matsuo}, {Pallottini}, {Sugahara}, \& {Vallini}}]{harikane20}
{Harikane}, Y., {Ouchi}, M., {Inoue}, A.~K., {et~al.} 2020,
  \bibinfo{title}{{Large Population of ALMA Galaxies at z > 6 with Very High [O
  III] 88 {\ensuremath{\mu}}m to [C II] 158 {\ensuremath{\mu}}m Flux Ratios:
  Evidence of Extremely High Ionization Parameter or PDR Deficit?},} \apj, 896,
  93, \dodoi{10.3847/1538-4357/ab94bd}

\bibitem[{Y. {Harikane} {et~al.}(2025){Harikane}, {Inoue}, {Ellis}, {Ouchi},
  {Nakazato}, {Yoshida}, {Ono}, {Sun}, {Sato}, {Ferrami}, {Fujimoto},
  {Kashikawa}, {McLeod}, {P{\'e}rez-Gonz{\'a}lez}, {Sawicki}, {Sugahara}, {Xu},
  {Yamanaka}, {Carnall}, {Cullen}, {Dunlop}, {Egami}, {Grogin}, {Isobe},
  {Koekemoer}, {Laporte}, {Lee}, {Magee}, {Matsuo}, {Matsuoka}, {Mawatari},
  {Nakajima}, {Nakane}, {Tamura}, {Umeda}, \& {Yanagisawa}}]{harikane25}
{Harikane}, Y., {Inoue}, A.~K., {Ellis}, R.~S., {et~al.} 2025,
  \bibinfo{title}{{JWST, ALMA, and Keck Spectroscopic Constraints on the UV
  Luminosity Functions at z {\ensuremath{\sim}} 7{\textendash}14: Clumpiness
  and Compactness of the Brightest Galaxies in the Early Universe},} \apj, 980,
  138, \dodoi{10.3847/1538-4357/ad9b2c}

\bibitem[{A.~I. {Hartley} {et~al.}(2023){Hartley}, {Nelson}, {Suess}, {Garcia},
  {Park}, {Hernquist}, {Bezanson}, {Nevin}, {Pillepich}, {Schechter},
  {Terrazas}, {Torrey}, {Wellons}, {Whitaker}, \& {Williams}}]{hartley23}
{Hartley}, A.~I., {Nelson}, E.~J., {Suess}, K.~A., {et~al.} 2023,
  \bibinfo{title}{{The first quiescent galaxies in TNG300},} \mnras, 522, 3138,
  \dodoi{10.1093/mnras/stad1162}

\bibitem[{T. {Hashimoto} {et~al.}(2019){Hashimoto}, {Inoue}, {Mawatari},
  {Tamura}, {Matsuo}, {Furusawa}, {Harikane}, {Shibuya}, {Knudsen}, {Kohno},
  {Ono}, {Zackrisson}, {Okamoto}, {Kashikawa}, {Oesch}, {Ouchi}, {Ota},
  {Shimizu}, {Taniguchi}, {Umehata}, \& {Watson}}]{hashimoto19}
{Hashimoto}, T., {Inoue}, A.~K., {Mawatari}, K., {et~al.} 2019,
  \bibinfo{title}{{Big Three Dragons: A z = 7.15 Lyman-break galaxy detected in
  [O III] 88 {\ensuremath{\mu}}m, [C II] 158 {\ensuremath{\mu}}m, and dust
  continuum with ALMA},} \pasj, 71, 71, \dodoi{10.1093/pasj/psz049}

\bibitem[{B. {Hatsukade} {et~al.}(2018){Hatsukade}, {Kohno}, {Yamaguchi},
  {Umehata}, {Ao}, {Aretxaga}, {Caputi}, {Dunlop}, {Egami}, {Espada},
  {Fujimoto}, {Hayatsu}, {Hughes}, {Ikarashi}, {Iono}, {Ivison}, {Kawabe},
  {Kodama}, {Lee}, {Matsuda}, {Nakanishi}, {Ohta}, {Ouchi}, {Rujopakarn},
  {Suzuki}, {Tamura}, {Ueda}, {Wang}, {Wang}, {Wilson}, {Yoshimura}, \&
  {Yun}}]{hatsukade18}
{Hatsukade}, B., {Kohno}, K., {Yamaguchi}, Y., {et~al.} 2018,
  \bibinfo{title}{{ALMA twenty-six arcmin$^{2}$ survey of GOODS-S at one
  millimeter (ASAGAO): Source catalog and number counts},} \pasj, 70, 105,
  \dodoi{10.1093/pasj/psy104}

\bibitem[{J.~M. {Helton} {et~al.}(2024){Helton}, {Sun}, {Woodrum}, {Hainline},
  {Willmer}, {Rieke}, {Rieke}, {Alberts}, {Eisenstein}, {Tacchella},
  {Robertson}, {Johnson}, {Baker}, {Bhatawdekar}, {Bunker}, {Chen}, {Egami},
  {Ji}, {Maiolino}, {Willott}, \& {Witstok}}]{helton24b}
{Helton}, J.~M., {Sun}, F., {Woodrum}, C., {et~al.} 2024,
  \bibinfo{title}{{Identification of High-redshift Galaxy Overdensities in
  GOODS-N and GOODS-S},} \apj, 974, 41, \dodoi{10.3847/1538-4357/ad6867}

\bibitem[{T. {Herard-Demanche} {et~al.}(2025){Herard-Demanche}, {Bouwens},
  {Oesch}, {Naidu}, {Decarli}, {Nelson}, {Brammer}, {Weibel}, {Xiao},
  {Stefanon}, {Walter}, {Matthee}, {Meyer}, {Wuyts}, {Reddy}, {Rowland}, {van
  Leeuwen}, {Haro}, {Dannerbauer}, {Shapley}, {Chisholm}, {van Dokkum},
  {Labbe}, {Illingworth}, {Schaerer}, \& {Shivaei}}]{hd25}
{Herard-Demanche}, T., {Bouwens}, R.~J., {Oesch}, P.~A., {et~al.} 2025,
  \bibinfo{title}{{Mapping dusty galaxy growth at z > 5 with FRESCO: detection
  of H{\ensuremath{\alpha}} in submm galaxy HDF850.1 and the surrounding
  overdense structures},} \mnras, 537, 788, \dodoi{10.1093/mnras/staf030}

\bibitem[{R. {Hill} {et~al.}(2024){Hill}, {Scott}, {McLeod}, {McLure},
  {Chapman}, \& {Dunlop}}]{hill24}
{Hill}, R., {Scott}, D., {McLeod}, D.~J., {et~al.} 2024, \bibinfo{title}{{An
  optimal ALMA image of the Hubble Ultra Deep Field in the era of JWST:
  obscured star formation and the cosmic far-infrared background},} \mnras,
  528, 5019, \dodoi{10.1093/mnras/stae346}

\bibitem[{J.~A. {Hodge} \& E. {da Cunha}(2020){Hodge} \& {da Cunha}}]{hodge20}
{Hodge}, J.~A., \& {da Cunha}, E. 2020, \bibinfo{title}{{High-redshift star
  formation in the Atacama large millimetre/submillimetre array era},} Royal
  Society Open Science, 7, 200556, \dodoi{10.1098/rsos.200556}

\bibitem[{J.~A. {Hodge} {et~al.}(2016){Hodge}, {Swinbank}, {Simpson}, {Smail},
  {Walter}, {Alexander}, {Bertoldi}, {Biggs}, {Brandt}, {Chapman}, {Chen},
  {Coppin}, {Cox}, {Dannerbauer}, {Edge}, {Greve}, {Ivison}, {Karim},
  {Knudsen}, {Menten}, {Rix}, {Schinnerer}, {Wardlow}, {Weiss}, \& {van der
  Werf}}]{hodge16}
{Hodge}, J.~A., {Swinbank}, A.~M., {Simpson}, J.~M., {et~al.} 2016,
  \bibinfo{title}{{Kiloparsec-scale Dust Disks in High-redshift Luminous
  Submillimeter Galaxies},} \apj, 833, 103, \dodoi{10.3847/1538-4357/833/1/103}

\bibitem[{T.~Y.-Y. {Hsiao} {et~al.}(2024){Hsiao}, {{\'A}lvarez-M{\'a}rquez},
  {Coe}, {Crespo G{\'o}mez}, {Abdurro'uf}, {Dayal}, {Larson}, {Bik},
  {Blanco-Prieto}, {Colina}, {P{\'e}rez-Gonz{\'a}lez}, {Costantin},
  {Prieto-Jim{\'e}nez}, {Adamo}, {Bradley}, {Conselice}, {Fujimoto}, {Furtak},
  {Hutchison}, {James}, {Jim{\'e}nez-Teja}, {Jung}, {Kokorev}, {Mingozzi},
  {Norman}, {Ricotti}, {Rigby}, {Sharon}, {Vanzella}, {Welch}, {Xu},
  {Zackrisson}, \& {Zitrin}}]{hsiao24a}
{Hsiao}, T. Y.-Y., {{\'A}lvarez-M{\'a}rquez}, J., {Coe}, D., {et~al.} 2024,
  \bibinfo{title}{{JWST MIRI Detections of H{\ensuremath{\alpha}} and [O III]
  and a Direct Metallicity Measurement of the z = 10.17 Lensed Galaxy
  MACS0647‑JD},} \apj, 973, 81, \dodoi{10.3847/1538-4357/ad6562}

\bibitem[{D.~H. {Hughes} {et~al.}(1998){Hughes}, {Serjeant}, {Dunlop},
  {Rowan-Robinson}, {Blain}, {Mann}, {Ivison}, {Peacock}, {Efstathiou}, {Gear},
  {Oliver}, {Lawrence}, {Longair}, {Goldschmidt}, \& {Jenness}}]{hughes98}
{Hughes}, D.~H., {Serjeant}, S., {Dunlop}, J., {et~al.} 1998,
  \bibinfo{title}{{High-redshift star formation in the Hubble Deep Field
  revealed by a submillimetre-wavelength survey},} \nat, 394, 241,
  \dodoi{10.1038/28328}

\bibitem[{A.~P.~S. {Hygate} {et~al.}(2023){Hygate}, {Hodge}, {da Cunha},
  {Rybak}, {Schouws}, {Inami}, {Stefanon}, {Graziani}, {Schneider}, {Dayal},
  {Bouwens}, {Smit}, {Bowler}, {Endsley}, {Gonzalez}, {Oesch}, {Stark},
  {Algera}, {Aravena}, {Barrufet}, {Ferrara}, {Fudamoto}, {Hilhorst}, {De
  Looze}, {Nanayakkara}, {Pallottini}, {Riechers}, {Sommovigo}, {Topping}, \&
  {van der Werf}}]{hygate23}
{Hygate}, A.~P.~S., {Hodge}, J.~A., {da Cunha}, E., {et~al.} 2023,
  \bibinfo{title}{{The ALMA REBELS Survey: discovery of a massive, highly
  star-forming, and morphologically complex ULIRG at z = 7.31},} \mnras, 524,
  1775, \dodoi{10.1093/mnras/stad1212}

\bibitem[{H. {Inami} {et~al.}(2022){Inami}, {Algera}, {Schouws}, {Sommovigo},
  {Bouwens}, {Smit}, {Stefanon}, {Bowler}, {Endsley}, {Ferrara}, {Oesch},
  {Stark}, {Aravena}, {Barrufet}, {da Cunha}, {Dayal}, {De Looze}, {Fudamoto},
  {Gonzalez}, {Graziani}, {Hodge}, {Hygate}, {Nanayakkara}, {Pallottini},
  {Riechers}, {Schneider}, {Topping}, \& {van der Werf}}]{inami22}
{Inami}, H., {Algera}, H. S.~B., {Schouws}, S., {et~al.} 2022,
  \bibinfo{title}{{The ALMA REBELS Survey: dust continuum detections at z >
  6.5},} \mnras, 515, 3126, \dodoi{10.1093/mnras/stac1779}

\bibitem[{Z. {Ji} {et~al.}(2024{\natexlab{a}}){Ji}, {Williams}, {Rieke}, {Lyu},
  {Alberts}, {Sun}, {Helton}, {Rieke}, {Shivaei}, {D'Eugenio}, {Tacchella},
  {Robertson}, {Zhu}, {Maiolino}, {Bunker}, {Sun}, \& {Willmer}}]{jiz24a}
{Ji}, Z., {Williams}, C.~C., {Rieke}, G.~H., {et~al.} 2024{\natexlab{a}},
  \bibinfo{title}{{Extended hot dust emission around the earliest massive
  quiescent galaxy},} arXiv e-prints, arXiv:2409.17233,
  \dodoi{10.48550/arXiv.2409.17233}

\bibitem[{Z. {Ji} {et~al.}(2024{\natexlab{b}}){Ji}, {Williams}, {Suess},
  {Tacchella}, {Johnson}, {Robertson}, {Alberts}, {Baker}, {Baum},
  {Bhatawdekar}, {Bonaventura}, {Boyett}, {Bunker}, {Carniani}, {Charlot},
  {Chen}, {Chevallard}, {Curtis-Lake}, {D'Eugenio}, {de Graaff}, {DeCoursey},
  {Egami}, {Eisenstein}, {Hainline}, {Hausen}, {Helton}, {Looser}, {Lyu},
  {Maiolino}, {Maseda}, {Nelson}, {Rieke}, {Rieke}, {Rix}, {Sandles}, {Sun},
  {{\"U}bler}, {Willmer}, {Willott}, \& {Witstok}}]{jiz24b}
{Ji}, Z., {Williams}, C.~C., {Suess}, K.~A., {et~al.} 2024{\natexlab{b}},
  \bibinfo{title}{{JADES: Rest-frame UV-to-NIR Size Evolution of Massive
  Quiescent Galaxies from Redshift z=5 to z=0.5},} arXiv e-prints,
  arXiv:2401.00934, \dodoi{10.48550/arXiv.2401.00934}

\bibitem[{R. {Kannan} {et~al.}(2022){Kannan}, {Garaldi}, {Smith}, {Pakmor},
  {Springel}, {Vogelsberger}, \& {Hernquist}}]{kannan22}
{Kannan}, R., {Garaldi}, E., {Smith}, A., {et~al.} 2022,
  \bibinfo{title}{{Introducing the THESAN project:
  radiation-magnetohydrodynamic simulations of the epoch of reionization},}
  \mnras, 511, 4005, \dodoi{10.1093/mnras/stab3710}

\bibitem[{D. {Kashino} {et~al.}(2023){Kashino}, {Lilly}, {Matthee}, {Eilers},
  {Mackenzie}, {Bordoloi}, \& {Simcoe}}]{kashino23}
{Kashino}, D., {Lilly}, S.~J., {Matthee}, J., {et~al.} 2023,
  \bibinfo{title}{{EIGER. I. A Large Sample of [O III]-emitting Galaxies at 5.3
  < z < 6.9 and Direct Evidence for Local Reionization by Galaxies},} \apj,
  950, 66, \dodoi{10.3847/1538-4357/acc588}

\bibitem[{R.~C. {Kennicutt} \& N.~J. {Evans}(2012){Kennicutt} \&
  {Evans}}]{ke12}
{Kennicutt}, R.~C., \& {Evans}, N.~J. 2012, \bibinfo{title}{{Star Formation in
  the Milky Way and Nearby Galaxies},} \araa, 50, 531,
  \dodoi{10.1146/annurev-astro-081811-125610}

\bibitem[{V. {Kokorev} {et~al.}(2025){Kokorev}, {Ch{\'a}vez Ortiz}, {Taylor},
  {Finkelstein}, {Arrabal Haro}, {Dickinson}, {Chisholm}, {Fujimoto},
  {Mu{\~n}oz}, {Endsley}, {Hu}, {Napolitano}, {Wilkins}, {Akins},
  {Amori{\'\i}n}, {Casey}, {Cheng}, {Cleri}, {Cole}, {Cullen}, {Daddi},
  {Davis}, {Donnan}, {Dunlop}, {Fern{\'a}ndez}, {Giavalisco}, {Grogin},
  {Hathi}, {Hirschmann}, {Kartaltepe}, {Koekemoer}, {Leung}, {Lucas}, {McLeod},
  {Papovich}, {Pentericci}, {P{\'e}rez-Gonz{\'a}lez}, {Somerville}, {Wang},
  {Yung}, \& {Zavala}}]{kokorev25a}
{Kokorev}, V., {Ch{\'a}vez Ortiz}, {\'O}.~A., {Taylor}, A.~J., {et~al.} 2025,
  \bibinfo{title}{{CAPERS Observations of Two UV-Bright Galaxies at z>10. More
  Evidence for Bursting Star Formation in the Early Universe},} arXiv e-prints,
  arXiv:2504.12504, \dodoi{10.48550/arXiv.2504.12504}

\bibitem[{N. {Laporte} {et~al.}(2021){Laporte}, {Zitrin}, {Ellis}, {Fujimoto},
  {Brammer}, {Richard}, {Oguri}, {Caminha}, {Kohno}, {Yoshimura}, {Ao},
  {Bauer}, {Caputi}, {Egami}, {Espada}, {Gonz{\'a}lez-L{\'o}pez}, {Hatsukade},
  {Knudsen}, {Lee}, {Magdis}, {Ouchi}, {Valentino}, \& {Wang}}]{laporte21}
{Laporte}, N., {Zitrin}, A., {Ellis}, R.~S., {et~al.} 2021,
  \bibinfo{title}{{ALMA Lensing Cluster Survey: a strongly lensed multiply
  imaged dusty system at z {\ensuremath{\geq}} 6},} \mnras, 505, 4838,
  \dodoi{10.1093/mnras/stab191}

\bibitem[{A. {Le{\'s}niewska} \& M.~J. {Micha{\l}owski}(2019){Le{\'s}niewska}
  \& {Micha{\l}owski}}]{lesniewska19}
{Le{\'s}niewska}, A., \& {Micha{\l}owski}, M.~J. 2019, \bibinfo{title}{{Dust
  production scenarios in galaxies at z {\ensuremath{\sim}}6-8.3},} \aap, 624,
  L13, \dodoi{10.1051/0004-6361/201935149}

\bibitem[{J. {Li} {et~al.}(2022){Li}, {Venemans}, {Walter}, {Decarli}, {Wang},
  \& {Cai}}]{lij22}
{Li}, J., {Venemans}, B.~P., {Walter}, F., {et~al.} 2022,
  \bibinfo{title}{{Spatially Resolved Molecular Interstellar Medium in a z =
  6.6 Quasar Host Galaxy},} \apj, 930, 27, \dodoi{10.3847/1538-4357/ac61d7}

\bibitem[{Q. {Li} {et~al.}(2025){Li}, {Conselice}, {Sarron}, {Harvey},
  {Austin}, {Adams}, {Trussler}, {Duan}, {Ferreira}, {Westcott}, {Harris},
  {Dole}, {Grogin}, {Frye}, {Koekemoer}, {Robertson}, {Windhorst}, {Polletta},
  {Hathi}, \& {Jansen}}]{liq25}
{Li}, Q., {Conselice}, C.~J., {Sarron}, F., {et~al.} 2025,
  \bibinfo{title}{{EPOCHS paper {\textendash} X. Environmental effects on
  Galaxy formation and protocluster Galaxy candidates at 4.5 < z < 10 from JWST
  observations},} \mnras, 539, 1796, \dodoi{10.1093/mnras/staf543}

\bibitem[{C.-F. {Lim} {et~al.}(2020){Lim}, {Wang}, {Smail}, {Scott}, {Chen},
  {Chang}, {Simpson}, {Toba}, {Shu}, {Clements}, {Greenslade}, {Ao}, {Babul},
  {Birkin}, {Chapman}, {Cheng}, {Cho}, {Dannerbauer},
  {Dudzevi{\v{c}}i{\={u}}t{\.{e}}}, {Dunlop}, {Gao}, {Goto}, {Ho}, {Hsu},
  {Hwang}, {Jeong}, {Koprowski}, {Lee}, {Lin}, {Lin}, {Micha{\l}owski},
  {Parsons}, {Sawicki}, {Shirley}, {Shim}, {Urquhart}, {Wang}, \&
  {Wang}}]{limc20a}
{Lim}, C.-F., {Wang}, W.-H., {Smail}, I., {et~al.} 2020,
  \bibinfo{title}{{SCUBA-2 Ultra Deep Imaging EAO Survey (Studies). III.
  Multiwavelength Properties, Luminosity Functions, and Preliminary Source
  Catalog of 450 {\ensuremath{\mu}}m Selected Galaxies},} \apj, 889, 80,
  \dodoi{10.3847/1538-4357/ab607f}

\bibitem[{X. {Lin} {et~al.}(2025){Lin}, {Fan}, {Wang}, {Sun}, {Champagne},
  {Egami}, {Kakiichi}, {Lyu}, {Tee}, {Yang}, {Bian}, {Bosman}, {Cai}, {Casey},
  {Decarli}, {Faisst}, {Fujimoto}, {Harish}, {Ilbert}, {Inoue}, {Jin},
  {Kartaltepe}, {Kocevski}, {Li}, {Liu}, {Liu}, {Schindler}, {Shuntov},
  {Tanaka}, {Vestergaard}, {Wu}, {Zhang}, \& {Zhang}}]{linx25b}
{Lin}, X., {Fan}, X., {Wang}, F., {et~al.} 2025, \bibinfo{title}{{Bridging
  Quasars and Little Red Dots: Insights into Broad-Line AGNs at $z=5-8$ from
  the First JWST COSMOS-3D Dataset},} arXiv e-prints, arXiv:2504.08039,
  \dodoi{10.48550/arXiv.2504.08039}

\bibitem[{A.~S. {Long} {et~al.}(2024{\natexlab{a}}){Long}, {Casey}, {McKinney},
  {Zavala}, {Akins}, {Cooper}, {Lambrides}, {Franco}, {Caputi}, {Champagne},
  {Man}, {Treister}, {Manning}, {Sanders}, {Talia}, {Aravena}, {Clements}, {da
  Cunha}, {Faisst}, {Gentile}, {Hodge}, {Brammer}, {Brusa}, {Finkelstein},
  {Fujimoto}, {Hayward}, {Ilbert}, {Jolly}, {Kartaltepe}, {Knudsen},
  {Koekemoer}, {Liu}, {Magdis}, {McCracken}, {Rhodes}, {Robertson}, {Scoville},
  {Sheth}, {Smolcic}, {Spilker}, {Taniguchi}, {Toft}, {Urry}, \&
  {Yun}}]{long24b}
{Long}, A.~S., {Casey}, C.~M., {McKinney}, J., {et~al.} 2024{\natexlab{a}},
  \bibinfo{title}{{The Extended Mapping Obscuration to Reionization with ALMA
  (Ex-MORA) Survey: 5$\sigma$ Source Catalog and Redshift Distribution},} arXiv
  e-prints, arXiv:2408.14546, \dodoi{10.48550/arXiv.2408.14546}

\bibitem[{A.~S. {Long} {et~al.}(2024{\natexlab{b}}){Long}, {Antwi-Danso},
  {Lambrides}, {Lovell}, {de la Vega}, {Valentino}, {Zavala}, {Casey},
  {Wilkins}, {Yung}, {Arrabal Haro}, {Bagley}, {Bisigello}, {Chworowsky},
  {Cooper}, {Cooper}, {Cooray}, {Croton}, {Dickinson}, {Finkelstein}, {Franco},
  {Gould}, {Hirschmann}, {Hutchison}, {Kartaltepe}, {Kocevski}, {Koekemoer},
  {Lucas}, {McKinney}, {Nere}, {Papovich}, {P{\'e}rez-Gonz{\'a}lez}, {Pirzkal},
  \& {Santini}}]{long24a}
{Long}, A.~S., {Antwi-Danso}, J., {Lambrides}, E.~L., {et~al.}
  2024{\natexlab{b}}, \bibinfo{title}{{Efficient NIRCam Selection of Quiescent
  Galaxies at 3 < z < 6 in CEERS},} \apj, 970, 68,
  \dodoi{10.3847/1538-4357/ad4cea}

\bibitem[{C.~C. {Lovell} {et~al.}(2023){Lovell}, {Roper}, {Vijayan}, {Seeyave},
  {Irodotou}, {Wilkins}, {Conselice}, {Fortuni}, {Kuusisto}, {Merlin},
  {Santini}, \& {Thomas}}]{lovell23}
{Lovell}, C.~C., {Roper}, W., {Vijayan}, A.~P., {et~al.} 2023,
  \bibinfo{title}{{First light and reionisation epoch simulations (FLARES) -
  VIII. The emergence of passive galaxies at z {\ensuremath{\geq}} 5},} \mnras,
  525, 5520, \dodoi{10.1093/mnras/stad2550}

\bibitem[{D. {Lutz} {et~al.}(2011){Lutz}, {Poglitsch}, {Altieri}, {Andreani},
  {Aussel}, {Berta}, {Bongiovanni}, {Brisbin}, {Cava}, {Cepa}, {Cimatti},
  {Daddi}, {Dominguez-Sanchez}, {Elbaz}, {F{\"o}rster Schreiber}, {Genzel},
  {Grazian}, {Gruppioni}, {Harwit}, {Le Floc'h}, {Magdis}, {Magnelli},
  {Maiolino}, {Nordon}, {P{\'e}rez Garc{\'\i}a}, {Popesso}, {Pozzi},
  {Riguccini}, {Rodighiero}, {Saintonge}, {Sanchez Portal}, {Santini}, {Shao},
  {Sturm}, {Tacconi}, {Valtchanov}, {Wetzstein}, \& {Wieprecht}}]{lutz11}
{Lutz}, D., {Poglitsch}, A., {Altieri}, B., {et~al.} 2011,
  \bibinfo{title}{{PACS Evolutionary Probe (PEP) - A Herschel key program},}
  \aap, 532, A90, \dodoi{10.1051/0004-6361/201117107}

\bibitem[{J. {Lyu} {et~al.}(2016){Lyu}, {Rieke}, \& {Alberts}}]{lyuj16}
{Lyu}, J., {Rieke}, G.~H., \& {Alberts}, S. 2016, \bibinfo{title}{{The
  Contribution of Host Galaxies to the Infrared Energy Output of z
  {\ensuremath{\gtrsim}} 5.0 Quasars},} \apj, 816, 85,
  \dodoi{10.3847/0004-637X/816/2/85}

\bibitem[{Z. {Ma} {et~al.}(2024){Ma}, {Sun}, {Cheng}, {Yan}, {Ling}, {Sun},
  {Foo}, {Egami}, {Diego}, {Cohen}, {Jansen}, {Summers}, {Windhorst},
  {D'Silva}, {Koekemoer}, {Coe}, {Conselice}, {Driver}, {Frye}, {Grogin},
  {Marshall}, {Nonino}, {Ortiz}, {Pirzkal}, {Robotham}, {Ryan}, {Willmer},
  {Adams}, {Hathi}, {Dole}, {Willner}, {Espada}, {Furtak}, {Hsiao}, {Li},
  {Chen}, {Jolly}, \& {Chen}}]{maz24}
{Ma}, Z., {Sun}, B., {Cheng}, C., {et~al.} 2024, \bibinfo{title}{{JWST View of
  Four Infant Galaxies at z = 8.31{\textendash}8.49 in the MACS J0416.1‑2403
  Field and Implications for Reionization},} \apj, 975, 87,
  \dodoi{10.3847/1538-4357/ad7b32}

\bibitem[{P. {Madau} \& M. {Dickinson}(2014){Madau} \& {Dickinson}}]{md14}
{Madau}, P., \& {Dickinson}, M. 2014, \bibinfo{title}{{Cosmic Star-Formation
  History},} \araa, 52, 415, \dodoi{10.1146/annurev-astro-081811-125615}

\bibitem[{C. {Mazzucchelli} {et~al.}(2019){Mazzucchelli}, {Decarli}, {Farina},
  {Ba{\~n}ados}, {Venemans}, {Strauss}, {Walter}, {Neeleman}, {Bertoldi},
  {Fan}, {Riechers}, {Rix}, \& {Wang}}]{mazzucchelli19}
{Mazzucchelli}, C., {Decarli}, R., {Farina}, E.~P., {et~al.} 2019,
  \bibinfo{title}{{Spectral Energy Distributions of Companion Galaxies to z
  {\ensuremath{\sim}} 6 Quasars},} \apj, 881, 163,
  \dodoi{10.3847/1538-4357/ab2f75}

\bibitem[{J. {McKinney} {et~al.}(2025){McKinney}, {Casey}, {Long}, {Cooper},
  {Manning}, {Franco}, {Akins}, {Lambrides}, {Gammon}, {Silva}, {Gentile},
  {Zavala}, {Amvrosiadis}, {Andika}, {Brinch}, {Champagne}, {Chartab},
  {Drakos}, {Faisst}, {Fujimoto}, {Gillman}, {Gozaliasl}, {Greve}, {Harish},
  {Hayward}, {Hirschmann}, {Ilbert}, {Kalita}, {Kartaltepe}, {Koekemoer},
  {Kokorev}, {Liu}, {Magdis}, {McCracken}, {Rhodes}, {Robertson}, {Talia},
  {Valentino}, \& {Vijayan}}]{mcKinney25a}
{McKinney}, J., {Casey}, C.~M., {Long}, A.~S., {et~al.} 2025,
  \bibinfo{title}{{SCUBADive. I. JWST+ALMA Analysis of 289 Submillimeter
  Galaxies in COSMOS-web},} \apj, 979, 229, \dodoi{10.3847/1538-4357/ada357}

\bibitem[{S.~G. {Murray} {et~al.}(2013){Murray}, {Power}, \&
  {Robotham}}]{murray13}
{Murray}, S.~G., {Power}, C., \& {Robotham}, A.~S.~G. 2013,
  \bibinfo{title}{{HMFcalc: An online tool for calculating dark matter halo
  mass functions},} Astronomy and Computing, 3, 23,
  \dodoi{10.1016/j.ascom.2013.11.001}

\bibitem[{R.~P. {Naidu} {et~al.}(2025){Naidu}, {Oesch}, {Brammer}, {Weibel},
  {Li}, {Matthee}, {Chisholm}, {Pollock}, {Heintz}, {Johnson}, {Shen},
  {Hviding}, {Leja}, {Tacchella}, {Ganguly}, {Witten}, {Atek}, {Belli}, {Bose},
  {Bouwens}, {Dayal}, {Decarli}, {de Graaff}, {Fudamoto}, {Giovinazzo},
  {Greene}, {Illingworth}, {Inoue}, {Kane}, {Labbe}, {Leonova},
  {Marques-Chaves}, {Meyer}, {Nelson}, {Roberts-Borsani}, {Schaerer}, {Simcoe},
  {Stefanon}, {Sugahara}, {Toft}, {van der Wel}, {van Dokkum}, {Walter},
  {Watson}, {Weaver}, \& {Whitaker}}]{naidu25b}
{Naidu}, R.~P., {Oesch}, P.~A., {Brammer}, G., {et~al.} 2025,
  \bibinfo{title}{{A Cosmic Miracle: A Remarkably Luminous Galaxy at
  $z_{\rm{spec}}=14.44$ Confirmed with JWST},} arXiv e-prints,
  arXiv:2505.11263.
\newblock \doarXiv{2505.11263}

\bibitem[{D. {Nelson} {et~al.}(2019){Nelson}, {Springel}, {Pillepich},
  {Rodriguez-Gomez}, {Torrey}, {Genel}, {Vogelsberger}, {Pakmor}, {Marinacci},
  {Weinberger}, {Kelley}, {Lovell}, {Diemer}, \& {Hernquist}}]{nelson19}
{Nelson}, D., {Springel}, V., {Pillepich}, A., {et~al.} 2019,
  \bibinfo{title}{{The IllustrisTNG simulations: public data release},}
  Computational Astrophysics and Cosmology, 6, 2,
  \dodoi{10.1186/s40668-019-0028-x}

\bibitem[{H.~T. {Nguyen} {et~al.}(2010){Nguyen}, {Schulz}, {Levenson},
  {Amblard}, {Arumugam}, {Aussel}, {Babbedge}, {Blain}, {Bock}, {Boselli},
  {Buat}, {Castro-Rodriguez}, {Cava}, {Chanial}, {Chapin}, {Clements},
  {Conley}, {Conversi}, {Cooray}, {Dowell}, {Dwek}, {Eales}, {Elbaz}, {Fox},
  {Franceschini}, {Gear}, {Glenn}, {Griffin}, {Halpern}, {Hatziminaoglou},
  {Ibar}, {Isaak}, {Ivison}, {Lagache}, {Lu}, {Madden}, {Maffei}, {Mainetti},
  {Marchetti}, {Marsden}, {Marshall}, {O'Halloran}, {Oliver}, {Omont}, {Page},
  {Panuzzo}, {Papageorgiou}, {Pearson}, {Perez Fournon}, {Pohlen}, {Rangwala},
  {Rigopoulou}, {Rizzo}, {Roseboom}, {Rowan-Robinson}, {Scott}, {Seymour},
  {Shupe}, {Smith}, {Stevens}, {Symeonidis}, {Trichas}, {Tugwell}, {Vaccari},
  {Valtchanov}, {Vigroux}, {Wang}, {Ward}, {Wiebe}, {Wright}, {Xu}, \&
  {Zemcov}}]{nguyen10}
{Nguyen}, H.~T., {Schulz}, B., {Levenson}, L., {et~al.} 2010,
  \bibinfo{title}{{HerMES: The SPIRE confusion limit},} \aap, 518, L5,
  \dodoi{10.1051/0004-6361/201014680}

\bibitem[{S. {Noll} {et~al.}(2009){Noll}, {Burgarella}, {Giovannoli}, {Buat},
  {Marcillac}, \& {Mu{\~n}oz-Mateos}}]{cigale09}
{Noll}, S., {Burgarella}, D., {Giovannoli}, E., {et~al.} 2009,
  \bibinfo{title}{{Analysis of galaxy spectral energy distributions from far-UV
  to far-IR with CIGALE: studying a SINGS test sample},} \aap, 507, 1793,
  \dodoi{10.1051/0004-6361/200912497}

\bibitem[{J.~B. {Oke} \& J.~E. {Gunn}(1983){Oke} \& {Gunn}}]{oke83}
{Oke}, J.~B., \& {Gunn}, J.~E. 1983, \bibinfo{title}{{Secondary standard stars
  for absolute spectrophotometry.},} \apj, 266, 713, \dodoi{10.1086/160817}

\bibitem[{S.~J. {Oliver} {et~al.}(2012){Oliver}, {Bock}, {Altieri}, {Amblard},
  {Arumugam}, {Aussel}, {Babbedge}, {Beelen}, {B{\'e}thermin}, {Blain},
  {Boselli}, {Bridge}, {Brisbin}, {Buat}, {Burgarella},
  {Castro-Rodr{\'\i}guez}, {Cava}, {Chanial}, {Cirasuolo}, {Clements},
  {Conley}, {Conversi}, {Cooray}, {Dowell}, {Dubois}, {Dwek}, {Dye}, {Eales},
  {Elbaz}, {Farrah}, {Feltre}, {Ferrero}, {Fiolet}, {Fox}, {Franceschini},
  {Gear}, {Giovannoli}, {Glenn}, {Gong}, {Gonz{\'a}lez Solares}, {Griffin},
  {Halpern}, {Harwit}, {Hatziminaoglou}, {Heinis}, {Hurley}, {Hwang}, {Hyde},
  {Ibar}, {Ilbert}, {Isaak}, {Ivison}, {Lagache}, {Le Floc'h}, {Levenson},
  {Faro}, {Lu}, {Madden}, {Maffei}, {Magdis}, {Mainetti}, {Marchetti},
  {Marsden}, {Marshall}, {Mortier}, {Nguyen}, {O'Halloran}, {Omont}, {Page},
  {Panuzzo}, {Papageorgiou}, {Patel}, {Pearson}, {P{\'e}rez-Fournon}, {Pohlen},
  {Rawlings}, {Raymond}, {Rigopoulou}, {Riguccini}, {Rizzo}, {Rodighiero},
  {Roseboom}, {Rowan-Robinson}, {S{\'a}nchez Portal}, {Schulz}, {Scott},
  {Seymour}, {Shupe}, {Smith}, {Stevens}, {Symeonidis}, {Trichas}, {Tugwell},
  {Vaccari}, {Valtchanov}, {Vieira}, {Viero}, {Vigroux}, {Wang}, {Ward},
  {Wardlow}, {Wright}, {Xu}, \& {Zemcov}}]{oliver12}
{Oliver}, S.~J., {Bock}, J., {Altieri}, B., {et~al.} 2012, \bibinfo{title}{{The
  Herschel Multi-tiered Extragalactic Survey: HerMES},} \mnras, 424, 1614,
  \dodoi{10.1111/j.1365-2966.2012.20912.x}

\bibitem[{A. {Pensabene} {et~al.}(2021){Pensabene}, {Decarli}, {Ba{\~n}ados},
  {Venemans}, {Walter}, {Bertoldi}, {Fan}, {Farina}, {Li}, {Mazzucchelli},
  {Novak}, {Riechers}, {Rix}, {Strauss}, {Wang}, {Wei{\ss}}, {Yang}, \&
  {Yang}}]{pensabene21}
{Pensabene}, A., {Decarli}, R., {Ba{\~n}ados}, E., {et~al.} 2021,
  \bibinfo{title}{{ALMA multiline survey of the ISM in two quasar
  host-companion galaxy pairs at z > 6},} \aap, 652, A66,
  \dodoi{10.1051/0004-6361/202039696}

\bibitem[{P.~G. {P{\'e}rez-Gonz{\'a}lez}
  {et~al.}(2024){P{\'e}rez-Gonz{\'a}lez}, {Rinaldi}, {Caputi},
  {{\'A}lvarez-M{\'a}rquez}, {Annunziatella}, {Langeroodi}, {Moutard},
  {Boogaard}, {Iani}, {Melinder}, {Costantin}, {{\"O}stlin}, {Colina}, {Greve},
  {Wright}, {Alonso-Herrero}, {Bik}, {Bosman}, {Crespo G{\'o}mez}, {Dicken},
  {Eckart}, {Garc{\'\i}a-Mar{\'\i}n}, {Gillman}, {G{\"u}del}, {Henning},
  {Hjorth}, {Jermann}, {Labiano}, {Meyer}, {Pei{\ensuremath{\beta}}ker}, {Pye},
  {Ray}, {Tikkanen}, {Walter}, \& {van der Werf}}]{pg24a}
{P{\'e}rez-Gonz{\'a}lez}, P.~G., {Rinaldi}, P., {Caputi}, K.~I., {et~al.} 2024,
  \bibinfo{title}{{A NIRCam-dark Galaxy Detected with the MIRI/F1000W Filter in
  the MIDIS/JADES Hubble Ultra Deep Field},} \apjl, 969, L10,
  \dodoi{10.3847/2041-8213/ad517b}

\bibitem[{A. {Pillepich} {et~al.}(2018){Pillepich}, {Nelson}, {Hernquist},
  {Springel}, {Pakmor}, {Torrey}, {Weinberger}, {Genel}, {Naiman}, {Marinacci},
  \& {Vogelsberger}}]{pillepich18}
{Pillepich}, A., {Nelson}, D., {Hernquist}, L., {et~al.} 2018,
  \bibinfo{title}{{First results from the IllustrisTNG simulations: the stellar
  mass content of groups and clusters of galaxies},} \mnras, 475, 648,
  \dodoi{10.1093/mnras/stx3112}

\bibitem[{E. {Pizzati} {et~al.}(2024){Pizzati}, {Hennawi}, {Schaye},
  {Schaller}, {Eilers}, {Wang}, {Frenk}, {Elbers}, {Helly}, {Mackenzie},
  {Matthee}, {Bordoloi}, {Kashino}, {Naidu}, \& {Yue}}]{pizzati24a}
{Pizzati}, E., {Hennawi}, J.~F., {Schaye}, J., {et~al.} 2024,
  \bibinfo{title}{{A unified model for the clustering of quasars and galaxies
  at z {\ensuremath{\approx}} 6},} \mnras, 534, 3155,
  \dodoi{10.1093/mnras/stae2307}

\bibitem[{G. {Popping} {et~al.}(2017){Popping}, {Somerville}, \&
  {Galametz}}]{popping17b}
{Popping}, G., {Somerville}, R.~S., \& {Galametz}, M. 2017,
  \bibinfo{title}{{The dust content of galaxies from z = 0 to z = 9},} \mnras,
  471, 3152, \dodoi{10.1093/mnras/stx1545}

\bibitem[{M. {Pudoka} {et~al.}(2025){Pudoka}, {Wang}, {Fan}, {Yang},
  {Champagne}, {Zhang}, {Rojas-Ruiz}, {Ba{\~n}ados}, {Belladitta}, {Bosman},
  {Eilers}, {Jin}, {Jun}, {Li}, {Liu}, {Mazzucchelli}, {Schindler}, {Wolf}, \&
  {Wu}}]{pudoka25}
{Pudoka}, M., {Wang}, F., {Fan}, X., {et~al.} 2025,
  \bibinfo{title}{{Lyman-Break Galaxies in the Mpc-Scale Environments Around
  Three $z\sim7.5$ Quasars With JWST Imaging},} arXiv e-prints,
  arXiv:2505.07932, \dodoi{10.48550/arXiv.2505.07932}

\bibitem[{D.~A. {Riechers} {et~al.}(2013){Riechers}, {Bradford}, {Clements},
  {Dowell}, {P{\'e}rez-Fournon}, {Ivison}, {Bridge}, {Conley}, {Fu}, {Vieira},
  {Wardlow}, {Calanog}, {Cooray}, {Hurley}, {Neri}, {Kamenetzky}, {Aguirre},
  {Altieri}, {Arumugam}, {Benford}, {B{\'e}thermin}, {Bock}, {Burgarella},
  {Cabrera-Lavers}, {Chapman}, {Cox}, {Dunlop}, {Earle}, {Farrah}, {Ferrero},
  {Franceschini}, {Gavazzi}, {Glenn}, {Solares}, {Gurwell}, {Halpern},
  {Hatziminaoglou}, {Hyde}, {Ibar}, {Kov{\'a}cs}, {Krips}, {Lupu}, {Maloney},
  {Martinez-Navajas}, {Matsuhara}, {Murphy}, {Naylor}, {Nguyen}, {Oliver},
  {Omont}, {Page}, {Petitpas}, {Rangwala}, {Roseboom}, {Scott}, {Smith},
  {Staguhn}, {Streblyanska}, {Thomson}, {Valtchanov}, {Viero}, {Wang},
  {Zemcov}, \& {Zmuidzinas}}]{riechers13}
{Riechers}, D.~A., {Bradford}, C.~M., {Clements}, D.~L., {et~al.} 2013,
  \bibinfo{title}{{A dust-obscured massive maximum-starburst galaxy at a
  redshift of 6.34},} \nat, 496, 329, \dodoi{10.1038/nature12050}

\bibitem[{G.~H. {Rieke} {et~al.}(2024){Rieke}, {Alberts}, {Shivaei}, {Lyu},
  {Willmer}, {P{\'e}rez-Gonz{\'a}lez}, \& {Williams}}]{riekeg24}
{Rieke}, G.~H., {Alberts}, S., {Shivaei}, I., {et~al.} 2024,
  \bibinfo{title}{{SMILES: A Prototype JWST Multiband Mid-infrared Survey},}
  \apj, 975, 83, \dodoi{10.3847/1538-4357/ad6cd2}

\bibitem[{G.~H. {Rieke} {et~al.}(2009){Rieke}, {Alonso-Herrero}, {Weiner},
  {P{\'e}rez-Gonz{\'a}lez}, {Blaylock}, {Donley}, \& {Marcillac}}]{rieke09}
{Rieke}, G.~H., {Alonso-Herrero}, A., {Weiner}, B.~J., {et~al.} 2009,
  \bibinfo{title}{{Determining Star Formation Rates for Infrared Galaxies},}
  \apj, 692, 556, \dodoi{10.1088/0004-637X/692/1/556}

\bibitem[{M.~J. {Rieke} {et~al.}(2023){Rieke}, {Kelly}, {Misselt},
  {Stansberry}, {Boyer}, {Beatty}, {Egami}, {Florian}, {Greene}, {Hainline},
  {Leisenring}, {Roellig}, {Schlawin}, {Sun}, {Tinnin}, {Williams}, {Willmer},
  {Wilson}, {Clark}, {Rohrbach}, {Brooks}, {Canipe}, {Correnti}, {DiFelice},
  {Gennaro}, {Girard}, {Hartig}, {Hilbert}, {Koekemoer}, {Nikolov}, {Pirzkal},
  {Rest}, {Robberto}, {Sunnquist}, {Telfer}, {Wu}, {Ferry}, {Lewis}, {Baum},
  {Beichman}, {Doyon}, {Dressler}, {Eisenstein}, {Ferrarese}, {Hodapp},
  {Horner}, {Jaffe}, {Johnstone}, {Krist}, {Martin}, {McCarthy}, {Meyer},
  {Rieke}, {Trauger}, \& {Young}}]{rieke23}
{Rieke}, M.~J., {Kelly}, D.~M., {Misselt}, K., {et~al.} 2023,
  \bibinfo{title}{{Performance of NIRCam on JWST in Flight},} \pasp, 135,
  028001, \dodoi{10.1088/1538-3873/acac53}

\bibitem[{P. {Rinaldi} {et~al.}(2025){Rinaldi}, {P{\'e}rez-Gonz{\'a}lez},
  {Rieke}, {Lyu}, {D'Eugenio}, {Wu}, {Carniani}, {Looser}, {Shivaei},
  {Boogaard}, {Diaz-Santos}, {Colina}, {{\"O}stlin}, {Alberts},
  {{\'A}lvarez-M{\'a}rquez}, {Annuziatella}, {Aravena}, {Bhatawdekar},
  {Bunker}, {Caputi}, {Charlot}, {Crespo G{\'o}mez}, {Curti}, {Eckart},
  {Gillman}, {Hainline}, {Kumari}, {Hjorth}, {Iani}, {Inami}, {Ji}, {Johnson},
  {Jones}, {Labiano}, {Maiolino}, {Melinder}, {Moutard}, {Pei{\ss}ker},
  {Rieke}, {Robertson}, {Scholtz}, {Tacchella}, {van der Werf}, {Walter},
  {Williams}, {Willott}, {Witstok}, {{\"U}bler}, \& {Zhu}}]{rinaldi25a}
{Rinaldi}, P., {P{\'e}rez-Gonz{\'a}lez}, P.~G., {Rieke}, G.~H., {et~al.} 2025,
  \bibinfo{title}{{Deciphering the Nature of Virgil: An Obscured AGN Lurking
  Within an Apparently Normal Lyman-{\ensuremath{\alpha}} Emitter During Cosmic
  Reionization},} arXiv e-prints, arXiv:2504.01852,
  \dodoi{10.48550/arXiv.2504.01852}

\bibitem[{B. {Robertson} {et~al.}(2024){Robertson}, {Johnson}, {Tacchella},
  {Eisenstein}, {Hainline}, {Arribas}, {Baker}, {Bunker}, {Carniani},
  {Cargile}, {Carreira}, {Charlot}, {Chevallard}, {Curti}, {Curtis-Lake},
  {D'Eugenio}, {Egami}, {Hausen}, {Helton}, {Jakobsen}, {Ji}, {Jones},
  {Maiolino}, {Maseda}, {Nelson}, {P{\'e}rez-Gonz{\'a}lez}, {Pusk{\'a}s},
  {Rieke}, {Smit}, {Sun}, {{\"U}bler}, {Whitler}, {Williams}, {Willmer},
  {Willott}, \& {Witstok}}]{robertson24}
{Robertson}, B., {Johnson}, B.~D., {Tacchella}, S., {et~al.} 2024,
  \bibinfo{title}{{Earliest Galaxies in the JADES Origins Field: Luminosity
  Function and Cosmic Star Formation Rate Density 300 Myr after the Big Bang},}
  \apj, 970, 31, \dodoi{10.3847/1538-4357/ad463d}

\bibitem[{Y. {Shen} {et~al.}(2024){Shen}, {Zhuang}, {Li}, {Burgasser}, {Fan},
  {Greene}, {Narayan}, {Shapley}, {Sun}, {Wang}, \& {Yang}}]{sheny24}
{Shen}, Y., {Zhuang}, M.-Y., {Li}, J., {et~al.} 2024, \bibinfo{title}{{NEXUS:
  the North ecliptic pole EXtragalactic Unified Survey},} arXiv e-prints,
  arXiv:2408.12713, \dodoi{10.48550/arXiv.2408.12713}

\bibitem[{L. {Silva} {et~al.}(1998){Silva}, {Granato}, {Bressan}, \&
  {Danese}}]{Silva98}
{Silva}, L., {Granato}, G.~L., {Bressan}, A., \& {Danese}, L. 1998,
  \bibinfo{title}{{Modeling the Effects of Dust on Galactic Spectral Energy
  Distributions from the Ultraviolet to the Millimeter Band},} \apj, 509, 103,
  \dodoi{10.1086/306476}

\bibitem[{J.~M. {Simpson} {et~al.}(2015){Simpson}, {Smail}, {Swinbank},
  {Chapman}, {Geach}, {Ivison}, {Thomson}, {Aretxaga}, {Blain}, {Cowley},
  {Chen}, {Coppin}, {Dunlop}, {Edge}, {Farrah}, {Ibar}, {Karim}, {Knudsen},
  {Meijerink}, {Micha{\l}owski}, {Scott}, {Spaans}, \& {van der
  Werf}}]{simpson15}
{Simpson}, J.~M., {Smail}, I., {Swinbank}, A.~M., {et~al.} 2015,
  \bibinfo{title}{{The SCUBA-2 Cosmology Legacy Survey: ALMA Resolves the
  Bright-end of the Sub-millimeter Number Counts},} \apj, 807, 128,
  \dodoi{10.1088/0004-637X/807/2/128}

\bibitem[{J.~M. {Simpson} {et~al.}(2019){Simpson}, {Smail}, {Swinbank},
  {Chapman}, {Chen}, {Geach}, {Matsuda}, {Wang}, {Wang}, {Yang}, {Ao},
  {Asquith}, {Bourne}, {Coogan}, {Coppin}, {Gullberg}, {Hine}, {Ho}, {Hwang},
  {Ivison}, {Kato}, {Lacaille}, {Lewis}, {Liu}, {Micha{\l}owski}, {Oteo},
  {Sawicki}, {Scholtz}, {Smith}, {Thomson}, \& {Wardlow}}]{simpson19}
{Simpson}, J.~M., {Smail}, I., {Swinbank}, A.~M., {et~al.} 2019,
  \bibinfo{title}{{The East Asian Observatory SCUBA-2 Survey of the COSMOS
  Field: Unveiling 1147 Bright Sub-millimeter Sources across 2.6 Square
  Degrees},} \apj, 880, 43, \dodoi{10.3847/1538-4357/ab23ff}

\bibitem[{I. {Smail} {et~al.}(1997){Smail}, {Ivison}, \& {Blain}}]{smail97}
{Smail}, I., {Ivison}, R.~J., \& {Blain}, A.~W. 1997, \bibinfo{title}{{A Deep
  Sub-millimeter Survey of Lensing Clusters: A New Window on Galaxy Formation
  and Evolution},} \apjl, 490, L5, \dodoi{10.1086/311017}

\bibitem[{G.~F. {Snyder} {et~al.}(2017){Snyder}, {Lotz}, {Rodriguez-Gomez},
  {Guimar{\~a}es}, {Torrey}, \& {Hernquist}}]{snyder17}
{Snyder}, G.~F., {Lotz}, J.~M., {Rodriguez-Gomez}, V., {et~al.} 2017,
  \bibinfo{title}{{Massive close pairs measure rapid galaxy assembly in mergers
  at high redshift},} \mnras, 468, 207, \dodoi{10.1093/mnras/stx487}

\bibitem[{M.~L. {Strandet} {et~al.}(2017){Strandet}, {Weiss}, {De Breuck},
  {Marrone}, {Vieira}, {Aravena}, {Ashby}, {B{\'e}thermin}, {Bothwell},
  {Bradford}, {Carlstrom}, {Chapman}, {Cunningham}, {Chen}, {Fassnacht},
  {Gonzalez}, {Greve}, {Gullberg}, {Hayward}, {Hezaveh}, {Litke}, {Ma},
  {Malkan}, {Menten}, {Miller}, {Murphy}, {Narayanan}, {Phadke}, {Rotermund},
  {Spilker}, \& {Sreevani}}]{strandet17}
{Strandet}, M.~L., {Weiss}, A., {De Breuck}, C., {et~al.} 2017,
  \bibinfo{title}{{ISM Properties of a Massive Dusty Star-forming Galaxy
  Discovered at z {\ensuremath{\sim}} 7},} \apjl, 842, L15,
  \dodoi{10.3847/2041-8213/aa74b0}

\bibitem[{F. {Sun} {et~al.}(2022){Sun}, {Egami}, {Fujimoto}, {Rawle}, {Bauer},
  {Kohno}, {Smail}, {P{\'e}rez-Gonz{\'a}lez}, {Ao}, {Chapman}, {Combes},
  {Dessauges-Zavadsky}, {Espada}, {Gonz{\'a}lez-L{\'o}pez}, {Koekemoer},
  {Kokorev}, {Lee}, {Morokuma-Matsui}, {Mu{\~n}oz Arancibia}, {Oguri},
  {Pell{\'o}}, {Ueda}, {Uematsu}, {Valentino}, {Van der Werf}, {Walth},
  {Zemcov}, \& {Zitrin}}]{sunf22a}
{Sun}, F., {Egami}, E., {Fujimoto}, S., {et~al.} 2022, \bibinfo{title}{{ALMA
  Lensing Cluster Survey: ALMA-Herschel Joint Study of Lensed Dusty
  Star-forming Galaxies across z ≃ 0.5 - 6},} \apj, 932, 77,
  \dodoi{10.3847/1538-4357/ac6e3f}

\bibitem[{F. {Sun} {et~al.}(2024){Sun}, {Helton}, {Egami}, {Hainline}, {Rieke},
  {Willmer}, {Eisenstein}, {Johnson}, {Rieke}, {Robertson}, {Tacchella},
  {Alberts}, {Baker}, {Bhatawdekar}, {Boyett}, {Bunker}, {Charlot}, {Chen},
  {Chevallard}, {Curtis-Lake}, {Danhaive}, {DeCoursey}, {Ji}, {Lyu},
  {Maiolino}, {Rujopakarn}, {Sandles}, {Shivaei}, {{\"U}bler}, {Willott}, \&
  {Witstok}}]{sunf24}
{Sun}, F., {Helton}, J.~M., {Egami}, E., {et~al.} 2024, \bibinfo{title}{{JADES:
  Resolving the Stellar Component and Filamentary Overdense Environment of
  Hubble Space Telescope (HST)-dark Submillimeter Galaxy HDF850.1 at z =
  5.18},} \apj, 961, 69, \dodoi{10.3847/1538-4357/ad07e3}

\bibitem[{F. {Sun} {et~al.}(2025{\natexlab{a}}){Sun}, {Wang}, {Yang},
  {Champagne}, {Decarli}, {Fan}, {Ba{\~n}ados}, {Cai}, {Colina}, {Egami},
  {Hennawi}, {Jin}, {Jun}, {Khusanova}, {Li}, {Li}, {Lin}, {Liu}, {Meyer},
  {Pudoka}, {Rieke}, {Shen}, {Tee}, {Venemans}, {Walter}, {Wu}, {Zhang}, \&
  {Zou}}]{sunf25a}
{Sun}, F., {Wang}, F., {Yang}, J., {et~al.} 2025{\natexlab{a}},
  \bibinfo{title}{{A SPectroscopic Survey of Biased Halos in the Reionization
  Era (ASPIRE): Spectroscopically Complete Census of Obscured Cosmic Star
  Formation Rate Density at z = 4{\textendash}6},} \apj, 980, 12,
  \dodoi{10.3847/1538-4357/ad9d0e}

\bibitem[{F. {Sun} {et~al.}(2025{\natexlab{b}}){Sun}, {Fudamoto}, {Lin},
  {Helton}, {Hsiao}, {Egami}, {Akhtarkavan}, {Bunker}, {Cai}, {DeCoursey},
  {Eisenstein}, {Fan}, {Harikane}, {Ji}, {Jin}, {Liu}, {Liu}, {Ma}, {Maiolino},
  {Ouchi}, {Tee}, {Wang}, {Willmer}, {Wu}, {Xu}, {Yang}, {Zhang}, \&
  {Zhu}}]{sunf25b}
{Sun}, F., {Fudamoto}, Y., {Lin}, X., {et~al.} 2025{\natexlab{b}},
  \bibinfo{title}{{Slitless Areal Pure-Parallel HIgh-Redshift Emission Survey
  (SAPPHIRES): Early Data Release of Deep JWST/NIRCam Images and Spectra in
  MACS J0416 Parallel Field},} arXiv e-prints, arXiv:2503.15587,
  \dodoi{10.48550/arXiv.2503.15587}

\bibitem[{Y. {Tamura} {et~al.}(2019){Tamura}, {Mawatari}, {Hashimoto}, {Inoue},
  {Zackrisson}, {Christensen}, {Binggeli}, {Matsuda}, {Matsuo}, {Takeuchi},
  {Asano}, {Sunaga}, {Shimizu}, {Okamoto}, {Yoshida}, {Lee}, {Shibuya},
  {Taniguchi}, {Umehata}, {Hatsukade}, {Kohno}, \& {Ota}}]{tamura19}
{Tamura}, Y., {Mawatari}, K., {Hashimoto}, T., {et~al.} 2019,
  \bibinfo{title}{{Detection of the Far-infrared [O III] and Dust Emission in a
  Galaxy at Redshift 8.312: Early Metal Enrichment in the Heart of the
  Reionization Era},} \apj, 874, 27, \dodoi{10.3847/1538-4357/ab0374}

\bibitem[{S. {Toft} {et~al.}(2014){Toft}, {Smol{\v{c}}i{\'c}}, {Magnelli},
  {Karim}, {Zirm}, {Michalowski}, {Capak}, {Sheth}, {Schawinski}, {Krogager},
  {Wuyts}, {Sanders}, {Man}, {Lutz}, {Staguhn}, {Berta}, {Mccracken}, {Krpan},
  \& {Riechers}}]{toft14}
{Toft}, S., {Smol{\v{c}}i{\'c}}, V., {Magnelli}, B., {et~al.} 2014,
  \bibinfo{title}{{Submillimeter Galaxies as Progenitors of Compact Quiescent
  Galaxies},} \apj, 782, 68, \dodoi{10.1088/0004-637X/782/2/68}

\bibitem[{C. {Turner} {et~al.}(2025){Turner}, {Tacchella}, {D'Eugenio},
  {Carniani}, {Curti}, {Glazebrook}, {Johnson}, {Lim}, {Looser}, {Maiolino},
  {Nanayakkara}, \& {Wan}}]{turner25a}
{Turner}, C., {Tacchella}, S., {D'Eugenio}, F., {et~al.} 2025,
  \bibinfo{title}{{Age-dating early quiescent galaxies: high star formation
  efficiency, but consistent with direct, higher-redshift observations},}
  \mnras, 537, 1826, \dodoi{10.1093/mnras/staf128}

\bibitem[{F. {Valentino} {et~al.}(2023){Valentino}, {Brammer}, {Gould},
  {Kokorev}, {Fujimoto}, {Jespersen}, {Vijayan}, {Weaver}, {Ito}, {Tanaka},
  {Ilbert}, {Magdis}, {Whitaker}, {Faisst}, {Gallazzi}, {Gillman},
  {Gim{\'e}nez-Arteaga}, {G{\'o}mez-Guijarro}, {Kubo}, {Heintz}, {Hirschmann},
  {Oesch}, {Onodera}, {Rizzo}, {Lee}, {Strait}, \& {Toft}}]{valentino23}
{Valentino}, F., {Brammer}, G., {Gould}, K. M.~L., {et~al.} 2023,
  \bibinfo{title}{{An Atlas of Color-selected Quiescent Galaxies at z > 3 in
  Public JWST Fields},} \apj, 947, 20, \dodoi{10.3847/1538-4357/acbefa}

\bibitem[{I.~F. {van Leeuwen} {et~al.}(2024){van Leeuwen}, {Bouwens}, {van der
  Werf}, {Hodge}, {Schouws}, {Stefanon}, {Algera}, {Aravena}, {Boogaard},
  {Bowler}, {da Cunha}, {Dayal}, {Decarli}, {Gonzalez}, {Inami}, {de Looze},
  {Sommovigo}, {Venemans}, {Walter}, {Barrufet}, {Ferrara}, {Graziani},
  {Hygate}, {Oesch}, {Palla}, {Rowland}, \& {Schneider}}]{vanleeuwen24}
{van Leeuwen}, I.~F., {Bouwens}, R.~J., {van der Werf}, P.~P., {et~al.} 2024,
  \bibinfo{title}{{Characterising the contribution of dust-obscured star
  formation at z {\ensuremath{\gtrsim}} 5 using 18 serendipitously identified
  [C II] emitters},} \mnras, 534, 2062, \dodoi{10.1093/mnras/stae2171}

\bibitem[{B.~P. {Venemans} {et~al.}(2019){Venemans}, {Neeleman}, {Walter},
  {Novak}, {Decarli}, {Hennawi}, \& {Rix}}]{venemans19}
{Venemans}, B.~P., {Neeleman}, M., {Walter}, F., {et~al.} 2019,
  \bibinfo{title}{{400 pc Imaging of a Massive Quasar Host Galaxy at a Redshift
  of 6.6},} \apjl, 874, L30, \dodoi{10.3847/2041-8213/ab11cc}

\bibitem[{B.~P. {Venemans} {et~al.}(2020){Venemans}, {Walter}, {Neeleman},
  {Novak}, {Otter}, {Decarli}, {Ba{\~n}ados}, {Drake}, {Farina}, {Kaasinen},
  {Mazzucchelli}, {Carilli}, {Fan}, {Rix}, \& {Wang}}]{venemans20}
{Venemans}, B.~P., {Walter}, F., {Neeleman}, M., {et~al.} 2020,
  \bibinfo{title}{{Kiloparsec-scale ALMA Imaging of [C II] and Dust Continuum
  Emission of 27 Quasar Host Galaxies at z {\ensuremath{\sim}} 6},} \apj, 904,
  130, \dodoi{10.3847/1538-4357/abc563}

\bibitem[{J.~D. {Vieira} {et~al.}(2010){Vieira}, {Crawford}, {Switzer}, {Ade},
  {Aird}, {Ashby}, {Benson}, {Bleem}, {Brodwin}, {Carlstrom}, {Chang}, {Cho},
  {Crites}, {de Haan}, {Dobbs}, {Everett}, {George}, {Gladders}, {Hall},
  {Halverson}, {High}, {Holder}, {Holzapfel}, {Hrubes}, {Joy}, {Keisler},
  {Knox}, {Lee}, {Leitch}, {Lueker}, {Marrone}, {McIntyre}, {McMahon}, {Mehl},
  {Meyer}, {Mohr}, {Montroy}, {Padin}, {Plagge}, {Pryke}, {Reichardt}, {Ruhl},
  {Schaffer}, {Shaw}, {Shirokoff}, {Spieler}, {Stalder}, {Staniszewski},
  {Stark}, {Vanderlinde}, {Walsh}, {Williamson}, {Yang}, {Zahn}, \&
  {Zenteno}}]{vieira10}
{Vieira}, J.~D., {Crawford}, T.~M., {Switzer}, E.~R., {et~al.} 2010,
  \bibinfo{title}{{Extragalactic Millimeter-wave Sources in South Pole
  Telescope Survey Data: Source Counts, Catalog, and Statistics for an 87
  Square-degree Field},} \apj, 719, 763, \dodoi{10.1088/0004-637X/719/1/763}

\bibitem[{F. {Walter} {et~al.}(2012){Walter}, {Decarli}, {Carilli}, {Bertoldi},
  {Cox}, {da Cunha}, {Daddi}, {Dickinson}, {Downes}, {Elbaz}, {Ellis}, {Hodge},
  {Neri}, {Riechers}, {Weiss}, {Bell}, {Dannerbauer}, {Krips}, {Krumholz},
  {Lentati}, {Maiolino}, {Menten}, {Rix}, {Robertson}, {Spinrad}, {Stark}, \&
  {Stern}}]{walter12}
{Walter}, F., {Decarli}, R., {Carilli}, C., {et~al.} 2012, \bibinfo{title}{{The
  intense starburst HDF{\,}850.1 in a galaxy overdensity at
  z{\,}{\ensuremath{\approx}}{\,}5.2 in the Hubble Deep Field},} \nat, 486,
  233, \dodoi{10.1038/nature11073}

\bibitem[{F. {Wang} {et~al.}(2019){Wang}, {Yang}, {Fan}, {Wu}, {Yue}, {Li},
  {Bian}, {Jiang}, {Ba{\~n}ados}, {Schindler}, {Findlay}, {Davies}, {Decarli},
  {Farina}, {Green}, {Hennawi}, {Huang}, {Mazzuccheli}, {McGreer}, {Venemans},
  {Walter}, {Dye}, {Lyke}, {Myers}, \& {Nunez}}]{wangf19a}
{Wang}, F., {Yang}, J., {Fan}, X., {et~al.} 2019, \bibinfo{title}{{Exploring
  Reionization-era Quasars. III. Discovery of 16 Quasars at 6.4
  {\ensuremath{\lesssim}} z {\ensuremath{\lesssim}} 6.9 with DESI Legacy
  Imaging Surveys and the UKIRT Hemisphere Survey and Quasar Luminosity
  Function at z {\ensuremath{\sim}} 6.7},} \apj, 884, 30,
  \dodoi{10.3847/1538-4357/ab2be5}

\bibitem[{F. {Wang} {et~al.}(2021){Wang}, {Fan}, {Yang}, {Mazzucchelli}, {Wu},
  {Li}, {Ba{\~n}ados}, {Farina}, {Nanni}, {Ai}, {Bian}, {Davies}, {Decarli},
  {Hennawi}, {Schindler}, {Venemans}, \& {Walter}}]{wangf21}
{Wang}, F., {Fan}, X., {Yang}, J., {et~al.} 2021, \bibinfo{title}{{Revealing
  the Accretion Physics of Supermassive Black Holes at Redshift z
  {\ensuremath{\sim}} 7 with Chandra and Infrared Observations},} \apj, 908,
  53, \dodoi{10.3847/1538-4357/abcc5e}

\bibitem[{F. {Wang} {et~al.}(2023){Wang}, {Yang}, {Hennawi}, {Fan}, {Sun},
  {Champagne}, {Costa}, {Habouzit}, {Endsley}, {Li}, {Lin}, {Meyer},
  {Schindler}, {Wu}, {Ba{\~n}ados}, {Barth}, {Bhowmick}, {Bieri}, {Blecha},
  {Bosman}, {Cai}, {Colina}, {Connor}, {Davies}, {Decarli}, {De Rosa}, {Drake},
  {Egami}, {Eilers}, {Evans}, {Farina}, {Haiman}, {Jiang}, {Jin}, {Jun},
  {Kakiichi}, {Khusanova}, {Kulkarni}, {Li}, {Liu}, {Loiacono}, {Lupi},
  {Mazzucchelli}, {Onoue}, {Pudoka}, {Rojas-Ruiz}, {Shen}, {Strauss}, {Tee},
  {Trakhtenbrot}, {Trebitsch}, {Venemans}, {Volonteri}, {Walter}, {Xie}, {Yue},
  {Zhang}, {Zhang}, \& {Zou}}]{wangf23}
{Wang}, F., {Yang}, J., {Hennawi}, J.~F., {et~al.} 2023, \bibinfo{title}{{A
  SPectroscopic Survey of Biased Halos in the Reionization Era (ASPIRE): JWST
  Reveals a Filamentary Structure around a z = 6.61 Quasar},} \apjl, 951, L4,
  \dodoi{10.3847/2041-8213/accd6f}

\bibitem[{F. {Wang} {et~al.}(in prep.){Wang}, {Yang}, {Hennawi}, {Fan}, {Sun},
  {Champagne}, {Costa}, {Habouzit}, {Endsley}, {Li}, {Lin}, {Meyer},
  {Schindler}, {Wu}, {Ba{\~n}ados}, {Barth}, {Bhowmick}, {Bieri}, {Blecha},
  {Bosman}, {Cai}, {Colina}, {Connor}, {Davies}, {Decarli}, {De Rosa}, {Drake},
  {Egami}, {Eilers}, {Evans}, {Farina}, {Haiman}, {Jiang}, {Jin}, {Jun},
  {Kakiichi}, {Khusanova}, {Kulkarni}, {Li}, {Liu}, {Loiacono}, {Lupi},
  {Mazzucchelli}, {Onoue}, {Pudoka}, {Rojas-Ruiz}, {Shen}, {Strauss}, {Tee},
  {Trakhtenbrot}, {Trebitsch}, {Venemans}, {Volonteri}, {Walter}, {Xie}, {Yue},
  {Zhang}, {Zhang}, \& {Zou}}]{wangfprep}
{Wang}, F., {Yang}, J., {Hennawi}, J.~F., {et~al.} in prep., \bibinfo{title}{{A
  SPectroscopic Survey of Biased Halos in the Reionization Era (ASPIRE):
  Program Overview},}

\bibitem[{T. {Wang} {et~al.}(2019){Wang}, {Schreiber}, {Elbaz}, {Yoshimura},
  {Kohno}, {Shu}, {Yamaguchi}, {Pannella}, {Franco}, {Huang}, {Lim}, \&
  {Wang}}]{wangt19}
{Wang}, T., {Schreiber}, C., {Elbaz}, D., {et~al.} 2019, \bibinfo{title}{{A
  dominant population of optically invisible massive galaxies in the early
  Universe},} \nat, 572, 211, \dodoi{10.1038/s41586-019-1452-4}

\bibitem[{D. {Watson} {et~al.}(2015){Watson}, {Christensen}, {Knudsen},
  {Richard}, {Gallazzi}, \& {Micha{\l}owski}}]{watson15}
{Watson}, D., {Christensen}, L., {Knudsen}, K.~K., {et~al.} 2015,
  \bibinfo{title}{{A dusty, normal galaxy in the epoch of reionization},} \nat,
  519, 327, \dodoi{10.1038/nature14164}

\bibitem[{L. {Whitler} {et~al.}(2025){Whitler}, {Stark}, {Topping},
  {Robertson}, {Rieke}, {Hainline}, {Endsley}, {Chen}, {Baker}, {Bhatawdekar},
  {Bunker}, {Carniani}, {Charlot}, {Chevallard}, {Curtis-Lake}, {Egami},
  {Eisenstein}, {Helton}, {Ji}, {Johnson}, {P{\'e}rez-Gonz{\'a}lez}, {Rinaldi},
  {Tacchella}, {Williams}, {Willmer}, {Willott}, \& {Witstok}}]{whitler25}
{Whitler}, L., {Stark}, D.~P., {Topping}, M.~W., {et~al.} 2025,
  \bibinfo{title}{{The $z rsim 9$ galaxy UV luminosity function from the JWST
  Advanced Deep Extragalactic Survey: insights into early galaxy evolution and
  reionization},} arXiv e-prints, arXiv:2501.00984,
  \dodoi{10.48550/arXiv.2501.00984}

\bibitem[{C.~C. {Williams} {et~al.}(2024){Williams}, {Alberts}, {Ji},
  {Hainline}, {Lyu}, {Rieke}, {Endsley}, {Suess}, {Sun}, {Johnson}, {Florian},
  {Shivaei}, {Rujopakarn}, {Baker}, {Bhatawdekar}, {Boyett}, {Bunker},
  {Cameron}, {Carniani}, {Charlot}, {Curtis-Lake}, {DeCoursey}, {de Graaff},
  {Egami}, {Eisenstein}, {Gibson}, {Hausen}, {Helton}, {Maiolino}, {Maseda},
  {Nelson}, {P{\'e}rez-Gonz{\'a}lez}, {Rieke}, {Robertson}, {Saxena},
  {Tacchella}, {Willmer}, \& {Willott}}]{williams24}
{Williams}, C.~C., {Alberts}, S., {Ji}, Z., {et~al.} 2024, \bibinfo{title}{{The
  Galaxies Missed by Hubble and ALMA: The Contribution of Extremely Red
  Galaxies to the Cosmic Census at 3 < z < 8},} \apj, 968, 34,
  \dodoi{10.3847/1538-4357/ad3f17}

\bibitem[{Y. {Yamaguchi} {et~al.}(2019){Yamaguchi}, {Kohno}, {Hatsukade},
  {Wang}, {Yoshimura}, {Ao}, {Caputi}, {Dunlop}, {Egami}, {Espada}, {Fujimoto},
  {Hayatsu}, {Ivison}, {Kodama}, {Kusakabe}, {Nagao}, {Ouchi}, {Rujopakarn},
  {Tadaki}, {Tamura}, {Ueda}, {Umehata}, {Wang}, \& {Yun}}]{yamaguchi19}
{Yamaguchi}, Y., {Kohno}, K., {Hatsukade}, B., {et~al.} 2019,
  \bibinfo{title}{{ALMA 26 arcmin$^{2}$ Survey of GOODS-S at 1 mm (ASAGAO):
  Near-infrared-dark Faint ALMA Sources},} \apj, 878, 73,
  \dodoi{10.3847/1538-4357/ab0d22}

\bibitem[{J. {Yang} {et~al.}(2023){Yang}, {Wang}, {Fan}, {Hennawi}, {Barth},
  {Ba{\~n}ados}, {Sun}, {Liu}, {Cai}, {Jiang}, {Li}, {Onoue}, {Schindler},
  {Shen}, {Wu}, {Bhowmick}, {Bieri}, {Blecha}, {Bosman}, {Champagne}, {Colina},
  {Connor}, {Costa}, {Davies}, {Decarli}, {De Rosa}, {Drake}, {Egami},
  {Eilers}, {Evans}, {Farina}, {Habouzit}, {Haiman}, {Jin}, {Jun}, {Kakiichi},
  {Khusanova}, {Kulkarni}, {Loiacono}, {Lupi}, {Mazzucchelli}, {Pan},
  {Rojas-Ruiz}, {Strauss}, {Tee}, {Trakhtenbrot}, {Trebitsch}, {Venemans},
  {Vestergaard}, {Volonteri}, {Walter}, {Xie}, {Yue}, {Zhang}, {Zhang}, \&
  {Zou}}]{yangj23}
{Yang}, J., {Wang}, F., {Fan}, X., {et~al.} 2023, \bibinfo{title}{{A
  SPectroscopic Survey of Biased Halos in the Reionization Era (ASPIRE): A
  First Look at the Rest-frame Optical Spectra of z > 6.5 Quasars Using JWST},}
  \apjl, 951, L5, \dodoi{10.3847/2041-8213/acc9c8}

\bibitem[{J. {Yang} {et~al.}(in prep.){Yang}, {Wang}, {Hennawi}, {Fan}, {Sun},
  {Champagne}, {Costa}, {Habouzit}, {Endsley}, {Li}, {Lin}, {Meyer},
  {Schindler}, {Wu}, {Ba{\~n}ados}, {Barth}, {Bhowmick}, {Bieri}, {Blecha},
  {Bosman}, {Cai}, {Colina}, {Connor}, {Davies}, {Decarli}, {De Rosa}, {Drake},
  {Egami}, {Eilers}, {Evans}, {Farina}, {Haiman}, {Jiang}, {Jin}, {Jun},
  {Kakiichi}, {Khusanova}, {Kulkarni}, {Li}, {Liu}, {Loiacono}, {Lupi},
  {Mazzucchelli}, {Onoue}, {Pudoka}, {Rojas-Ruiz}, {Shen}, {Strauss}, {Tee},
  {Trakhtenbrot}, {Trebitsch}, {Venemans}, {Volonteri}, {Walter}, {Xie}, {Yue},
  {Zhang}, {Zhang}, \& {Zou}}]{yangjprep}
{Yang}, J., {Wang}, F., {Hennawi}, J.~F., {et~al.} in prep., \bibinfo{title}{{A
  SPectroscopic Survey of Biased Halos in the Reionization Era (ASPIRE): Quasar
  Host},}

\bibitem[{J.~A. {Zavala} {et~al.}(2018){Zavala}, {Monta{\~n}a}, {Hughes},
  {Yun}, {Ivison}, {Valiante}, {Wilner}, {Spilker}, {Aretxaga}, {Eales},
  {Avila-Reese}, {Ch{\'a}vez}, {Cooray}, {Dannerbauer}, {Dunlop}, {Dunne},
  {G{\'o}mez-Ruiz}, {Micha{\l}owski}, {Narayanan}, {Nayyeri}, {Oteo}, {Rosa
  Gonz{\'a}lez}, {S{\'a}nchez-Arg{\"u}elles}, {Schloerb}, {Serjeant}, {Smith},
  {Terlevich}, {Vega}, {Villalba}, {van der Werf}, {Wilson}, \&
  {Zeballos}}]{zavala18}
{Zavala}, J.~A., {Monta{\~n}a}, A., {Hughes}, D.~H., {et~al.} 2018,
  \bibinfo{title}{{A dusty star-forming galaxy at z = 6 revealed by strong
  gravitational lensing},} Nature Astronomy, 2, 56,
  \dodoi{10.1038/s41550-017-0297-8}

\bibitem[{J.~A. {Zavala} {et~al.}(2025){Zavala}, {Castellano}, {Akins}, {Bakx},
  {Burgarella}, {Casey}, {Ch{\'a}vez Ortiz}, {Dickinson}, {Finkelstein},
  {Mitsuhashi}, {Nakajima}, {P{\'e}rez-Gonz{\'a}lez}, {Arrabal Haro},
  {Bergamini}, {Buat}, {Backhaus}, {Calabr{\`o}}, {Cleri},
  {Fern{\'a}ndez-Arenas}, {Fontana}, {Franco}, {Grillo}, {Giavalisco},
  {Grogin}, {Hathi}, {Hirschmann}, {Ikeda}, {Jung}, {Kartaltepe}, {Koekemoer},
  {Larson}, {McKinney}, {Papovich}, {Rosati}, {Saito}, {Santini}, {Terlevich},
  {Terlevich}, {Treu}, \& {Yung}}]{zavala25}
{Zavala}, J.~A., {Castellano}, M., {Akins}, H.~B., {et~al.} 2025,
  \bibinfo{title}{{A luminous and young galaxy at z = 12.33 revealed by a
  JWST/MIRI detection of H{\ensuremath{\alpha}} and [O III]},} Nature
  Astronomy, 9, 155, \dodoi{10.1038/s41550-024-02397-3}

\bibitem[{F. {Ziparo} {et~al.}(2023){Ziparo}, {Ferrara}, {Sommovigo}, \&
  {Kohandel}}]{ziparo23}
{Ziparo}, F., {Ferrara}, A., {Sommovigo}, L., \& {Kohandel}, M. 2023,
  \bibinfo{title}{{Blue monsters. Why are JWST super-early, massive galaxies so
  blue?},} \mnras, 520, 2445, \dodoi{10.1093/mnras/stad125}

\bibitem[{A. {Zitrin} {et~al.}(2015){Zitrin}, {Labb{\'e}}, {Belli}, {Bouwens},
  {Ellis}, {Roberts-Borsani}, {Stark}, {Oesch}, \& {Smit}}]{zitrin15}
{Zitrin}, A., {Labb{\'e}}, I., {Belli}, S., {et~al.} 2015,
  \bibinfo{title}{{Lyman{\ensuremath{\alpha}} Emission from a Luminous z = 8.68
  Galaxy: Implications for Galaxies as Tracers of Cosmic Reionization},} \apjl,
  810, L12, \dodoi{10.1088/2041-8205/810/1/L12}

\end{thebibliography}
\bibliographystyle{aasjournalv7}




\end{document}